\pdfoutput=1
\documentclass[journal=jcisd8,manuscript=article]{achemso}

\usepackage{amsmath}
\usepackage{amssymb}
\usepackage{graphicx}
\usepackage{bm}
\usepackage[version=4]{mhchem}
\usepackage{algpseudocode}
\usepackage{booktabs}

\makeatletter
\providecommand{\@mcitecorrectmaxwidthbibitem}{0}
\providecommand{\@mcitecorrectmaxwidthsubitem}{0}
\makeatother

\csname mciteErrorOnUnknownfalse\endcsname
\AtBeginDocument{\singlespacing}

\SectionNumbersOn

\title{Shape-Constrained Bayesian Active Learning of Self-Limiting Saturation Curves}

\author{Pouyan Navabi}
\affiliation{Department of Chemistry, University of Illinois Chicago, Chicago, Illinois 60607, United States}

\author{Christos G. Takoudis}
\email{takoudis@uic.edu}
\affiliation{Department of Chemistry, University of Illinois Chicago, Chicago, Illinois 60607, United States}
\alsoaffiliation{Department of Chemical Engineering, University of Illinois Chicago, Chicago, Illinois 60607, United States}
\alsoaffiliation{Department of Biomedical Engineering, University of Illinois Chicago, Chicago, Illinois 60607, United States}

\keywords{saturation curves, active learning, Bayesian inference, monotone regression, I-splines, adsorption kinetics}

\begin{document}

\begin{abstract}
Saturation curves, monotone responses that rise from zero to a plateau, recur across gas adsorption, enzyme kinetics, and self-limiting growth. Learning one from a few costly, noisy measurements is a common bottleneck. We present a Bayesian active-learning platform built on a penalized monotone I-spline. Non-negative weights make every posterior draw rise from zero and never fall, a survey measurement anchors the plateau in place of a flatness prior, and the input reaching any saturation level is read from the posterior by level crossing, assuming no kinetic form. Each measurement is placed where the surrogate is least certain, which beats random designs and edges a log-spaced grid of the same budget. One fixed configuration recovers five kinetically distinct regimes from twenty measurements to a saturation error of $0.037$ to $0.041$, below the output noise, and averages $0.045$ on $250$ curves from an unseen distribution. Running unconstrained Gaussian processes (GPs) and two unpenalized monotone bases through the same loop separates each ingredient's contribution. Logarithmic resolution and the roughness penalty carry the accuracy, while monotonicity guarantees a level crossing unique on every draw. The penalty keeps the surrogate the most accurate model at every noise level, and its second-order form centers the recommended input for saturation where a first-order penalty reads late (a bias of $+0.32$~s with $94\%$ coverage, against $+0.44$~s and $64\%$ for the matched unconstrained GP). Because the recommended input lies where the curve is nearly flat, no method can pin it down precisely, and we report it as a cost-indexed quantile with the probability of falling short.
\end{abstract}

\section{Introduction}

Many measured responses in the physical and life sciences are \emph{saturation curves} that rise monotonically from zero as an input is increased, then level off as the system self-limits. Surface coverage versus pressure in ideal monolayer gas adsorption follows the Langmuir isotherm.\cite{1} Enzyme reaction rate versus substrate concentration follows Michaelis--Menten kinetics.\cite{2} Receptor occupancy versus ligand concentration follows the Hill equation.\cite{3} Sensor and detector outputs approach a plateau as the stimulus grows. The response of a self-limiting deposition saturates with reagent exposure.\cite{4,5} In each case the same questions recur, namely the shape of the curve, the input at which the response reaches a chosen fraction of saturation, and, because each measurement is costly, how a limited experimental budget is best spent answering them. A method that learns the curve data-efficiently while respecting its defining structure, monotonicity and a self-limiting plateau, and that chooses each next measurement from its own uncertainty instead of from a fixed schedule, is therefore broadly useful. Choosing the next input from the current fit also closes the experimental loop, a step toward autonomous, self-driving laboratories\cite{6,7} in which the instrument proposes and runs its own experiments. In such a loop no operator has to stop between runs to read the curve and pick the next input, and a surrogate that pools noisy points lets each point be a short run, so with in situ metrology the campaign can finish in a fraction of the reactor time a hand-measured curve takes. How such a campaign differs from a curve measured by hand is set out for atomic layer deposition (ALD) in Supporting Information (SI) Sec.~\ref{sec:si-cost}, and what the recommended input then costs in production in SI Sec.~\ref{sec:si-quantile}.

Such loops have moved in the past few years from proposals to working systems. A community perspective describes autonomous experimentation platforms in which a machine-learning model selects each experiment from the results so far,\cite{8} Bayesian optimization (BO) has reached better reaction outcomes than expert chemists with fewer experiments,\cite{9} and Bayesian design of experiments is advocated within catalysis research itself.\cite{10} Thin-film deposition has followed the same path. Gaussian-process (GP) BO has tuned the plasma pretreatment, temperature, and pulse time of an ALD process in a few iterations,\cite{11} neural networks trained on simulated reactor growth profiles have predicted the dose time that saturates growth across a cross-flow reactor,\cite{12} BO has steered sputter deposition toward target film properties in the presence of process fluctuations,\cite{13} agents built on reasoning language models have been set to find the saturating dose times of a simulated ALD process without being told that it is self-limiting,\cite{14} and recent perspectives survey what machine learning can and cannot yet do for ALD process development.\cite{15,16} The optimizers in these studies are generic GPs, black-box networks, or language models, and the monotone, self-limiting structure of the saturation curve is not part of them.

A natural surrogate for such a curve embeds a parametric kinetic form into the covariance kernel of a GP and selects each next input where the surrogate is least certain. Our own earlier active-learning loop built a Langmuir adsorption model into the kernel in exactly this way.\cite{17} That established the premise on which the present work rests, that a saturation curve can be recovered data-efficiently from a handful of well-chosen measurements, but it also exposed the limitations we set out to remove. Building one kinetic form into the kernel commits the surrogate to a single regime, so a response whose uptake is sigmoidal, logarithmic, or dispersive instead of Langmuir-like is fit with the wrong curvature. Even within one family, a warped stationary kernel does not guarantee that every posterior draw is non-decreasing. A natural halfway step is to keep the Langmuir form but allow a spread of rate constants, since first-order kinetics on sites with different rates gives a non-negative mixture of saturating exponentials, a family that Bernstein's theorem\cite{18} characterizes exactly and that also serves, as a sum of Langmuir pathways, as the simulated process in the agent study cited above.\cite{14} Such a mixture has the guarantees we want, a pinned origin and monotonicity for any non-negative weights, but its rate is steepest at the origin and only falls, so it reaches only concave curves and a cooperative onset with an inflection lies outside it. The monotone spline basis used here represents those mixtures to within its approximation error and also reaches the sigmoidal and kinked shapes that lie outside them, so we adopt it as the surrogate and use exponential mixtures only to generate held-out test curves (Sec.~\ref{sec:heldout}). The two shortcomings of the kernel and the narrowness of the mixture together motivate the model-free, monotone surrogate developed here.

\begin{figure}[htbp]
\centering
\includegraphics[width=\textwidth]{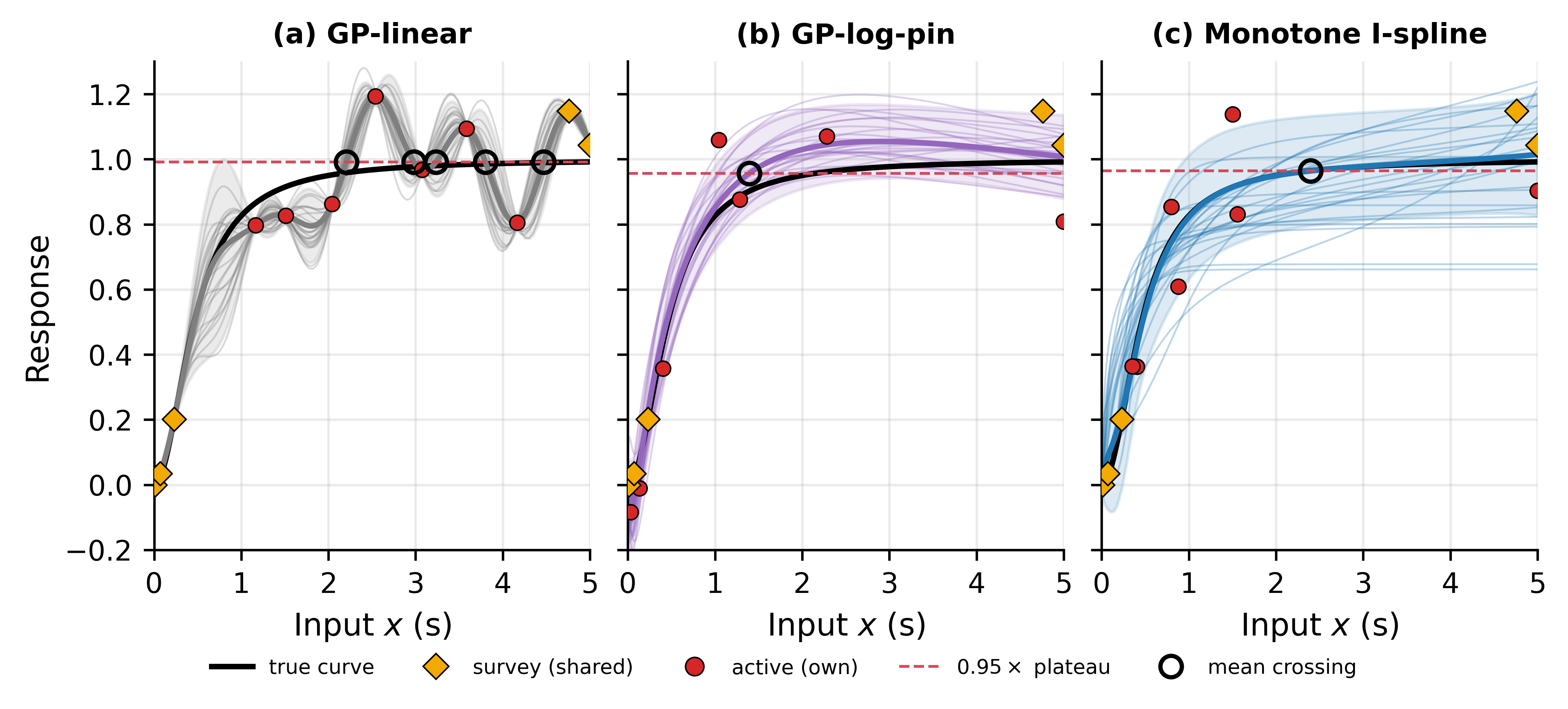}
\caption{The recommended input is read off the curve by level crossing, and only the monotone surrogate keeps that read-out well defined, as shown here on a sigmoidal (Sips) saturation curve. Each surrogate runs its own active-learning loop from the same five-point survey (gold diamonds), and its seven active points (red) are chosen by its own uncertainty sampling. Each panel shows the true curve (black), the posterior mean (bold), the $95\%$ band (shaded), a sample of posterior draws (thin), and $0.95\times$ that surrogate's own inferred plateau (dashed) with the mean's crossings of it (rings). (a) An unconstrained GP on the raw input axis (GP-linear, Sec.~\ref{sec:comparators}) wiggles across the plateau and meets the level at several different inputs, so the read-out is multi-valued. (b) The resolution- and origin-matched GP (GP-log-pin) overshoots, and its draws visibly turn back down, so the posterior is non-monotone. (c) For the monotone I-spline, every draw rises from the pinned origin and crosses the level exactly once, so the read-out is always unique.}
\label{fig:gp_compare}
\end{figure}

GP regression is the usual surrogate for such loops because it provides analytic uncertainty.\cite{9,19} Stationary kernels, however, encode no monotonicity, so a GP fit to sparse, noisy saturation data frequently dips between observations (Fig.~\ref{fig:gp_compare}). Such dips corrupt one of the quantities read from the curve, the input at a chosen saturation level, because that input is read from the posterior one draw at a time, as the crossing of the level, and a draw that dips can cross it several times. Run through the loop used here, the unconstrained mean is non-monotone in most fits, and the loop places $12$--$46\%$ of its active measurements where its own mean is falling (SI Sec.~\ref{sec:si-gpfail}). Those misled selections cost little curve accuracy (Sec.~\ref{sec:whymono}), and the main problem is the read-out. The GP also cannot be repaired afterward, because isotonic projection of the mean leaves step artifacts,\cite{20} a running-maximum guard can propagate one upward excursion across the domain, and projecting the draws onto the monotone cone\cite{21} keeps the GP's overconfident spread, so it does not restore a calibrated crossing posterior (Sec.~\ref{sec:whymono}).

We build monotonicity into the surrogate from the start. The response is expanded in Ramsay's integrated splines (I-splines),\cite{22} whose non-negative-weighted combinations are monotone by construction, and a random walk on the logarithms of the weights keeps them positive (Sec.~\ref{sec:priors_sec}), so that the posterior mean and every individual posterior draw are globally monotone functions. The input attaining any saturation level is then read directly from that posterior by level crossing, with no parametric kinetic form assumed.

The components we draw on are individually well established. Our contribution is their combination and systematic test for self-limiting curves. Since Ramsay's I-spline basis,\cite{22} shape-constrained regression splines have been developed extensively,\cite{23} including Bayesian versions that place priors directly on the constrained coefficients so that the whole posterior respects the shape,\cite{24,25} and our surrogate is of this kind. The GP literature reaches monotonicity by other routes, namely virtual observations of a non-negative slope at selected inputs, which still allow dips between those inputs,\cite{26} post-hoc projection of the draws,\cite{21} or finite-dimensional approximations in which linear constraints on the knot coefficients make the curve non-decreasing everywhere on the domain.\cite{27,28} Fully Bayesian versions of the virtual-slope approach have been sampled by Markov-chain Monte Carlo (MCMC) for computer experiments.\cite{29} The finite-dimensional approximations are the closest relatives of our surrogate. Constrained GP regression has since been surveyed as a whole,\cite{30} and the finite-dimensional construction has been given efficient samplers\cite{31} and extended to large-scale problems.\cite{32}

Steering data collection with such a model is also established. In dose-finding trials the continual reassessment method fits a monotone dose--toxicity curve and chooses the next dose from it.\cite{33} In adaptive psychophysics, a GP kept nearly monotone by the virtual-slope approach has been run with several sampling rules, among them sampling where the posterior variance is largest, the uncertainty-sampling rule we adopt.\cite{34} There, a rule aimed at the detection threshold, the stimulus at which the response crosses a set level, locates that threshold faster. The MaxMod algorithm chooses knots instead of measurements, inserting them one at a time into an exactly constrained GP, each where it most changes the fit, with a convergence guarantee for a fixed data set,\cite{35} an adaptive-knot alternative to the fixed-knot-plus-penalty basis we use. Benchmark-dose (BMD) analysis reads off the dose at which a fitted dose--response curve reaches a set response level, and this has been done draw by draw on a monotone Bayesian spline posterior,\cite{36} which is the level-crossing read-out we report. A recent semi-parametric BMD method derives lower confidence limits on that dose from a closely related model, a monotone additive model with penalized B-splines.\cite{37} Locating the input at which a function crosses a level is also a recognized target for sequential design in its own right. Excursion-set and contour-location methods choose new samples to pin down where that crossing lies,\cite{38,39,40} and modern Bayesian experimental design provides the information-theoretic framework within which such acquisition rules are understood.\cite{41} Closer to the present setting, the measurement of adsorption isotherms has itself been treated as a design problem. Bayesian design of experiments has been used to choose the measurements that most reduce the parameter uncertainty of a Langmuir isotherm,\cite{42} D-optimal designs have been derived for the standard isotherm families, each tied to the parametric form it assumes,\cite{43} model-based design has been used to select isotherm measurements and to discriminate among candidate models,\cite{44} and uncertainty-based active learning with a GP surrogate has chosen the points at which simulated isotherms are evaluated.\cite{45} Autonomous model-based experimental design in flow chemistry plans and runs experiments and refits an empirical model of the yield as the results come in,\cite{46} and active learning has proposed the most informative feeds for identifying which reactions in a catalytic network are active.\cite{47} These designs are built around an assumed parametric model or around an unconstrained GP, whereas the loop developed here uses a monotone nonparametric surrogate that commits to neither a kinetic form nor an unconstrained curve. As the dose-finding, psychophysics, and BMD work above shows, a monotone surrogate steering data collection, and a level-crossing read-out taken from it, are not themselves new.

The present framework adds a surrogate whose structure matches the full defining physics of a self-limiting response, beyond monotonicity. The origin is pinned exactly ($f(0)=0$ for every draw), which the constrained-GP constructions above do not build in, the plateau is observed at the largest input $x_{\max}$ and need not be assumed, and the logarithmic knots supply resolution where the curve bends, so that one surrogate serves five kinetically distinct regimes without retuning. To our knowledge a shape-constrained monotone surrogate has not before been tested systematically across kinetically distinct self-limiting regimes. The five-family development set and the held-out benchmark of Sec.~\ref{sec:heldout}, on which the general claim of this paper rests, together provide such a test.

This paper makes five contributions.
\begin{enumerate}
\item The surrogate itself (Secs.~\ref{sec:model_sec}--\ref{sec:priors_sec}), with the guarantees just stated holding for every posterior draw.
\item A one-feature-at-a-time ablation against unconstrained GPs (Sec.~\ref{sec:whymono}, Table~\ref{tab:gp}) showing that multi-scale logarithmic resolution and the roughness penalty account for the curve accuracy, while the bare shape constraint does not. Matched for resolution and origin, the GP's error is $1.26\times$ the penalized monotone surrogate's, whereas monotonicity alone, without the penalty, adds no accuracy. The constraint instead guarantees a level crossing unique on every draw.
\item The finding that the roughness penalty and its order resolve two problems. The penalty itself closes the high-noise reversal that the bare monotone constraint suffers (Fig.~\ref{fig:noise}), and its second-order form, which leaves a smoothly decaying rate unpenalized, lets the fit ease into the plateau at nearly the true pace, so the read-out is well centered (Table~\ref{tab:readout}) where a first-order penalty reads late (SI Table~\ref{tab:si-model-ablation}).
\item A conditioning analysis showing that recovering the input at $95\%$ of plateau is intrinsically ill conditioned, because the curve is nearly flat there, an amplification that every model inherits, ours included. We pair it with a reporting convention (a cost-indexed quantile of the crossing posterior alongside the under-saturation risk) so that the residual choice is stated as a cost decision and no default is left hidden.
\item A held-out benchmark of $250$ curves from an unseen distribution, on which the benefit of the penalized monotone surrogate over the matched GP lies chiefly in the reliability of the tail, with only a modest lead in mean accuracy.
\end{enumerate}
We use a standard acquisition rule, uncertainty sampling, and do not propose a new one. It beats non-adaptive random designs outright and edges a strong log-spaced grid, and because the design is nested an experimenter can keep adding points and stop as soon as the uncertainty band is tight enough, whereas a fixed grid must be rebuilt to grow. We have no mechanism for the edge over the grid (Sec.~\ref{sec:acq_compare}), so we report it as an empirical observation. All results are on synthetic ground truths under an explicit noise model, so every claim can be checked against a known curve, and we are specific throughout about which claims that supports.

\section{Methodology}

We first describe the regression surrogate, which is the primary contribution of this work and is agnostic to the underlying kinetics. The specific saturation curves used to generate test data are deferred to Sec.~\ref{sec:kinetics}, since the method is designed to learn any monotone saturation curve and does not fit a particular kinetic form. A glossary of the statistical terms used below and a table of the notation are given in the SI (Sec.~\ref{sec:si-terms} and Table~\ref{tab:si-notation} there). The construction is developed below in words, and the production model is written out as equations where each part is introduced, namely the curve, the likelihood, the prior, the acquisition rule, the read-out, and the noise model. The mathematics behind those choices, the spline recursion, the classical form of the penalty and what it leaves free, and the formulas behind the posterior summaries and read-outs, is collected in SI Secs.~\ref{sec:si-blocks}--\ref{sec:si-posterior}.

\subsection{Monotone I-spline surrogate}
\label{sec:model_sec}

We do not assume any algebraic formula for the saturation curve, no Langmuir form and no exponential. Instead we build the curve from a fixed set of simple, pre-computed building-block functions and let the data decide how much of each to use. This is the spline idea, in which the blocks are chosen once and only their weights are unknown. Changing the weights bends the curve into almost any shape, much as mixing different amounts of a few stock solutions yields a range of concentrations, and because the curve depends on the weights through a simple sum, matching it to the measurements is an easier and more robust problem than fitting a stiff nonlinear formula.

The key is choosing blocks that encode the physics we know must hold. A saturation curve must start at zero, since zero input gives zero response, must never decrease, since more input cannot undo the response, and must level off once the system self-limits. We choose blocks that guarantee the first two properties for any non-negative weights. The third is not imposed at all. The data establish it, because a measurement is always included at the largest input, so the plateau is observed and need not be assumed.

The blocks are constructed in two steps, starting from a set of $P$ \emph{M-splines} $M_m(x)$, localized non-negative bumps that are each active only over a limited window of input and zero elsewhere.\cite{22} Integrating each bump from the origin gives the blocks we actually use, the \emph{I-splines} $I_m(x)=\int_0^{x} M_m(u)\,du$, each of which is a smooth non-decreasing step rising from $I_m(0)=0$. We model the saturating response as a non-negative combination of these steps with no intercept,
\begin{equation}
f(x) = \sum_{m=1}^{P} w_m\, I_m(x), \qquad w_m \geq 0 .
\label{eq:model}
\end{equation}
Two physical properties then hold exactly for any non-negative weights. First, zero input yields zero response, $f(0)=0$, because every block starts at zero. We call this property the \emph{pinned origin}. Second, because each step is the running total of its bump, the slope of the curve is the weighted sum of the bumps themselves, a sum of non-negative terms that can never be negative. The curve therefore cannot turn downward anywhere. Monotonicity is thus an arithmetic guarantee, which does not rely on a penalty that only encourages it and which no fitting step can break. The recursion that defines the bumps and the closed form of their integrals are given in SI Sec.~\ref{sec:si-blocks}.

\begin{figure}[htbp]
\centering
\includegraphics[width=\textwidth]{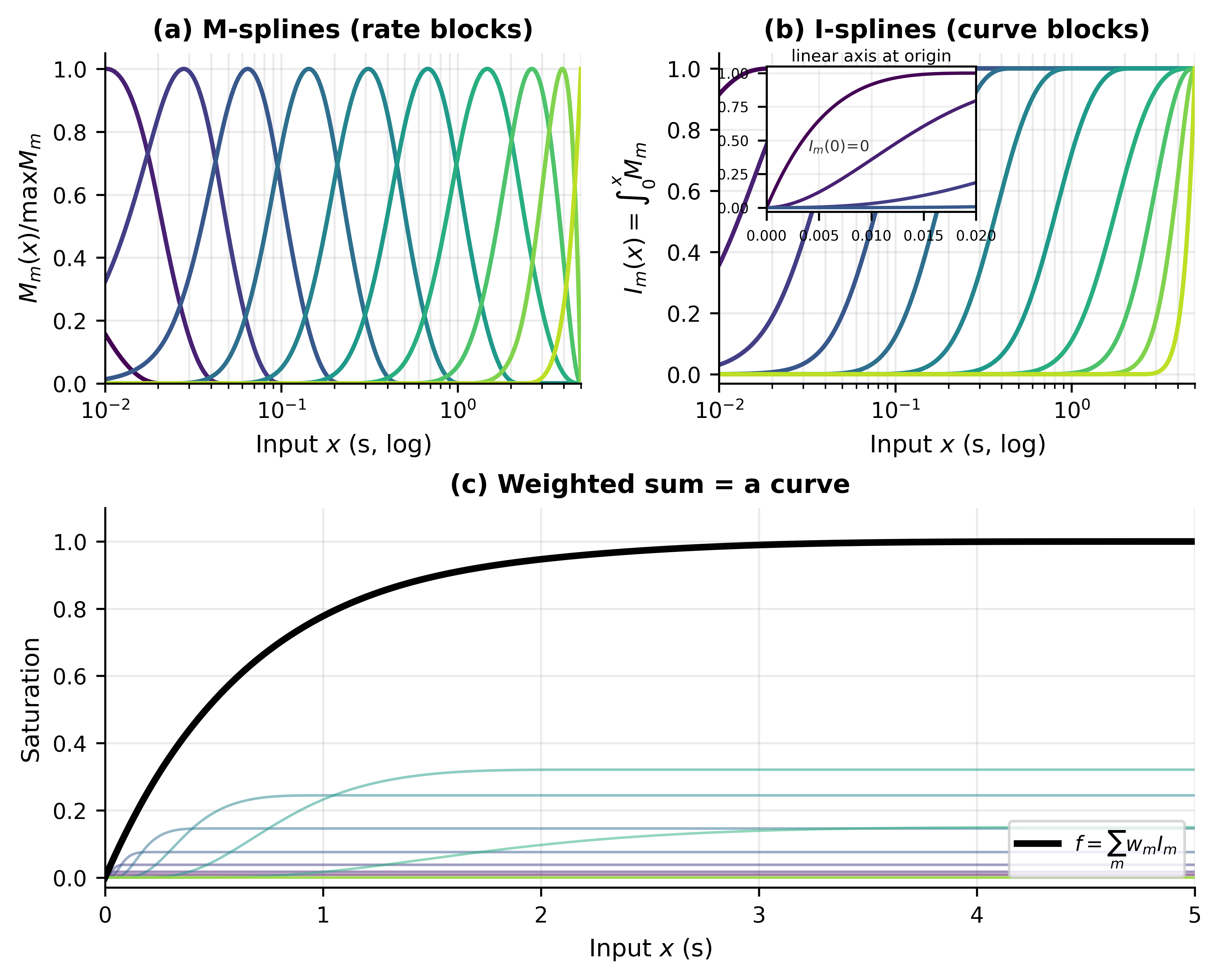}
\caption{The spline building blocks, drawn with the coarse $P=11$ basis (seven log-spaced interior knots) for legibility. The surrogate uses $P=25$. (a) M-splines, non-negative unit-area bumps tiling the input axis, dense near the origin, on a log-$x$ axis with each bump scaled to unit peak. (b) I-splines, the running integral of each bump, smooth steps rising from $0$ to $1$ at staggered inputs, with an inset on a linear axis showing that every step starts at $I_m(0)=0$. (c) A non-negative weighted sum $f(x)=\sum_m w_m I_m(x)$ (bold) with the scaled steps faint beneath. The weights are the best non-negative least-squares fit of the basis to the Langmuir curve of the main campaign.}
\label{fig:basis}
\end{figure}

Figure~\ref{fig:basis} shows the construction on a coarse version of the basis. Panel (a) shows the rate blocks, a row of overlapping bumps that tile the input axis. They are narrow and closely spaced near the origin, where the curve rises fast, and wide and sparse toward the plateau, so on the logarithmic axis of the panel they look evenly spaced. Panel (b) is the running total of each bump, a step that is zero before its window, rises once while the input crosses the window, and holds at one after it. Panel (c) is a weighted sum of the steps, and both guarantees can be seen in it. The weights are the only part of the curve that the data determine, and the staggered rise points of the steps give the sum the freedom to follow almost any monotone shape.

In practice, the bumps are built from standard B-splines\cite{48} and each I-spline is the exact closed-form integral of its B-spline, so the whole basis is pre-computed once and reused at every iteration. The interior knot points that separate the bump windows are placed on a logarithmic grid, packing them closely near the origin where saturation curves bend most sharply and spreading them toward the plateau where the curve is nearly flat. The left boundary knots are placed at the origin $x=0$ instead of at the first measurement, which makes the pinned origin hold exactly even though no data are collected below the minimum input $x_{\min}$, and the right boundary knots sit at the largest input $x_{\max}$.

\begin{figure}[htbp]
\centering
\includegraphics[width=\textwidth]{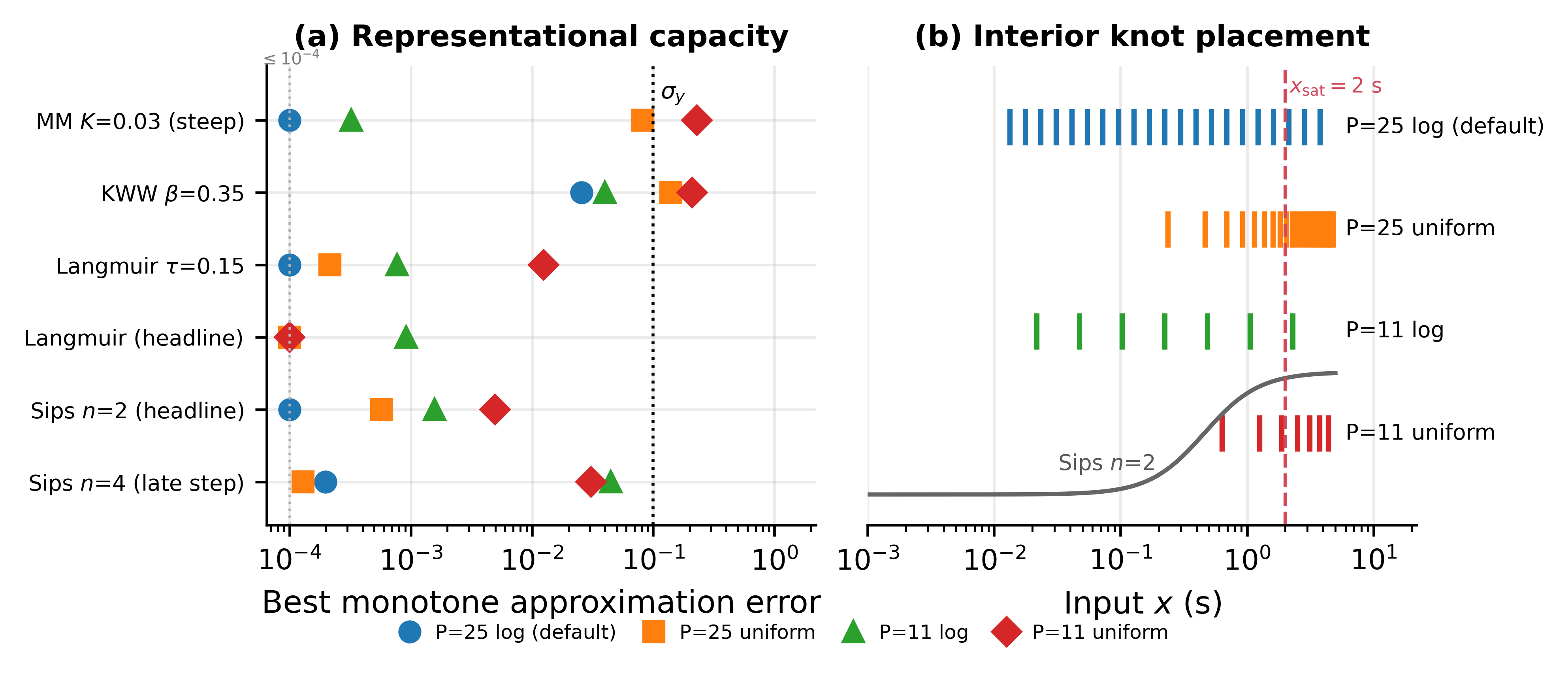}
\caption{Why the knots are log-spaced, measured without a prior, a sampler, or a seed. (a) Best monotone approximation error of the noise-free ground truth onto four I-spline bases (non-negative least squares, scored by the saturation error of Sec.~\ref{sec:metrics}), one marker per configuration. The dotted vertical line marks $\sigma_y=0.1$ for scale, and markers at the left edge are at or below the plotting resolution. (b) The interior knots of each configuration on a logarithmic input axis, with the Sips $n=2$ curve of the main campaign for scale and the true $x_{\rm sat}=2.0$~s dashed. Curves marked headline in (a) are those of the main campaign, and MM is Michaelis--Menten. The logarithmic grids pack knots near the origin and through the onset, where a uniform grid of the same size is sparse.}
\label{fig:knots}
\end{figure}

The logarithmic placement is a design choice not forced by the monotonicity construction, and the relevant question is representational, namely how close a monotone curve the basis can express at all once the knots are fixed, independent of prior, sampler, noise, or seed. The answer is the best non-negative-weighted approximation (non-negative least squares\cite{49} of the noise-free ground truth onto the design matrix), and Fig.~\ref{fig:knots} reports it for four knot configurations. The pattern is clear on the fast-rising shapes that dominate self-limiting saturation, where the default $P=25$ logarithmic grid is closer than a uniform grid of the same size by a factor that ranges from about five on the fast Kohlrausch--Williams--Watts (KWW) rise to nearly three orders of magnitude on a steep Michaelis--Menten rise ($0.0001$ or below, against $0.08$), because the innermost knots of a uniform grid sit far out along the axis, past where those curves have already done half their rise (Fig.~\ref{fig:knots}b). The $P=25$ log basis also resolves the step-like late onset of the Sips $n=4$ shape of Sec.~\ref{sec:kinetics} (best approximation $0.0002$, against $0.045$ for the coarse $P=11$ log grid), whose rise falls where a log grid is coarsest. Most self-limiting responses concentrate their curvature near the origin, so the logarithmic grid is the right default. The many knots supply the resolution, and the roughness penalty of Sec.~\ref{sec:priors_sec} keeps them from fitting noise. This best-approximation quantity is the representational error of the basis, as distinct from the statistical error of a fit to noisy data, and the noise-free sweeps of Sec.~\ref{sec:per_isotherm} measure it together with the small smoothing the fit adds.

\subsection{The roughness penalty and the saturated value}
\label{sec:priors_sec}

This section states the statistical model we fit, namely the likelihood and the prior that carries the roughness penalty and keeps the weights positive, and then explains how the saturated value is set by the data instead of by a prior. We do not look for a single best set of weights. We use Bayesian inference, which involves three things. The prior states what we assume about the weights before any measurement, for example that they are positive, so the curve only rises, and that they change smoothly from one block to the next without sudden jumps. The likelihood says how well a given curve matches the measurements in hand. The posterior is the two put together, the prior updated by the data. It is a whole range of weight sets that fit the data, some more likely than others, and this range lets us draw uncertainty bands around the curve.

The prior itself carries no data, and it is the same whatever we go on to measure. All the measurements, the initial survey and every point the active loop adds later, enter through the likelihood, and each one sharpens the posterior. The survey therefore does not set the prior. It is simply the first batch of data, and we recompute the posterior on all the data gathered so far each time a new point comes in.

The likelihood assumes each measurement lands near the true curve, with Gaussian (bell-curve) noise of a size $\sigma$ that the fit learns along with the weights, so a curve that runs close to all the points is more likely than one that misses them. Written out,
\begin{equation}
p(\{y_i\}\mid \bm{w},\sigma)=\prod_{i=1}^{N}\mathcal{N}\!\left(y_i \mid f(x_i),\,\sigma^2\right),
\label{eq:likelihood}
\end{equation}
where $y_i$ is the response measured at input $x_i$, $N$ is the number of measurements so far, $\bm{w}=(w_1,\dots,w_P)$ collects the weights, $f$ is the curve of Eq.~(\ref{eq:model}), and $\mathcal{N}(y\mid\mu,\sigma^2)$ is the Gaussian density with mean $\mu$ and variance $\sigma^2$.

The prior has two jobs, to keep every weight positive, because a negative weight would let the curve dip, and to stop the curve from bending around to fit the noise, because we have $P=25$ building blocks but only a handful of points. Working with the logarithms $\log w_m$ as the free parameters handles both, since the exponential of any log-weight is positive and the smoothing acts on the same log-weights. Instead of limiting how large the weights are, we limit how sharply their trend bends from one block to the next, so a fine set of blocks cannot chase noise. We do this with a second-order random walk on the log-weights. Each new log-weight is the straight-line continuation of the previous two plus a small random bend, so the weights trace a straight line in $\log w$ and bend only where the data ask. How large a bend is allowed is set by a scale $\tau_\eta$, which the model learns from the data instead of fixing. Written out, together with the two starting values the walk needs, the prior is
\begin{align}
\log w_1 \sim \mathcal{N}\!\left(\log(1/P),\,1.5^2\right), \quad
\log w_2 - \log w_1 \sim \mathcal{N}(0,\,0.5^2), \quad
\tau_\eta \sim \mathrm{HalfNormal}(s_\tau), \nonumber\\
\log w_{m+1} = 2\log w_m - \log w_{m-1} + \tau_\eta\, z_m, \quad z_m \sim \mathcal{N}(0,1), \qquad
\sigma \sim \mathrm{HalfNormal}(s_\sigma),
\label{eq:priors}
\end{align}
where $\sim$ means ``is drawn from'' and $\mathrm{HalfNormal}(s)$ is a Gaussian of scale $s$ cut at zero and kept on the positive side. The first line says where the walk starts, with the level $\log w_1$ centered on each block carrying $1/P$ of the plateau, at a wide spread of $1.5$ in log units that leaves the real height to the data, and the first step of mean zero and spread $0.5$, so no direction is assumed. The first line also gives the bend scale $\tau_\eta$ its half-normal prior. The second line is the walk itself, run for $m=2,\dots,P-1$, with the bend $\tau_\eta z_m$ of inferred size, and the last term is the noise scale of Eq.~(\ref{eq:likelihood}). Besides the starting spreads of $1.5$ and $0.5$, the numbers fixed in advance are the half-normal scales $s_\tau$ and $s_\sigma$, both set to $0.3$.

Because every weight is positive, the two guarantees of Sec.~\ref{sec:model_sec}, that the curve only rises and starts from the pinned origin, hold for every sampled curve as well as for the average.

The smoothness the walk adds has a useful shape, since the penalty acts on curvature and a straight line in the log-weights costs nothing. On our log-spaced knot grid a straight log-weight line turns out to be a power-law rate, $f'(x)\propto x^{\xi-1}$, with an exponent $\xi$ the fit is free to pick (derived in SI Sec.~\ref{sec:si-pspline}). This is the behavior a saturation curve needs, because the rise can ease into the plateau at whatever power the data show, so one setting handles rates as different as the Elovich $1/x$, the KWW tail, and the Langmuir onset (Sec.~\ref{sec:kinetics}) with no retuning.

A power that fades ($\xi<1$) has a rate that dies away, which is how a curve eases toward its plateau. A power that grows ($\xi>1$) never levels off, and the penalty allows it just as freely. The fit is held down by the readings at $x_{\max}$, so the data rule a growing power out even though the penalty does not.

Because $\tau_\eta$ is learned instead of fixed, the fit smooths itself to match the data. Noisy data pull $\tau_\eta$ down and iron out wiggles that an unpenalized fit would keep, while a sharp onset lets it grow. The fixed scale $s_\tau$ only says how large a bend to expect before seeing the data. On this grid it lets the power in the rate drift by under one unit from one knot to the next and rarely by more than two, which is room enough to turn from the onset into the plateau over a few knots and too little for the weights to zigzag between neighbors. SI Sec.~\ref{sec:si-pspline} gives the arithmetic, and SI Sec.~\ref{sec:si-compute} the values of $\tau_\eta$ the data return, all well below the prior's $95\%$ point.

This is the second-order penalized spline (P-spline) of Eilers and Marx,\cite{50} in the Bayesian form that learns its own smoothing,\cite{51} adapted here to keep the weights positive. The order of the penalty matters, since a first-order penalty would hold the log-weights flat instead of straight, which gives a $1/x$ rate that levels off only grudgingly and reads the input late. The gentle approach to the plateau that the second-order penalty allows keeps the read-out well centered (Sec.~\ref{sec:pulse_time}), and the penalty itself removes the high-noise reversal that the unpenalized monotone basis suffers (Sec.~\ref{sec:noisesweep}). We keep the first-order penalty only as a comparison arm (SI Table~\ref{tab:si-model-ablation}).

The noise level $\sigma$ is learned from the data instead of being assumed. Its scale $s_\sigma=0.3$ sits well above the expected scatter (the output noise $\sigma_y=0.1$ of Sec.~\ref{sec:noise}) while keeping the sampler from writing real signal off as noise. The knots are fixed in advance and never adapted to the measurements of a fit, so no fit carries a model choice to correct for, and the only scales we learn are $\tau_\eta$ and $\sigma$.

At the $N=20$ budget used throughout, the data drive the fit more than the priors do, though the roughness-prior scale $s_\tau$ still has some pull. On measured data every response is divided by the largest observed value before fitting, so that the priors act on the scale they were set for, and the fit is multiplied back into original units afterward. The synthetic ground truths used here all have saturated value $A=1$, so they already sit on that scale and are not rescaled. On real data the divisor is a single noisy reading, but because the fit is scaled back by the same number, the reported curve and read-out depend on it only through the prior scales. We report how much the results depend on $s_\tau$ and $s_\sigma$ in SI Sec.~\ref{sec:si-priors}, and return to the divisor in Sec.~\ref{sec:limitations}.

A saturation curve has to level off somewhere, and we let the data decide where. The plateau value is $f(x_{\max})=\sum_m w_m$, since every I-spline equals one at $x_{\max}$, and because the curve only rises this is also its highest point. The survey always measures at the largest input $x_{\max}$ (Sec.~\ref{sec:setup}), so we see the plateau directly.

The usual alternative, a prior that pushes the slope to zero at the edge, forces the fit flat there even when the data are still rising, so we avoid it. Our own prior on $f(x_{\max})$ is in fact very loose. Under the walk the spread of $\log w_m$ grows quickly along the sequence, so the plateau is barely pinned down beforehand, and the penalty does not force the level-off either, since it leaves a smooth rise alone and fights only a sudden late jump. The plateau is therefore set by the data, which lets us drop the usual flatness prior and still recover a self-limiting curve.

\subsection{Drawing the curve and its uncertainty}
\label{sec:posterior}

We need the full distribution of plausible curves and cannot rely on a single best fit. Because the non-negativity constraint rules out a closed-form solution, we sample the distribution numerically with the No-U-Turn Sampler (NUTS),\cite{52} an efficient MCMC method implemented in NumPyro.\cite{53} The result is a collection of $S$ weight sets $\{\bm{w}^{(s)}\}_{s=1}^{S}$ drawn in proportion to their posterior probability. Each weight set defines one candidate curve $f^{(s)}(x)$. One can picture the $S=500$ draws used here as a bundle of candidate saturation curves, all passing through the origin and threading near the measured points, but differing where data are sparse. The price of the second-order penalty is that this posterior is not easy to sample, because its log-weights are a double cumulative sum with an inferred step scale, which gives the prior heavy tails and the posterior a pronounced funnel.\cite{54} The production configuration pays for that in divergent transitions, and we report the diagnostics, the cause, and a remedy that removes most of them without changing any reported quantity in SI Sec.~\ref{sec:si-diag}.

This bundle gives both the fitted curve and its uncertainty. At any input $x$ we read off the value of every candidate curve and summarize the spread. The fitted curve $\mu_\ast(x)$ is the point-by-point average of the candidate curves, and the uncertainty $\sigma_\ast(x)$ is their standard deviation (s.d.) at that input (SI Sec.~\ref{sec:si-posterior}). Where the candidate curves agree closely, $\sigma_\ast(x)$ is small, and where they fan out, typically between measurements or below the first one, it is large. The shaded band in the figures is $\mu_\ast(x)\pm 1.96\sigma_\ast(x)$, a nominal $95\%$ band (its empirical coverage is assessed in Sec.~\ref{sec:calibration}). No candidate curve dips, so neither the fitted curve nor any draw can show a spurious local maximum.

\subsection{Choosing the next experiment}
\label{sec:acq}

Each measurement is expensive, so every new one should be as informative as possible about the saturation curve. We use uncertainty sampling, a simple and widely used active-learning rule.\cite{55,56} We place the next measurement where the surrogate is currently least sure of the curve, at the input of maximum posterior s.d.,
\begin{equation}
x_{N+1} = \arg\max_{x}\;\sigma_\ast(x).
\label{eq:acq}
\end{equation}
Because $\sigma_\ast(x)$ is the spread of the candidate curves about their mean, this rule sends each run to the part of the curve the current data leave most uncertain, and drives that uncertainty down where it is largest. We then re-sample the posterior, recompute $\sigma_\ast$, and repeat. How much this active choice beats picking inputs at random, using the same surrogate and the same starting survey, is measured in Sec.~\ref{sec:acq_compare}. The loop can command any input in $[x_{\min},x_{\max}]$ on a grid of $0.001$~s, the resolution at which an input is physically set, so it can propose and re-measure anything, right down to the near-origin onset where the steepest curves are already rising.

Uncertainty sampling on a flexible surrogate tends to revisit the same inputs, and it does here. A finished $N=20$ design has $12.9$ distinct inputs on average when readings within $0.02$~s are counted as one, with $4.7$ of the $20$ readings landing at $x_{\max}$. The posterior is least sure there because $x_{\max}$ sits at the edge, with data on one side only, and its plateau value $f(x_{\max})=\sum_m w_m$ depends on every weight, so a repeat reading at $x_{\max}$ does not tighten the plateau the way a repeat reading tightens the curve at an interior point. This clustering means the nominal budget covers fewer distinct inputs than it appears to (SI Sec.~\ref{sec:si-cost}). The repeated readings are also free replication, which SI Sec.~\ref{sec:si-outside} reuses to spot drift.

The acquisition rule is generic, and the monotone surrogate slots into a standard active-learning loop without changes.

\subsection{Model-free saturation read-out}
\label{sec:readout}

The posterior over curves is the primary result, and any operating quantity can be read from it. The one most often needed is a saturation input, the smallest input at which the response reaches a set fraction $\alpha$ of its own saturated value, written $x_{\rm sat}(\alpha)$. When the aim is to run a process at saturation the conventional choice is $\alpha=0.95$, and we use it throughout as the working example. We read this straight off the fitted curves, with no Langmuir or other parametric form assumed.

Each candidate curve carries its own plateau, the value $f^{(s)}(x_{\max})$ it reaches at the far edge of the window. That edge is measured repeatedly, because the survey visits $x_{\max}$ and the acquisition keeps returning to it (Sec.~\ref{sec:acq}). Working from the final posterior, we find the first input at which each curve reaches $\alpha$ times its own plateau, located by interpolation. That input is the curve's crossing.

Monotonicity makes this crossing well defined, because a curve that starts at zero and rises throughout, as every draw does, meets any level below its plateau exactly once, so the crossing exists for every $\alpha<1$ and is the only one. A non-monotone fit could meet the same level several times and leave nothing single to report.

Doing this for every draw turns the crossing into a posterior for $x_{\rm sat}(\alpha)$. We report its median as the estimate, written $\hat{x}_{\rm sat}(\alpha)$ or $\hat{x}_\alpha$ for short, and its spread as the uncertainty, and the same draws answer any $\alpha$ at no extra cost. A practitioner who needs a safety margin can instead read off the one-sided $95$th percentile. In each case the level is measured against the plateau the curve actually reaches inside the window, with no asymptote extrapolated past it, which keeps the read-out model-free and puts gentle and sharp curves on the same footing.

\paragraph{Why the $95\%$ input is hard to pin down.} The input used to set a process at saturation, the $95\%$ point, is also the hardest feature of the curve to locate, and it is worth seeing why before choosing how to report it. Near saturation the curve is almost flat, so reading an input off it means inverting a nearly horizontal curve, where a small error in the fitted height becomes a large error in the recovered input. Two height errors enter, a vertical misfit $\Delta y$, because the fitted curve may sit a little above or below the target at the true crossing, and a plateau misestimate $\Delta A$, because the read-out measures the level against the inferred plateau in place of the true saturated value $A$. Expanding the fitted curve about the crossing turns these two vertical errors into the horizontal one we care about. To first order the input shift is the combined height error, $\alpha\,\Delta A-\Delta y$, divided by the slope of the fitted curve at the crossing (SI Sec.~\ref{sec:si-cond}). We call this the conditioning relation, and its denominator explains the difficulty, since saturation is by definition where that slope is small, so a small height error is magnified into a large shift in the input. The sign can go either way, because a fit that under-reads the plateau while approaching it from below may cross early or late, depending on which term wins.

This magnification belongs to the question itself. Any method that reads an input off a saturating curve near its plateau inherits it, whether the curve is a spline, a GP, or a fitted Langmuir form. The method affects only how large the two height errors are.

In practice the slope that does the magnifying is the fitted slope, which on a curve the fit rounds off, such as a sharp cap, can differ from the true slope. We therefore evaluate the relation with the fitted mean $\mu_\ast$, which is always available, whereas the true curve is not available on real data. Section~\ref{sec:pulse_time} and Fig.~\ref{fig:readout}b check it against the shifts actually observed, and report the amplification each family carries.

\paragraph{From an uncertain crossing to a recommendation.} Because the $95\%$ input is uncertain, and uncertain in a lopsided way, no single number is automatically the one to report. The question is which point of the crossing posterior to hand over, and that depends on what it costs to err in each direction. Recommending too short an input (under-saturating) and too long an input (over-dosing) rarely carry the same penalty. If they did, the median crossing would be the natural report. When they differ, the right report is a higher or lower quantile of the posterior, shifted toward the safer side so that the costlier mistake becomes unlikely. The pinball loss\cite{57} makes this exact by charging an error at one rate when the recommendation is too short and at another when it is too long, and the report that minimizes it is precisely a quantile of the crossing posterior (SI Sec.~\ref{sec:si-posterior}). The whole family of sensible recommendations is then a single dial,
\begin{equation}
\hat{x}_q(\alpha) = Q_q\!\left[\,x_{\rm sat}^{(s)}(\alpha)\,\right],
\qquad
\frac{q}{1-q}=\frac{\text{cost of under-saturating by one second}}{\text{cost of over-dosing by one second}},
\label{eq:quantile}
\end{equation}
where $Q_q$ is the $q$ quantile taken over the posterior draws $x_{\rm sat}^{(s)}$. Because $q$ follows from the ratio of these two costs, a practitioner can set it from a cost they are able to judge, with no abstract penalty weight to choose (the recurring cost of a second of input is estimated in SI Sec.~\ref{sec:si-quantile}). The neutral choice $q=0.5$ gives the median crossing, and $q=0.95$ gives a conservative one-sided limit. It mirrors the $95\%$ lower limit of BMD analysis, which errs toward the smaller dose,\cite{36,37} while ours errs toward the longer input, making under-saturation unlikely at the price of some over-dosing. We report the whole curve of $\hat{x}_q$ against $q$ (Fig.~\ref{fig:readout}c), so that the degree of conservatism is stated explicitly instead of being left as a hidden default.

\paragraph{Reporting in response units.} The ill-conditioning belongs to the input axis and does not affect the response axis. It is therefore often clearer to report in the units the practitioner acts in, namely how far up the plateau the response has climbed at a chosen input. For any input $x$ they name, and for each posterior draw, we form the attained fraction $\theta(x)$, the response at $x$ divided by that draw's own plateau, and from it the under-saturation risk $R(x)$, the posterior probability that the response there still falls short of the target $\alpha$ (SI Sec.~\ref{sec:si-posterior}). Forming them draw by draw keeps the plateau uncertainty inside the answer. Each draw's attained fraction rises from $0$ at the origin to $1$ at $x_{\max}$, because the draw starts at zero and is divided by its own value at $x_{\max}$. Because of that, ``a draw is still below $\alpha$ at $x$'' is the same statement as ``$x$ lies below that draw's crossing,'' so $R(x)=\Pr(x_{\rm sat}>x)$ is simply the crossing posterior read as a survival curve. It holds the same information as the input read-out, in friendlier units. Instead of an interval on an ill-conditioned input, one can say for a typical Sips fit at $N=20$ ``at $2.3$~s the response is expected to have reached $96\%$ of its plateau, just past the $95\%$ target, yet $3$ in $10$ of the plausible curves are still short of it.'' The $96\%$ says how far up the plateau the response is expected to be, and the $3$ in $10$ how likely it is to fall short of the target.

Two properties make $R(x)$ easy to use, and one caveat comes with it. It decreases as $x$ grows, so a target risk fixes a unique smallest input, and $R(x_{\max})=0$ by construction, so it always returns an answer inside the window. That last property is also a blind spot, because $R(x)$ trusts the in-window plateau and so cannot tell when the window itself was too short to reach saturation. For that we add a separate check, the window-sufficiency statistic $r=[f(x_{\max})-f(x_{\max}/2)]/f(x_{\max})$, the share of the plateau that the curve gains over the last doubling of input, computed for each posterior draw and averaged over the draws. A curve that has leveled off gains almost nothing over its last doubling, so a large $r$ means the curve is still rising at the window's edge and the window should be widened. How well $r$ catches a short window, and why its cutoff depends on the window length, is shown in Sec.~\ref{sec:limitations}, with the details in SI Secs.~\ref{sec:si-window} and~\ref{sec:si-outside}.

\subsection{Noise model}
\label{sec:noise}

To mimic real experimental limits we corrupt the ground-truth saturation curve $f$, the curve the surrogate of Eq.~(\ref{eq:model}) is to recover, with two independent sources of error. One is finite precision in setting the input, so the commanded input is jittered by Gaussian noise of s.d.\ $\sigma_x$ before the curve is evaluated, and a jittered input that falls past either edge of the window is clamped to that edge. The other is output noise, Gaussian with s.d.\ $\sigma_y$, added to the response afterward. The recorded input is always the commanded one, since the jittered input is not observed, so the surrogate sees the jitter as extra apparent scatter, larger where the curve is steep, and absorbs it into the single inferred noise term $\sigma$ of Sec.~\ref{sec:priors_sec}. Written out, each measurement follows
\begin{equation}
\begin{gathered}
\tilde{x} = \min\!\bigl(\max(x + \varepsilon_x,\, x_{\min}),\, x_{\max}\bigr), \qquad
y_{\rm obs}(x) = f(\tilde{x}) + \varepsilon_y, \\[3pt]
\varepsilon_x \sim \mathcal{N}(0,\sigma_x^2), \qquad \varepsilon_y \sim \mathcal{N}(0,\sigma_y^2)\ \ \text{independent},
\end{gathered}
\label{eq:noise}
\end{equation}
where $x$ is the commanded input, $\tilde{x}$ the input actually realized after the jitter $\varepsilon_x$ and the clamp to the window, $y_{\rm obs}$ the recorded response, $\varepsilon_y$ the output noise, and $\sigma_x=0.01$~s and $\sigma_y=0.1$ the two noise scales used unless stated otherwise. Figure~\ref{alg:main} summarizes the complete active-learning loop in pseudocode.

\begin{figure}[htbp]
\caption{The monotone I-spline active-learning loop in pseudocode.}
\label{alg:main}
\centering
\begin{minipage}{0.94\textwidth}
\hrule height 0.8pt
\vspace{3pt}
\begin{algorithmic}[1]
\Require window $[x_{\min},x_{\max}]$; survey size $N_0$; active budget $N_a$; saturation fraction $\alpha$
\Ensure monotone posterior for $f$; saturation input $\hat{x}_{\rm sat}(\alpha)$
\State build log-spaced interior knots on $[x_{\min},x_{\max}]$, boundary knots at $0$ and $x_{\max}$; pre-compute I-spline basis
\State $\mathcal{D} \gets$ measurements at $x_{\min}$, $x_{\max}$, and $N_0-2$ log-spaced stratified interior points
\Statex \hspace{\algorithmicindent} \emph{(the point at $x_{\max}$ anchors the plateau with a measurement, so no flatness prior is needed)}
\For{$n = 1, \dots, N_a$}
  \State sample $\{\bm{w}^{(s)},\tau_\eta^{(s)},\sigma^{(s)}\} \sim p(\bm{w},\tau_\eta,\sigma \mid \mathcal{D})$ by NUTS \Comment{likelihood and prior, Sec.~\ref{sec:priors_sec}}
  \State $\mu_\ast(x), \sigma_\ast(x) \gets$ mean and s.d.\ of the monotone draws $f^{(s)}$ \Comment{Sec.~\ref{sec:posterior}}
  \State $x_{N+1} \gets \arg\max_x \sigma_\ast(x)$ \Comment{uncertainty sampling, Eq.~(\ref{eq:acq})}
  \State $\mathcal{D} \gets \mathcal{D} \cup \{(x_{N+1},\, \mathrm{measure}(x_{N+1}))\}$ \Comment{noise model, Sec.~\ref{sec:noise}}
\EndFor
\State sample the posterior once more on the final $\mathcal{D}$
\State \Return curve estimate $\mu_\ast \pm 1.96\sigma_\ast$; median first crossing $\hat{x}_{\rm sat}(\alpha)$ \Comment{Sec.~\ref{sec:readout}}
\end{algorithmic}
\vspace{3pt}
\hrule height 0.8pt
\end{minipage}
\end{figure}

\subsection{Comparison surrogates as a design space}
\label{sec:comparators}

The monotone surrogate differs from a textbook GP regressor along three separate axes and adds a roughness penalty on top, and we treat these as a small design space to be explored one feature at a time. The axes are (i) the input resolution, linear or logarithmic, (ii) whether the origin is pinned, so that the curve starts at zero response, and (iii) whether the curve is monotone. The penalized I-spline surrogate sits at one corner (log, pinned, monotone). Three unconstrained GP surrogates occupy three of the other corners, changing one axis at a time, so reading across them isolates what each choice contributes.

GP-linear is a GP with a constant mean and a radial-basis-function kernel on the raw input axis (linear, unpinned, unconstrained). GP-log puts the same kernel on the log-warped input $u=\log(1+x/x_{\min})$, which is logarithmic above $x_{\min}$ and finite at the origin (log, unpinned, unconstrained). GP-log-pin adds a nearly noiseless pseudo-observation at $(0,0)$ so the surrogate passes through the origin as the I-spline does (log, pinned, unconstrained). GP-log-pin is the resolution- and origin-matched comparator, matching the I-spline on every axis but the monotonicity constraint. Two further arms (comparison models run through the identical loop) remove the roughness penalty from the monotone basis itself, at the production resolution ($P=25$) and at the coarse resolution of Fig.~\ref{fig:basis} ($P=11$), with independent half-normal weights of one inferred scale in place of the random walk of Eq.~(\ref{eq:priors}) (SI Sec.~\ref{sec:si-surrogates}). The gap between GP-log-pin and the unpenalized basis at the same resolution is then the test of what the constraint alone contributes, and the further gap to the penalized I-spline is the penalty's (Table~\ref{tab:gp}).

Every GP arm shares the survey design, budget, acquisition rule, noise realizations, and seeds with the I-spline campaign, and sets its lengthscale, amplitude, and noise by maximum marginal likelihood over a grid of candidate values. The grid is wide enough that the chosen values land in its interior in every fit of the two log-axis arms and in $82\%$ of GP-linear fits, the remainder being genuine short-lengthscale optima, so its limits do not constrain the comparison (the exact ranges are given in SI Sec.~\ref{sec:si-compute}). The saturation read-out of Sec.~\ref{sec:readout} is applied to every arm in the same way, each referenced to its own inferred plateau, the posterior value at $x_{\max}$ where the survey always measures. No arm is handed $A$, so each must recover it from its own fit, and a misjudged plateau feeds directly into the read-out through the plateau term of the conditioning relation (Sec.~\ref{sec:readout}).

Fixing those three hyperparameters at their best-likelihood values means each GP arm treats them as known and carries none of their uncertainty forward, whereas the I-spline averages over its roughness and noise scales. Part of the coverage gap in Sec.~\ref{sec:calibration} is therefore due to this difference in how hyperparameter uncertainty is handled and cannot be credited to the monotonicity constraint. That makes the gap an upper bound on what the constraint itself contributes, and the caveat does not apply to the curve-accuracy comparison, which scores only the posterior mean. Section~\ref{sec:whymono} examines which axis drives accuracy and what the constraint buys once the axes are matched.

\subsection{Evaluation metrics}
\label{sec:metrics}

A root-mean-square (RMS) error on $f(x)$ over the whole domain is a poor figure of merit here. Most of the domain is the flat plateau, so such an error mostly rewards pinning the saturated level and is nearly blind to the rising region. We instead measure error along the rise, and anchor every metric at the same reference points. For a saturation fraction $\alpha$, let $x_\alpha$ be the input where the true curve reaches $\alpha A$, the true value of $x_{\rm sat}(\alpha)$. All three metrics are evaluated at these true crossings.

The basic quantity is the vertical misfit there, how far the fitted response sits from the level it should reach, $\mu_\ast(x_\alpha)-\alpha A$. We report it two ways and pair it with a third metric on the input axis. (i) The saturation error across levels is its single-number summary, the RMS of that misfit over a fine set of nineteen levels $\alpha\in\{5,10,\dots,95\}\%$, in response units ($A=1$). This is the one accuracy score we quote and compare across methods, and the output noise $\sigma_y=0.1$ is the natural scale to read it against. Throughout, pooled means averaged over all five families and every seed of the campaign in question. (ii) The level-wise response error $\Delta y(\alpha)=\mu_\ast(x_\alpha)-\alpha A$ is the same misfit left signed and reported at five representative levels $\alpha\in\{5,25,50,75,95\}\%$ instead of squared and averaged over levels, so one can see where along the rise the fit runs high or low, which the single number hides. (iii) The level-wise input error $\Delta x(\alpha)=\hat{x}_\alpha-x_\alpha$ is the horizontal counterpart. It measures how far the median read-out $\hat{x}_\alpha$ of Sec.~\ref{sec:readout} lands from the true crossing $x_\alpha$, so it is the input error at the true level, whereas (ii) is the response error at the true input. Response errors share the response units ($A=1$) and input errors are in seconds. Both (ii) and (iii) are signed, so systematic bias stays visible and does not average away.

\section{Kinetic Models}
\label{sec:kinetics}

The surrogate deliberately assumes nothing about the regime beyond monotonicity and the pinned origin. A real process need not follow any one textbook law. Its uptake can combine several mechanisms, or sit between the standard forms, so no single empirical function is guaranteed to describe it. We want a surrogate that can represent any monotone saturation curve, whatever shape it takes, without committing to a chosen family. To test that, we use five ground-truth saturation curves. They come from distinct corners of the physical and life sciences and cover four qualitatively distinct shapes, namely concave, sigmoidal, logarithmic, and dispersive (Langmuir and Michaelis--Menten are both concave, so five families give four shapes). We use these five because they are known and well established. They serve as a benchmark for the method and do not define its scope. All five are normalized to the same saturated value $A=1$ and the same $95\%$ saturation input $x_{\rm sat}=2.0$~s on the domain $[0.010,5.0]$~s, so that differences in recovery reflect curve shape alone. The input is measured in seconds throughout, and the method does not care what it physically means. All five are observed through the same noise model (Sec.~\ref{sec:noise}), and none is favored by the method. The headline shape parameters are $n=2$, $\gamma=1.69$, and $\beta=0.55$, and the remaining scale of each family follows from $x_{\rm sat}=2.0$~s (SI Sec.~\ref{sec:si-metrics}). Their disciplinary origins and domain-specific forms are collected in SI Sec.~\ref{sec:si-families}.

\subsection{Langmuir}
\begin{equation}
f(x)=A\bigl(1-e^{-x/\tau}\bigr).
\label{eq:iso_langmuir}
\end{equation}
The concave, ideal self-limiting form, with time constant $\tau$. It is the first-order kinetics of Langmuir adsorption on a fixed number of sites\cite{1} and the canonical saturation shape, and it serves as the reference against which the others are compared.

\subsection{Michaelis--Menten}
\begin{equation}
f(x)=\frac{A\,x}{K+x}.
\label{eq:iso_dissoc}
\end{equation}
A rectangular hyperbola, the Michaelis--Menten law of enzyme kinetics\cite{2,58} (algebraically the Langmuir isotherm in pressure), included for its shape. The half-saturation input $K$ sets the approach, and a small $K$ leaves much of the curve just below saturation, a demanding test of plateau recovery.

\subsection{Sips ($n=2$)}
\begin{equation}
f(x)=\frac{A\,(x/\tau_S)^{n}}{1+(x/\tau_S)^{n}},\qquad n>1.
\label{eq:iso_sips}
\end{equation}
Sigmoidal for exponent $n>1$, with a zero initial slope, an inflection, and a cooperative rise. In this regime it is the Hill equation of pharmacology,\cite{3} with $\tau_S$ the half-maximal effective concentration (EC$_{50}$) and $n$ the Hill coefficient. We keep the label ``Sips''\cite{59} for the shared algebra, although Sips's derivation needs an exponent between zero and one, where the curve is concave. We use only the $n>1$ branch. It is the shape a concave-only model cannot represent at all, an S-shaped onset, which the non-negative-weighted rate bumps accommodate by starting near zero, rising, then decaying. We include it as a deliberate stress shape. A cooperative, sigmoidal onset is plausible in dose--response settings, and the surrogate is not told whether to expect one.

\subsection{Elovich}
\begin{equation}
f(x)=\min\!\Bigl(A,\ \tfrac{A}{\gamma}\ln\!\bigl(1+x/\tau_E\bigr)\Bigr).
\label{eq:iso_elovich}
\end{equation}
A logarithmic rise\cite{60} whose rate falls off as $1/x$ ($A/\gamma$ the slope per unit of $\ln(1+x/\tau_E)$), hard-capped at the saturated value. The cap in Eq.~(\ref{eq:iso_elovich}) makes the otherwise-unbounded curve self-limiting and manufactures a slope discontinuity, so this is a deliberately non-physical worst case, the one ground truth with a kink.

\subsection{Kohlrausch--Williams--Watts (stretched exponential)}
\begin{equation}
f(x)=A\bigl(1-e^{-(x/\tau_p)^{\beta}}\bigr),\qquad \beta<1.
\label{eq:iso_kww}
\end{equation}
The stretched exponential of dispersive relaxation ($\beta<1$),\cite{61,62,63} commonly read as a spread of rate constants.\cite{62} Smaller $\beta$ gives a faster initial rise followed by a longer, slower tail, the opposite curvature emphasis to the sigmoidal Sips case.

\section{Results and Discussion}

\subsection{Setup}
\label{sec:setup}

We evaluate the surrogate on the five saturation curves of Sec.~\ref{sec:kinetics}, each observed through the noise model of Sec.~\ref{sec:noise} with input jitter $\sigma_x=0.01$~s and output noise $\sigma_y=0.1$. We use cubic M-splines ($k=4$), and hence quartic I-splines, with twenty-one log-spaced interior knots ($P=25$ basis functions), penalized as in Sec.~\ref{sec:priors_sec}, drawing $S=500$ posterior samples after $500$ warmup steps per fit. The basis order $k=4$ and count $P=25$ are justified in SI Sec.~\ref{sec:si-metrics}. The initial design is a survey of $N_0=5$ points, with one at the minimum input $x_{\min}$, one at the maximum input $x_{\max}$, and three interior points placed by a log-spaced stratified sample (the interval $[x_{\min},x_{\max}]$ is split into three equal strata in $\log x$, and one point is drawn uniformly in $\log x$ within each). The two endpoints anchor the onset and the saturated value, so the plateau is fixed by data and does not depend on a flatness prior. The interior survey points coarsely map the rise. We then run $N_a=15$ active iterations, bringing the budget to $N=20$ measurements. The surrogate, its priors, and its knots are held fixed throughout, and only the ground-truth curve changes. Further details of the setup are given in SI Sec.~\ref{sec:si-compute}.

Three complementary studies share this surrogate. The first is an acquisition comparison (Sec.~\ref{sec:acq_compare}) over $R=50$ independent seeds, which pits uncertainty sampling, Eq.~(\ref{eq:acq}), against three non-adaptive baselines at equal budget (uniform random, log-uniform random, and a log-spaced grid). Every design observes $x_{\max}$, so all inherit the same data-anchored plateau. The second is a per-level accuracy study (Sec.~\ref{sec:per_isotherm}), which reports the level-wise response and input errors of the $N=20$ uncertainty-sampling fits (mean $\pm$ one s.d.\ over the 50 seeds). The third is a set of noise-free regime sweeps (Sec.~\ref{sec:per_isotherm}), in which each family is swept over three regimes (its time scale for Langmuir and Michaelis--Menten, its shape parameter for the other three) and fit once at $\sigma_x=\sigma_y=0$, so the residual error tracks the representational error of the basis. The baseline definitions and the swept parameter values are in SI Sec.~\ref{sec:si-metrics}.

Throughout this section, arms are compared by the ratio of their pooled saturation errors (Sec.~\ref{sec:metrics}), so $2\times$ means the other arm's error is twice ours. Because every arm runs on the same $250$ combinations of family and seed, each difference is tested with a paired Wilcoxon signed-rank test, which asks whether the $250$ run-by-run differences are centered on zero. A $p$ value of $10^{-15}$ then means a difference this consistent would essentially never arise by chance. These $p$ values measure consistency across seeds within the fixed set of five ground truths. Because the five truths were chosen instead of sampled, they say nothing about other curve families, for which the held-out benchmark of Sec.~\ref{sec:heldout} is the test.

\subsection{One fixed surrogate recovers five kinetic regimes}
\label{sec:per_isotherm}

\begin{figure}[htbp]
\centering
\includegraphics[width=\columnwidth]{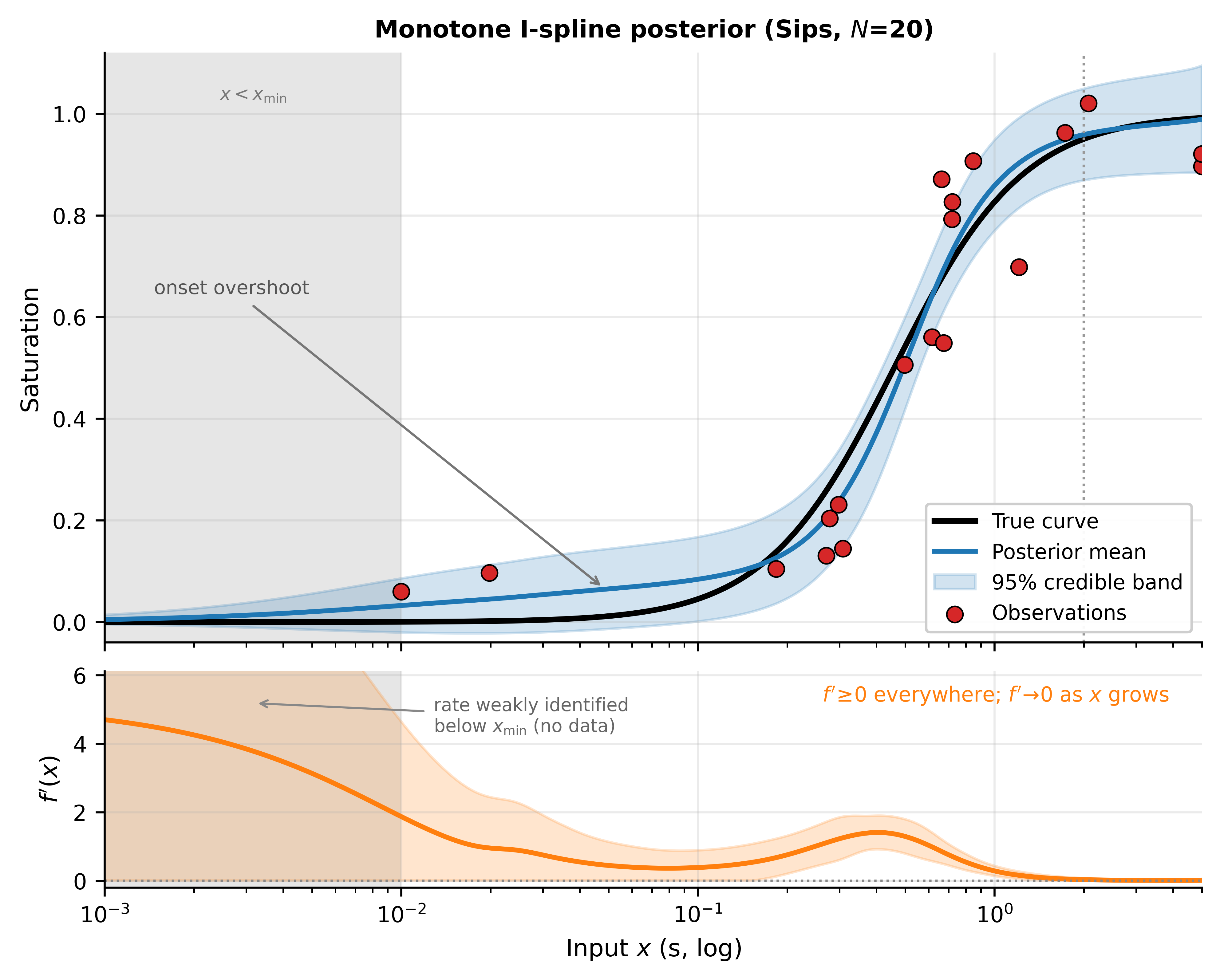}
\caption{A single monotone I-spline fit on the sigmoidal Sips curve at $N=20$, on a logarithmic input axis. The top panel shows the posterior mean, the $95\%$ band, the true curve, and the twenty observations, with the region below $x_{\min}$ shaded and the true $x_{\rm sat}=2.0$~s dotted. The bottom panel shows the posterior rate $f'(x)$, non-negative by construction (Sec.~\ref{sec:model_sec}) and decaying to zero once the measurement at $x_{\max}$ pins the plateau.}
\label{fig:final_fit}
\end{figure}

This section shows that the one fixed configuration of Sec.~\ref{sec:setup} recovers all five regimes to within the measurement noise, first on a single fit, then across the fifty-seed campaign and against parametric fits, and then in the noise-free limit. Figure~\ref{fig:final_fit} shows what a single fit looks like on the sigmoidal Sips curve, the shape a concave-only model cannot represent. The posterior mean rounds the cooperative onset upward, the overshoot marked in the figure and quantified at the $5\%$ level in Table~\ref{tab:isotherms}(b), then follows the true curve up the rise and onto the plateau, and the derivative panel makes the structural guarantee visible. The rate $f'(x)$ is non-negative everywhere, rises to its main peak across the onset, and decays to zero at the tail once the measurement at $x_{\max}$ pins the plateau. The rate should not be read below $x_{\min}$, where no data constrain it and the fixed knot grid shows a narrow near-origin feature that the true onset does not have. The response there is unaffected, since it starts from the pinned origin and stays small.

Across the fifty-seed campaign every regime is recovered to within the measurement noise (Table~\ref{tab:isotherms}). The saturation error, the RMS gap between the fitted and true response at nineteen saturation levels (Sec.~\ref{sec:metrics}), averages $0.039$ across families and ranges only from $0.037$ to $0.041$, below the output-noise level $\sigma_y=0.1$. The error that remains is smaller than the scatter in a single reading. The inferred plateau is within a few percent of the truth in every regime, and the recommended input delivers between $0.92$ and $1.00$ of the true plateau against a target of $0.95$, straddling it instead of sitting on one side. The five curves are genuinely different. Langmuir and Michaelis--Menten are concave, Sips is S-shaped, Elovich is kinked, and KWW has a long dispersive tail, yet the same model follows each of them. A surrogate built around a single kinetic form would instead force that form's curvature onto the others.

A fair objection is that someone who already knows the mechanism could fit its formula and do better. A parametric comparison shows otherwise (SI Sec.~\ref{sec:si-parametric}). Fitting each curve's own true form, an oracle that assumes the answer in advance, only ties the model-free fit (pooled saturation error $0.038$ against $0.039$), because at twenty noisy points the error is set by parameter uncertainty, and its read-out of the input is no more reliable. Fitting the wrong form, the realistic case when the mechanism is unknown, costs up to three times the error. Assuming no form therefore gives up nothing while removing the risk of guessing wrong.

The residual error comes from fitting noisy data, and the basis itself is close to exact. In the noise-free sweeps, fourteen of the fifteen regime variants are reproduced with a saturation error below $0.007$, far under the noise level, including the step-like Sips $n=4$ onset. The one exception is the strongly dispersive KWW curve with $\beta=0.35$ ($0.046$), whose $25\%$ level lies below the first measurable input, so that no model-free fit to data inside the window can resolve it (SI Sec.~\ref{sec:si-perfamily}).

The small errors that remain under noise have one reproducible pattern. The smooth basis rounds the onset up and approaches the plateau low, so the response error at the $5\%$ level is positive in four of the five families (KWW, whose $5\%$ crossing lies well below the window, is instead biased low there) and negative at the $95\%$ level in all five, with near-zero errors between (Table~\ref{tab:isotherms}(b)). The pattern is reproducible across seeds at several times its standard error, so it is a \emph{smoothing bias}, a systematic error that the penalized fit makes under noise and that is the same in every seed, and it appears in every regime. The input errors stay small through the middle of the rise, while at $95\%$ they grow, change sign from family to family, and their error bars widen sharply. That widening is the first sign of the conditioning problem that Sec.~\ref{sec:pulse_time} takes up.

\begin{table}[htbp]
\centering
\caption{Recovery and level-wise errors at $N=20$ under uncertainty sampling, as mean $\pm$ one s.d.\ over the $R=50$ seeds ($A=1$). (a) The saturation error across levels, the inferred plateau $f(x_{\max})$, and the delivered fraction $\theta_{\rm true}(\hat{x}_{\rm sat})$ of the true plateau at the recommended input (target $0.95$, a validation diagnostic, since it references the known true plateau $A$). (b) The level-wise response error $\Delta y$ and input error $\Delta x$ (in seconds) of Sec.~\ref{sec:metrics} at the five operating levels. Entries with a vanishing spread at the $5\%$ level are read-outs held at $x_{\min}$, because the true crossing lies below the window (SI Sec.~\ref{sec:si-metrics}).}
\label{tab:isotherms}
\footnotesize

{\centering(a) Overall recovery\par}
\vspace{2pt}
\begin{tabular}{lccc}
\toprule
curve & sat. error & $f(x_{\max})$ & $\theta_{\rm true}(\hat{x}_{\rm sat})$ \\
\midrule
Langmuir                           & $0.040\pm0.015$ & $0.998\pm0.047$ & $0.969\pm0.030$ \\
Michaelis--Menten                  & $0.037\pm0.015$ & $0.978\pm0.043$ & $0.920\pm0.041$ \\
Sips ($n=2$)                  & $0.038\pm0.015$ & $0.990\pm0.041$ & $0.936\pm0.026$ \\
Elovich                            & $0.041\pm0.014$ & $1.013\pm0.041$ & $0.995\pm0.014$ \\
KWW                                & $0.039\pm0.011$ & $0.988\pm0.039$ & $0.953\pm0.040$ \\
\midrule
mean                               & $0.039$ & $0.993$ & $0.955$ \\
\bottomrule
\end{tabular}

\vspace{8pt}
{\centering(b) Level-wise errors\par}
\vspace{2pt}
\setlength{\tabcolsep}{4pt}
\begin{tabular}{lccccc}
\toprule
curve & $5\%$ & $25\%$ & $50\%$ & $75\%$ & $95\%$ \\
\midrule
\multicolumn{6}{@{}l}{Response error $\Delta y(\alpha)$ (response units, $A=1$)}\\
Langmuir          & $+0.033\pm0.023$ & $+0.016\pm0.036$ & $+0.014\pm0.046$ & $-0.002\pm0.035$ & $-0.041\pm0.034$ \\
Michaelis--Menten & $+0.010\pm0.017$ & $+0.009\pm0.042$ & $+0.005\pm0.040$ & $-0.003\pm0.045$ & $-0.017\pm0.031$ \\
Sips ($n=2$)      & $+0.050\pm0.024$ & $+0.017\pm0.035$ & $+0.001\pm0.038$ & $-0.003\pm0.039$ & $-0.013\pm0.037$ \\
Elovich           & $+0.034\pm0.022$ & $-0.002\pm0.041$ & $+0.019\pm0.040$ & $+0.019\pm0.032$ & $-0.064\pm0.032$ \\
KWW               & $-0.031\pm0.005$ & $-0.008\pm0.040$ & $+0.021\pm0.036$ & $+0.005\pm0.039$ & $-0.034\pm0.032$ \\
\addlinespace
\midrule
\multicolumn{6}{@{}l}{Input error $\Delta x(\alpha)$ (seconds)}\\
Langmuir          & $-0.012\pm0.008$ & $-0.011\pm0.038$ & $-0.015\pm0.070$ & $+0.009\pm0.169$ & $+0.60\pm0.66$ \\
Michaelis--Menten & $+0.005\pm0.000$ & $-0.002\pm0.007$ & $-0.004\pm0.020$ & $-0.001\pm0.114$ & $-0.43\pm0.81$ \\
Sips ($n=2$)      & $-0.058\pm0.021$ & $-0.014\pm0.030$ & $-0.004\pm0.044$ & $-0.008\pm0.099$ & $-0.13\pm0.42$ \\
Elovich           & $-0.017\pm0.015$ & $+0.014\pm0.056$ & $-0.020\pm0.079$ & $-0.019\pm0.183$ & $+1.05\pm0.63$ \\
KWW               & $+0.009\pm0.000$ & $+0.002\pm0.008$ & $-0.015\pm0.029$ & $-0.016\pm0.134$ & $+0.51\pm0.89$ \\
\bottomrule
\end{tabular}
\end{table}

\subsection{Where the accuracy comes from}
\label{sec:whymono}

The surrogate differs from a textbook GP along three axes at once, input resolution, the pinned origin, and monotonicity, and it adds a roughness penalty on top. To learn which of these earns the accuracy of Table~\ref{tab:isotherms}, the ablation of Sec.~\ref{sec:comparators} runs three unconstrained GP arms and two unpenalized monotone bases through the identical loop, changing one design choice at a time (Table~\ref{tab:gp}). The three GP arms are a GP on the raw input axis (GP-linear), the same GP on a logarithmic axis (GP-log), and that GP with the origin pinned at zero (GP-log-pin). The two monotone bases carry no roughness penalty, one at the production resolution ($P=25$) and one coarser ($P=11$). The accuracy comes from resolution first and from the roughness penalty second, and the monotonicity constraint contributes almost nothing to it. Its contribution is a read-out that is always defined and unique.

Resolution is the largest single factor, and moving the GP from a linear to a logarithmic axis removes most of its deficit, from $2.44\times$ down to $1.36\times$ the penalized surrogate's error, because a single global lengthscale on the raw axis cannot follow the steep onsets, and pinning the origin closes a little more ($1.26\times$). Our own Langmuir-kernel predecessor\cite{17} fits the same pattern, since its plateau-compressing input warp acts much as the logarithmic axis does here. On this evidence the warp supplied resolution, and the correctness of the kinetic form behind it mattered less. The penalty is the other source, and it only shows once resolution is matched. Without the penalty the fine monotone basis, whose weights are each drawn independently from a half-normal prior so that neighboring weights are not tied together, follows the noise and is $1.6\times$ worse than GP-log-pin, whereas the GP tunes its smoothness to the data by choosing the lengthscale that best explains them. The coarse $P=11$ basis avoids that overfitting only by giving up resolution. With the penalty the same fine basis becomes the most accurate arm, $2.0\times$ better than its unpenalized twin and $1.26\times$ better than GP-log-pin, with the lower error on $66\%$ of the $250$ matched runs. Every arm in Table~\ref{tab:gp} differs from the penalized surrogate at $p<10^{-10}$ in the paired test.

Most of the lead over GP-log-pin comes from the acquisition, since each arm in Table~\ref{tab:gp} drives its own loop. The GP's band is nearly flat, because the marginal likelihood picks lengthscales of about half the warped domain, so its uncertainty sampling cycles among a few inputs and ends above what the same GP reaches on a log-spaced grid. Fitted to the very designs and readings the I-spline collected, GP-log-pin scores $1.07\times$ our error on the uncertainty-sampling designs and ties us on a log-spaced grid (SI Sec.~\ref{sec:si-gpmatched}). The part of the gap that disappears on identical designs is the price of the flat band, which is itself a property of the unconstrained stationary kernel on this problem. In absolute terms the lead over GP-log-pin is modest, a pooled saturation error of $0.039$ against $0.049$. It is also smaller than the movement the prior scales alone can produce in the fitted curve. Refitting the penalized surrogate with each of its two prior scales multiplied or divided by five, alone or together, shifts the fitted curve at the point where it moves most by $0.035$ to $0.066$ at $N=20$ depending on the family, almost entirely through the roughness-prior scale (SI Sec.~\ref{sec:si-priors}). The size of the lead over GP-log-pin should therefore be read as contingent on that scale staying within a factor of a few of its adopted value. Comparisons in which every arm shares the same surrogate and priors, such as the acquisition study of Sec.~\ref{sec:acq_compare}, are far less exposed, because a change in the prior moves all of their arms together.

\begin{table}[htbp]
\centering
\caption{GP ablation and the penalty decomposition (Secs.~\ref{sec:comparators} and~\ref{sec:whymono}), as the saturation error across levels at $N=20$ averaged over the same $R=50$ seeds at $\sigma_y=0.1$. Reading left to right isolates resolution and then the origin. Comparing unpen.\ $P{=}25$ with GP-log-pin isolates monotonicity at matched resolution, and the last two columns isolate the roughness penalty.}
\label{tab:gp}
\footnotesize
\setlength{\tabcolsep}{4pt}
\begin{tabular}{lcccccc}
\toprule
curve & GP-linear & GP-log & GP-log-pin & unpen.\ $P{=}11$ & unpen.\ $P{=}25$ & penalized \\
\midrule
Langmuir          & $0.075$ & $0.052$ & $0.054$ & $0.062$ & $0.085$ & $0.040$ \\
Michaelis--Menten & $0.144$ & $0.042$ & $0.039$ & $0.046$ & $0.047$ & $0.037$ \\
Sips ($n=2$)      & $0.067$ & $0.059$ & $0.050$ & $0.080$ & $0.126$ & $0.038$ \\
Elovich           & $0.061$ & $0.062$ & $0.062$ & $0.071$ & $0.094$ & $0.041$ \\
KWW               & $0.130$ & $0.050$ & $0.040$ & $0.037$ & $0.038$ & $0.039$ \\
\midrule
pooled            & $0.095$ & $0.053$ & $0.049$ & $0.059$ & $0.078$ & $\mathbf{0.039}$ \\
ratio to penalized& $2.44$  & $1.36$  & $1.26$  & $1.52$ & $2.00$ & n/a \\
paired $p$ (vs.\ pen.)& $2\times10^{-38}$ & $5\times10^{-16}$ & $3\times10^{-11}$ & $3\times10^{-27}$ & $4\times10^{-33}$ & n/a \\
\bottomrule
\end{tabular}
\end{table}

The constraint's benefit is structural and exact, and it is the reason the read-out of Sec.~\ref{sec:readout} is well posed. Across every refit in this study all draws are non-decreasing across the window and the pinned origin holds to machine precision, with no projection or guard applied at any stage, so the level-crossing read-out is unique on every draw. The unconstrained arms never reach this state, since even matched for resolution and origin the GP-log-pin posterior mean still dips somewhere in $70\%$ of fits, and the loop places about one in eight of its active measurements where its own mean is falling (full inventory in SI Sec.~\ref{sec:si-gpfail}). Those misled selections cost the GP no measurable curve accuracy, so the constraint contributes well-posedness, and well-posedness cannot be restored after the fact. Replacing the GP-log-pin mean, or every one of its draws, by the nearest non-decreasing curve (isotonic regression for the mean, the Lin and Dunson projection\cite{21} for the draws) leaves the saturation error at the raw GP's value and covers the true saturation input in only $61\%$ of intervals against our $94\%$. Projection keeps the overconfident spread of the raw GP, which covers only $64\%$. The only cost of building the constraint in is that the posterior has no closed form and must be sampled, about $2.5$~s per fit (SI Sec.~\ref{sec:si-compute}), which puts a full $20$-measurement determination at well under a minute of sampler time against the minutes to hours a physical measurement takes (SI Sec.~\ref{sec:si-cost}).

\begin{figure}[htbp]
\centering
\includegraphics[width=\textwidth]{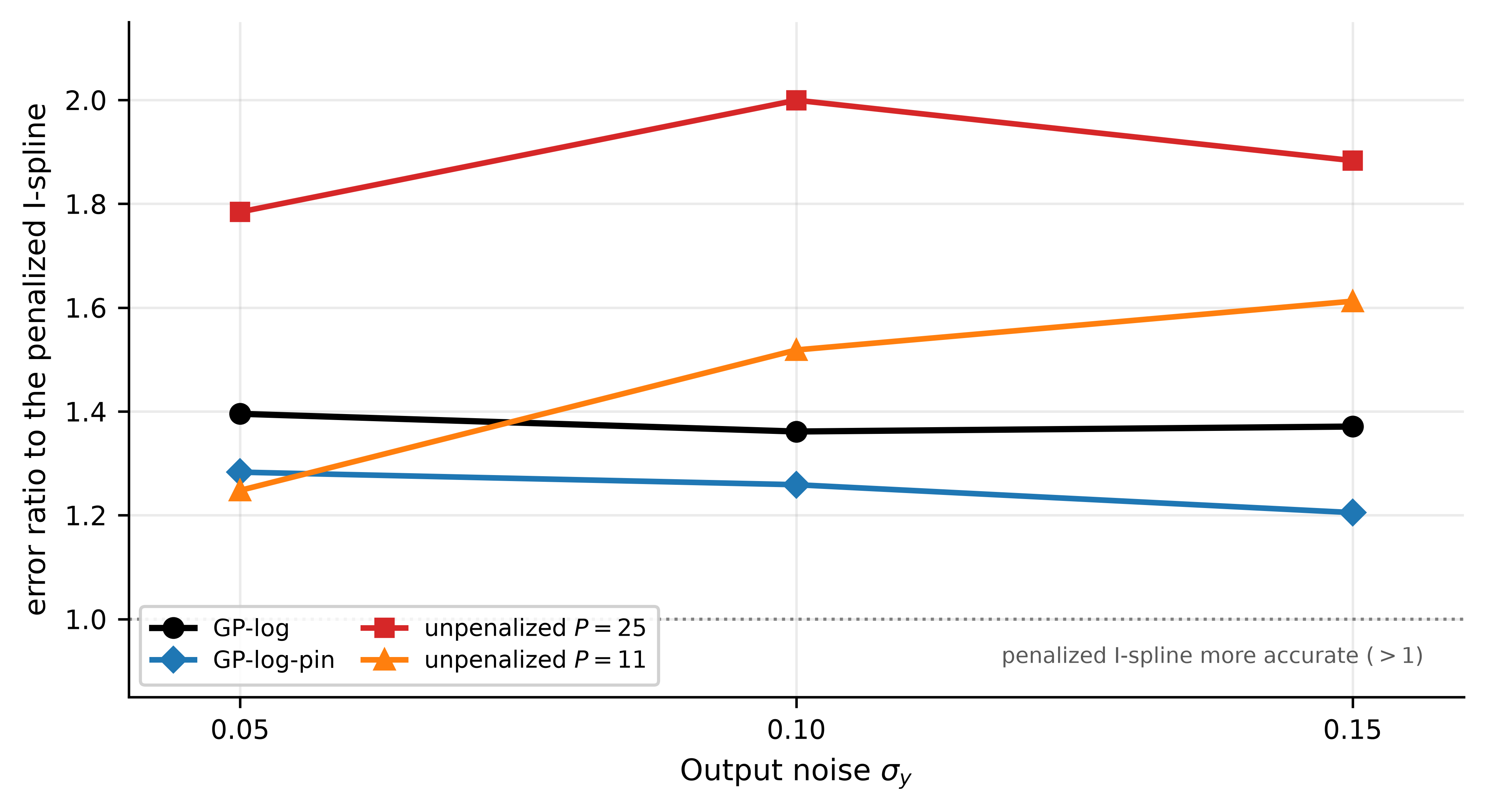}
\caption{Each arm of Table~\ref{tab:gp} other than GP-linear, as its saturation error at $N=20$ pooled over families divided by the penalized surrogate's, against the output-noise level (values in SI Table~\ref{tab:si-noise}). The two GP ratios stay flat while both unpenalized ratios end higher than they start, and the unpenalized $P=11$ line crosses above GP-log's near $\sigma_y=0.08$, the reversal of the bare constraint.}
\label{fig:noise}
\end{figure}

\paragraph{The same trend at every noise level.}
\label{sec:noisesweep}
Every result above is taken at one noise level, $\sigma_y=0.1$. To see whether the ranking of Table~\ref{tab:gp} holds as the scatter grows, we repeated the same comparison at output-noise levels of $0.05$, $0.1$, and $0.15$ with everything else fixed, for every arm of Table~\ref{tab:gp} except GP-linear, which is already far behind at the base level. Noise is also where a shape constraint is expected to matter most. A constraint is a form of regularization, a restriction that keeps the fit from following scatter, and regularization should pay off in proportion to the scatter it has to remove. Figure~\ref{fig:noise} plots each arm's error as a ratio to the penalized surrogate's against the noise level, and SI Table~\ref{tab:si-noise} lists the errors behind it.

The two GP ratios stay flat, about $1.4\times$ for GP-log and $1.2$ to $1.3\times$ for the origin-matched GP-log-pin, so the penalized surrogate's lead holds as the noise grows. Both unpenalized ratios end higher than they start, so the bare constraint loses ground exactly where regularization was expected to help, the opposite of what that argument predicts. The coarse $P=11$ basis is the unpenalized arm to watch, because its few knots act as a crude fixed smoother, so at the lowest level it is slightly more accurate than GP-log, and as the noise grows it falls behind, its line crossing above GP-log's at a noise just below a tenth of the plateau (near $\sigma_y=0.08$ in Fig.~\ref{fig:noise}). This is the reversal a fixed basis with no control over its own smoothness suffers once it starts fitting scatter. No line comes near one, so the penalized surrogate is the most accurate arm at every level, and it stays about twice as accurate as its own unpenalized twin throughout. On identical designs the lead over GP-log-pin shrinks to a few percent at every level (SI Sec.~\ref{sec:si-gpmatched}), so here too most of the margin is the GP's acquisition, while GP-log, without the pinned origin, trails on the same data by a margin that grows with the noise. As the data get noisier, the penalty keeps the monotone surrogate ahead. The constraint on its own does not.

The reason is smoothness adaptation, which the GP has and the unpenalized basis lacks. The GP re-estimates its lengthscale and noise level at every step, so as the noise grows it attributes more of the scatter to noise and returns a smoother mean, whereas the unpenalized monotone basis has no comparable control. Its inferred weight scale sets the height of the curve, which the measurements at $x_{\max}$ fix anyway, and neither it nor the inferred noise level says anything about the curve's roughness. The roughness penalty supplies exactly that control through its inferred scale $\tau_\eta$, which is why the penalized surrogate stays ahead of the GP across the whole range. SI Sec.~\ref{sec:si-noise} rules out the two other candidate explanations, the plateau anchor and the noise prior, and gives the breakdown by family.

The guarantee, meanwhile, does not depend on the noise level at all. None of the three monotone arms produces a non-monotone draw at any level, whereas the unconstrained mean dips somewhere in most fits at every level, with a deepest dip that grows with the noise.

\subsection{Reading the saturation input from the posterior}
\label{sec:pulse_time}

Once the curve is recovered, the quantity most often asked of it is a saturation input, and when the aim is to set a process at saturation that is the input at $95\%$ of the plateau, $x_{\rm sat}(0.95)$, read from the posterior by level crossing (Sec.~\ref{sec:readout}). Table~\ref{tab:isotherms}(b) shows how well that read-out does at each level. From the $25\%$ level through $75\%$ saturation the predicted inputs are within about $0.02$~s of the truth in every regime (and within $0.06$~s at the $5\%$ level). At the $95\%$ level the error grows to between $-0.4$ and $+1.1$~s across the five regimes, a mix of over- and under-shoot with a pooled bias of $+0.32$~s.

The sign of that error is where the order of the penalty shows. The production surrogate penalizes the second differences of the log-weights (Sec.~\ref{sec:priors_sec}), which leaves a smoothly decaying rate unpenalized. Refitting the same designs with a first-order penalty, which penalizes changes in the level of the log-weights and so favors a flat log-weight profile, reads the input late, by a median of $+1.8$~s over the $250$ runs (SI Table~\ref{tab:si-model-ablation}). On the log-spaced knots a flat log-weight profile is a $1/x$ rate, so the fit keeps rising into the plateau, levels off only grudgingly, and reaches the crossing late. The second-order penalty lets the rate roll off at nearly the true pace, so the inferred plateau stays calibrated ($0.98$--$1.01$, Table~\ref{tab:isotherms}(a)) and the crossing is neither raised nor delayed. The residual bias, being mixed in sign, largely averages away across regimes.

\begin{figure}[htbp]
\centering
\includegraphics[width=\textwidth]{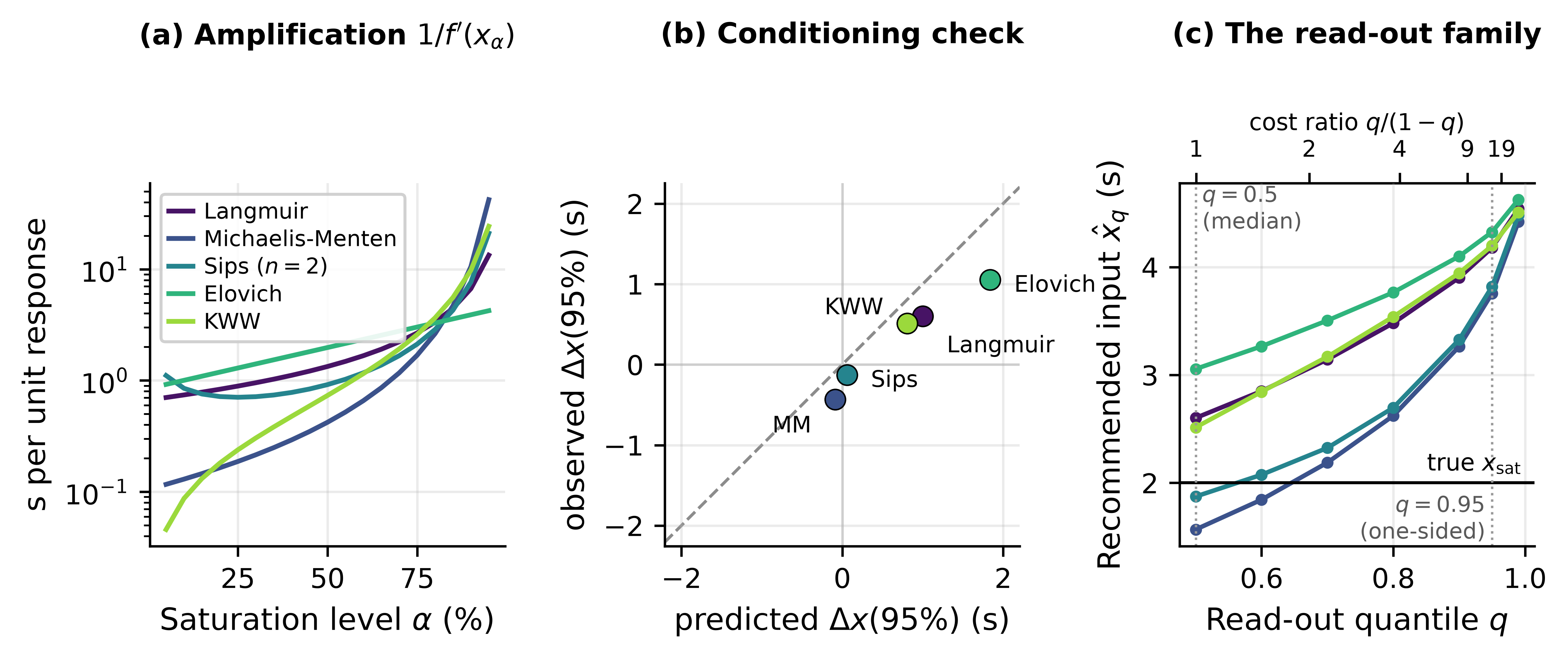}
\caption{Reading the recommended input out of the posterior. (a) The amplification $1/f'(x_\alpha)$ of the five true curves against saturation level. (b) The input shift at $95\%$ predicted by the conditioning relation against the observed one, one point per family (MM is Michaelis--Menten), with the dashed line marking equality (SI Sec.~\ref{sec:si-cond}). (c) The recommended input $\hat{x}_q$ of Eq.~(\ref{eq:quantile}) against the read-out quantile $q$, one line per family, with the truth as the solid line. The top axis gives the cost ratio $q/(1-q)$ that sets $q$, and the dotted vertical lines mark the median ($q=0.5$) and the one-sided limit ($q=0.95$).}
\label{fig:readout}
\end{figure}

The jump between the $75\%$ and $95\%$ levels is a property of the question itself, and it follows from the conditioning relation of Sec.~\ref{sec:readout}. Near saturation the curve is almost flat, so a small error in fitted height becomes a large error in the recovered input, by a factor equal to the reciprocal of the slope at the crossing. How large that amplification is depends on how gently the curve approaches its plateau, and on any smoothly saturating curve it is typically an order of magnitude larger at the $95\%$ level than in the middle of the rise. On the five benchmark curves the $95\%$-level amplification itself spans an order of magnitude, from the abruptly capped Elovich form to the slowly saturating Michaelis--Menten form (Fig.~\ref{fig:readout}a), and on the fitted curves, the ones the read-out actually inverts, it lies in a narrower band (SI Table~\ref{tab:si-cond}). Every method inherits this amplification and controls only the size of the height error being amplified. Under the second-order penalty that error is small, which is why even the worst family (Elovich, $+1.05$~s) stays within about a second. The conditioning relation also predicts the order of magnitude of each family's shift from its plateau and level errors, and its sign in every family except Sips, whose shift lies within seed scatter (Fig.~\ref{fig:readout}b, SI Sec.~\ref{sec:si-cond}).

\paragraph{Coverage and calibration.}
\label{sec:calibration}
A posterior is only useful if its uncertainty is honest, so we check both the read-out and the curve for coverage, how often a nominal $95\%$ interval actually contains the truth. For the read-out the interval is the central $95\%$ of the crossing posterior. It covers the true $x_{\rm sat}=2.0$~s in $94\%$ of runs pooled over the five families, close to the nominal level, with the kinked Elovich curve lowest and Sips highest (Table~\ref{tab:readout}). The unconstrained GP-log-pin arm, read by the identical convention, covers the truth in only $64\%$ of runs and is biased later ($+0.44$~s against $+0.32$~s). Its posterior is narrower and prone to dips, so it is confidently wrong about where the crossing lies. Part of that gap is due to the GP treating its fitted hyperparameters as known (Sec.~\ref{sec:comparators}), so the comparison bounds the constraint's own contribution from above, but the direction is not in doubt. The monotone surrogate is at once the most accurate arm and the one whose crossing is honestly covered, which is why its read-out is the one we recommend.

The credible band on the curve is honest to the same degree. It contains the true curve at $94\%$ of input points against a nominal $95\%$, and at $96\%$ over the region where data exist, where no family stands out. The small shortfall sits entirely below the first measurable input, where every draw starts from the pinned origin and the band is almost closed. Across the window the band is if anything wider than the run-to-run scatter of the fit, so the coverage it loses comes from the smoothing bias of Sec.~\ref{sec:per_isotherm}, a systematic error that no variance band can express. The band's proper use is to say how much to trust the curve. It does not say where the curve is wrong, because its width is only weakly related to where the error falls, so it should be read as a coverage statement instead of a map of the error. Two approximations mean the nominal $95\%$ should be read as approximate. The likelihood assumes one noise level everywhere, although the input jitter makes the true scatter larger where the curve is steep. The band is also drawn with equal room above and below the mean, although near the origin and the plateau the plausible curves are boxed in on one side, so their spread is lopsided. Both approximations, the per-family coverage values, the width-to-error relation, and the comparison with the unpenalized basis are given in SI Sec.~\ref{sec:si-window}.

\paragraph{Which point to report.}
The recommendation is one point on the crossing posterior, Eq.~(\ref{eq:quantile}), and which point to report is a cost decision that does not correct for bias. The median ($q=0.5$) is close to the truth and well covered, and the one-sided limit ($q=0.95$) is deliberately conservative. Figure~\ref{fig:readout}c shows the whole family of $\hat{x}_q$ against $q$ for the five curves, and SI Sec.~\ref{sec:si-quantile} shows how a practitioner sets $q$ from the ratio of the cost of under-saturating to the cost of over-dosing. Because the bias of the median read-out is a few tenths of a second, the recurring overshoot is a minor cost (SI Sec.~\ref{sec:si-quantile}). The under-saturation risk $R(x)$ of Sec.~\ref{sec:readout} carries the same information in the units the practitioner acts in. The one caveat is that the median read-out and its coverage hold for any window that brackets saturation but fail if the window is too short, the failure the window-sufficiency statistic $r$ is designed to catch (SI Table~\ref{tab:window} and Sec.~\ref{sec:si-window}).

\begin{table}[htbp]
\centering
\caption{The saturation input $x_{\rm sat}(0.95)$ read off the I-spline and GP-log-pin posteriors by the same level-crossing convention, at $N=20$ over $50$ seeds per family, with each surrogate driving its own uncertainty-sampling loop. Bias is each run's median crossing minus the true $2.0$~s, averaged over the runs, and coverage is the fraction of runs whose nominal $95\%$ two-sided interval contains it, with a binomial standard error\cite{64} of up to $7$ percentage points per family and up to $3$ for the pooled row over all $250$ runs.}
\label{tab:readout}
\footnotesize
\setlength{\tabcolsep}{4pt}
\begin{tabular}{lcccc}
\toprule
 & \multicolumn{2}{c}{I-spline} & \multicolumn{2}{c}{GP-log-pin} \\
\cmidrule(lr){2-3}\cmidrule(lr){4-5}
curve & bias (s) & coverage & bias (s) & coverage \\
\midrule
Langmuir          & $+0.60$ & $94\%$ & $+0.88$ & $48\%$ \\
Michaelis--Menten & $-0.43$ & $98\%$ & $-0.60$ & $82\%$ \\
Sips ($n=2$)      & $-0.13$ & $100\%$ & $-0.05$ & $84\%$ \\
Elovich           & $+1.05$ & $78\%$ & $+1.38$ & $32\%$ \\
KWW               & $+0.51$ & $98\%$ & $+0.58$ & $74\%$ \\
\midrule
pooled            & $+0.32$ & $94\%$ & $+0.44$ & $64\%$ \\
\bottomrule
\end{tabular}
\end{table}

\subsection{Choosing where to measure}
\label{sec:acq_compare}

With the surrogate's accuracy and read-out established, we ask how much the adaptive loop gains over a non-adaptive design of the same budget. The input axis is logarithmic in character and all five curves do their work in the first fraction of the window, so uniform random selection is too weak a baseline, and a design that respects the log scale is the fair yardstick. To keep a fixed grid from knowing in advance where to look, we randomize the saturation location for each seed ($x_{\rm sat}$ log-uniform in $[0.2,4.0]$~s, with each family's scale re-solved so that only the position of the rise varies). Uncertainty sampling beats log-uniform random selection by $1.19\times$ pooled and uniform random selection by $1.79\times$, and holds a smaller but consistent edge over a log-spaced grid ($1.13\times$ pooled, $1.12\times$ in the fixed-location campaign), winning on the three families whose rise is spread out (Langmuir, Sips, and Elovich) and tying on the two fast-rising ones (Michaelis--Menten and KWW). A log-spaced grid remains a strong baseline, because it builds in the near-origin resolution that a bounded log-scaled curve rewards. The full table and three robustness controls (the candidate-set geometry, the survey size, and an unconstrained acquisition driver) are in SI Sec.~\ref{sec:si-controls}.

The case for the adaptive rule rests on two further properties. The first is reliability, since uncertainty sampling reaches an error below the output-noise level by $N=20$ in 100\% of seeds on every family, whereas the two random designs reach it in only 68--94\% of seeds on the sharply structured Sips and Elovich curves. The second is nesting, which no fixed grid offers, because every measurement stays in the design. The experimenter can therefore stop once the band is tight enough and resume if it is not, whereas a log-spaced grid's $N$-point design is not a subset of its $2N$-point design and extending it means starting over. On these curves the error falls below the output-noise level at a median of $5$ to $9$ measurements, well inside the twenty-point budget. We do not have a mechanism for the small edge over the grid. Uncertainty sampling needs the band only to say where the data leave the curve undetermined, and the band does this reliably, because it is wide in the gaps between measurements and narrow near them, so the rule fills the largest gap on the surrogate's own scale. Saying where the fitted mean is actually wrong is a different job, and the band does it only weakly (Sec.~\ref{sec:calibration}), because much of the remaining error is the smoothing bias, which is the same in every seed and which no variance band can express. The edge therefore does not come from the surrogate pointing at its own mistakes, and we report it as an observation. The contribution we claim is the surrogate that makes the band and the read-out trustworthy, whichever design supplies its data.

\begin{figure}[htbp]
\centering
\includegraphics[width=\textwidth]{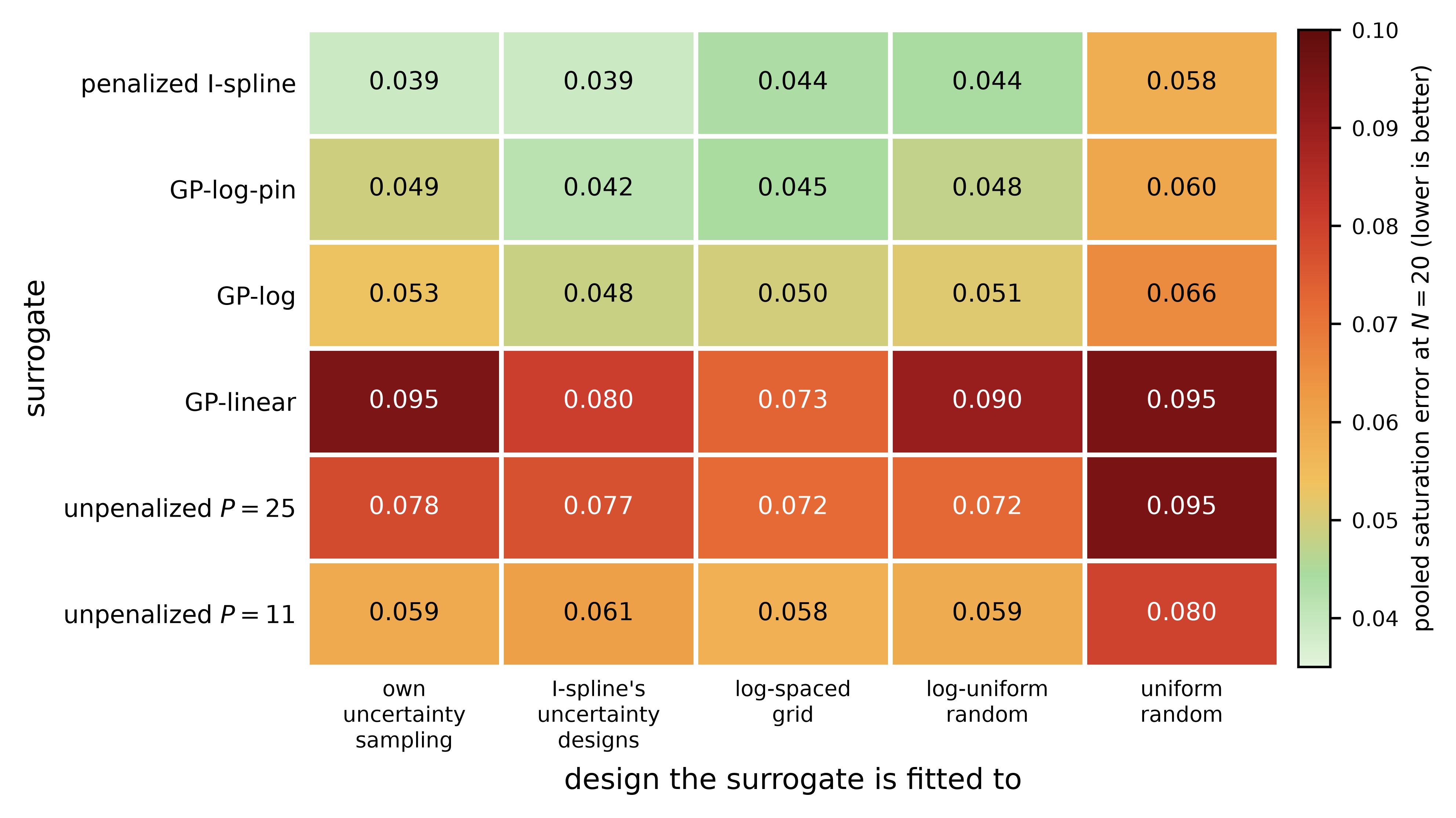}
\caption{Surrogate and design on one grid. Each cell is the pooled saturation error at $N=20$ over five families and $50$ seeds for one surrogate fitted to one kind of design, and lighter is better. In the first column each surrogate drives its own uncertainty-sampling loop, and in the other columns the surrogates are fitted to the designs the I-spline campaign stored (SI Sec.~\ref{sec:si-gpmatched}).}
\label{fig:scorecard}
\end{figure}

Figure~\ref{fig:scorecard} puts the surrogate and the design on one grid. Each cell is the pooled saturation error at $N=20$ of one surrogate fitted to one kind of design, with every other surrogate fitted to the I-spline's stored designs in every column but the first (SI Sec.~\ref{sec:si-gpmatched}). Reading down a column, the penalized I-spline, GP-log-pin, and GP-log differ little on the same data, GP-linear and the unpenalized bases trail them by up to twofold, and the penalized I-spline is the lightest cell in every column. Reading along a log-axis GP row, the first cell, that arm driving its own loop, is darker than its cell on any log-scaled design, so what the GP loses it loses in the acquisition. Reading along the penalized I-spline row, its own uncertainty sampling is its lightest cell, ahead of the grid and both random designs. The unpenalized bases and GP-linear behave like the log-axis GPs in this respect, since none of them does better under its own uncertainty sampling than on a log-spaced grid, so the penalized I-spline is the only surrogate whose own loop is its best design.

\subsection{Beyond the development set}
\label{sec:heldout}

Everything so far is measured on the five families that shaped the surrogate. The spline order, the number of knots, and their logarithmic spacing were all chosen with those curves in view, so a score on them alone is a configuration graded against the curves that selected it. To separate the surrogate from the curves that shaped it, we froze the configuration, holding the basis, its knots, the prior scales, and the sampler settings exactly as they were set on the five families, and ran it once on ground truths drawn from a distribution it had never seen.

Two generators cover the two kinds of shape a self-limiting response can take. A mixture of up to eight exponentials spans the shapes whose rate only falls, from Langmuir through dispersive, Elovich-like rises, and a Hill sigmoid supplies the cooperative onsets. Each curve is rescaled to a random saturation location, as in Sec.~\ref{sec:acq_compare}, so neither the shape nor the position of the rise is fixed. We drew $200$ mixtures and $50$ Hill curves. All $250$ are monotone, start at zero, and saturate inside the window. The set is not entirely new, though, since a one-component mixture is a Langmuir curve, and the position of the rise accounts for more of what separates a held-out curve from the nearest development curve than its shape does. The benchmark therefore tests generalization mostly in rise location and in shape only partially (SI Sec.~\ref{sec:si-heldout}).

Each held-out curve was measured under the standard noise model (Sec.~\ref{sec:noise}), and two surrogates each ran their own uncertainty-sampling loop to $N=20$ on it once, the frozen configuration and GP-log-pin, the strongest of the unconstrained arms in Table~\ref{tab:gp} and therefore the comparator reported here (GP-log was run as well and trails further, SI Sec.~\ref{sec:si-heldout}). The frozen configuration does not overfit its development set. Its saturation error averaged over the $250$ held-out curves is only slightly above its average over the five it was tuned on (Table~\ref{tab:heldout} against Table~\ref{tab:isotherms}), and it leads GP-log-pin on both kinds of shape. Fitted to the I-spline's own held-out designs, GP-log-pin still trails, by more than on the development set, and exceeds the output-noise level on about twice as many curves, so on the held-out curves a larger part of the lead is the surrogate itself (SI Sec.~\ref{sec:si-gpmatched}).

In mean accuracy that lead is modest, and in reliability, the property a practitioner facing an unfamiliar system cares about most, it is clear. The I-spline exceeds the output-noise level on a few curves in a hundred and GP-log-pin on about three times as many, and the largest I-spline error over the $250$ curves is below the largest GP-log-pin error.

One possible confound is that the saturation location is drawn log-uniformly, the same measure the log-spaced basis is matched to, so part of the lead could be that alignment. Repeating the whole benchmark with the location drawn uniformly (lower block of Table~\ref{tab:heldout}) leaves the mean advantage unchanged and the reliability edge intact, with no I-spline curve above the noise level while GP-log-pin still has some.

The modest mean lead over GP-log-pin found in Sec.~\ref{sec:whymono} therefore holds on curves drawn from a distribution instead of chosen by hand, and on these curves the lead shows chiefly in the reliability of the tail.

\begin{table}[htbp]
\centering
\caption{Saturation error at $N=20$ on the $250$ held-out curves, each surrogate driving its own uncertainty-sampling loop once per curve, for the frozen I-spline and for GP-log-pin. The last column is the fraction of curves whose error exceeds the output noise $\sigma_y=0.1$, and the lower block draws the saturation location uniformly instead of log-uniformly.}
\label{tab:heldout}
\begin{tabular}{llccc}
\toprule
generator & surrogate & mean & max & $>\sigma_y$ \\
\midrule
mixture ($200$)   & I-spline        & $0.045$ & $0.171$ & $3\%$ \\
                  & GP-log-pin      & $0.057$ & $0.194$ & $9\%$ \\
\midrule
Hill ($50$)       & I-spline        & $0.044$ & $0.112$ & $2\%$ \\
                  & GP-log-pin      & $0.060$ & $0.142$ & $10\%$ \\
\midrule
all ($250$)       & I-spline        & $0.045$ & $0.171$ & $3\%$ \\
                  & GP-log-pin      & $0.058$ & $0.194$ & $9\%$ \\
\midrule
\multicolumn{5}{l}{\emph{$x_{\rm sat}$ drawn uniformly (control)}} \\
all ($250$)       & I-spline        & $0.040$ & $0.092$ & $0\%$ \\
                  & GP-log-pin      & $0.053$ & $0.184$ & $7\%$ \\
\bottomrule
\end{tabular}
\end{table}

\subsection{Limitations and outlook}
\label{sec:limitations}

\paragraph{When an assumption fails.} Every ground truth in the benchmark rises without falling and levels off inside the window, and every measurement is clean, with Gaussian scatter and no outlier or drift. In SI Sec.~\ref{sec:si-outside} we break the window and the clean-measurement assumptions in four ways and watch what happens, and for each case we give a check that can be run on real data, where the truth is unknown. Two of the four cases matter, and a third weakness needs no test to see.

The first case is a window that ends before the curve has leveled off. The method takes the response at $x_{\max}$ as the plateau. If the curve is still rising there, the method still reports a plateau, a crossing, and a narrow interval, and it is wrong without any sign of trouble in the fit itself. When we cut the window short on the benchmark curves, the recommended input falls far short of the target. The check for this is the window-sufficiency statistic $r$ of Sec.~\ref{sec:readout}, which is small when the window is long enough and several times larger when it is not. A cutoff set on the benchmark catches most short windows and rarely flags a long one. The cutoff is not a fixed number, though. The fitted curve levels off gradually, so $r$ grows as the window is made tighter even when the window is still long enough, and a cutoff set on a long window raises false alarms on a tight one. The cutoff must therefore be read against the length of the window relative to the recommended input, and the released tool prints $r$ together with the values to expect for windows of different lengths.

The second case is a bad reading at the plateau. The read-out is measured against the fitted plateau, so the readings at $x_{\max}$ carry more weight than any others. When one of them is pushed up by a few times the noise level, the fitted plateau rises and the recommended input moves late, in the direction that the read-out's sensitivity to the plateau predicts and by about half the size a linear sensitivity would give (SI Sec.~\ref{sec:si-outside}). Forcing extra clean readings at $x_{\max}$ did not fix it. The loop already places several there on its own, so that test is not decisive, but it shows that a monotone fit at its right edge is pulled toward the largest reading there. The fix that helps is a likelihood that tolerates outliers. Replacing the Gaussian likelihood by a Student-$t$ likelihood\cite{65} cuts the effect of the bad reading on the plateau by about a third and on the read-out by nearly half, costs almost nothing on clean data (SI Table~\ref{tab:si-studentt}), and keeps every guarantee, because monotonicity lives in the prior and the likelihood has no part in it. This is the setting we recommend for field data, and the released tool switches to it with one flag. Another route around the read-out's dependence on the fitted plateau is to make the saturation input itself a parameter of the model, inferred with the rest, so that it need not be read off afterward.

The other two cases are mild for the read-out, since a slow drift in the instrument's response over the campaign, added as a straight-line trend across the measurements, triples the saturation error but barely moves the recommended input, with the drift rate recoverable from the loop's repeated measurements (SI Sec.~\ref{sec:si-outside}), and an exact repeat of a reading changes nothing.

The third weakness comes from the guarantee itself. A response that rises, peaks, and then falls, because some loss process works against the uptake, has a falling part that the monotone prior cannot represent. The fit will pass a rising curve through such data, and nothing inside the model warns that it did, so the method should be used only where the response is known in advance to rise without falling.

\paragraph{What was held fixed, and what it costs.} Three choices in the model are fixed before any data arrive. The first is the knots, whose number and places do not move during the loop. Many log-spaced knots give the basis enough detail to follow every shape in the benchmark, including the step-like Sips $n=4$ onset, and the roughness penalty stops the fine basis from fitting noise, so detail and smoothing work together. An onset sharper than any tested here could still fall between the knots of a fixed grid. A natural extension is to insert knots where they most change the fit, as the MaxMod algorithm does for a constrained GP on a fixed data set,\cite{35} and to repeat this as measurements arrive. The second is the pinned origin, a hard rule that some saturating responses break, for example an assay with a baseline offset. Relaxing it costs one parameter, an intercept $f_0\ge0$ with a half-normal prior, $f(x)=f_0+\sum_m w_m I_m(x)$, which leaves the slope, and so the monotonicity, untouched and gives back the present model at $f_0=0$. The third is the scaling of the response in Sec.~\ref{sec:priors_sec}, which changes nothing in the results reported here because every ground truth has $A=1$. On measured data the released tool divides every reading by the largest one, a single noisy value that is biased high because it is the maximum. The fit is scaled back by the same number, so this divisor reaches the reported curve only through the prior scales, and a shift of a few percent in those scales has only a modest pull at the budgets used here (SI Sec.~\ref{sec:si-priors}). Dividing instead by the mean of the repeated readings at $x_{\max}$, which the loop makes on its own (Sec.~\ref{sec:acq}), would remove the bias.

The roughness penalty also has a sampling cost, since its hierarchical prior gives the posterior the funnel shape that Hamiltonian samplers find hard, and with NumPyro's default uniform starting point the production runs show divergent transitions. Starting the chains at the prior median, the default in the released tool for measured data, removes most of them and leaves every reported number unchanged at its quoted precision, so we report the campaigns as run and give the diagnostics and the paired comparison in SI Sec.~\ref{sec:si-diag}.

\paragraph{Outlook and a practical recipe.} The five-family study is a controlled comparison, in which every strategy, surrogate, and noise level sees the same ground truths and the same seeds, so the differences between them are paired, but on its own the study says nothing about saturation curves in general. The held-out benchmark extends the claim to a whole distribution of shapes, with the caveat about shape coverage given in Sec.~\ref{sec:heldout}. Both assume one input in a fixed environment. Richer settings, with several inputs or with regimes that change with the environment, would need a richer basis but leave the monotonicity machinery as it is, since it rests only on non-negative M-splines carrying non-negative weights. A product of one I-spline basis per input, with non-negative weights, rises in each input and would map a saturation surface. On the acquisition side, a rule that aims at where the fit is likely to be wrong, or at $x_{\rm sat}$ directly, is the change most likely to give adaptive selection a clear margin over a log-spaced grid, and the rules developed for locating a contour\cite{38,39,40} are the natural place to start.

On a new system, choose $x_{\max}$ generously so that the window reaches past saturation, use the Student-$t$ likelihood if a bad reading is plausible, and run the survey, including $x_{\min}$ and $x_{\max}$. Add points by uncertainty sampling and stop when the band is narrow enough (on the benchmark the error falls below the noise level well within the evaluation budget, Sec.~\ref{sec:acq_compare}). Check $r$ against the values expected for a window of that length. Report $\hat{x}_q$ at the $q$ that the cost ratio of Eq.~(\ref{eq:quantile}) implies, together with the under-saturation risk $R(x)$ at the intended input. Every step is in the released implementation.

\section{Conclusions}

We have presented a shape-constrained Bayesian active-learning platform for saturation curves, built on a penalized monotone I-spline and tested throughout on synthetic ground truths. Non-negative weights, kept positive through a random walk on their logarithms, make every posterior draw rise from the pinned origin without falling, and a second-order roughness penalty on the log-weights, with an inferred scale, adapts the smoothing to the noise. A survey measurement at the largest input anchors the plateau in place of a flatness prior, and the saturation input is read from the posterior by level crossing, with no kinetic form assumed.

An ablation that runs unconstrained GPs and two unpenalized monotone bases through the same loop, changing one feature at a time, traces the accuracy to two sources, the multi-scale logarithmic resolution and the roughness penalty. Matched for resolution and origin, the bare shape constraint adds no accuracy, since without the penalty the fine monotone basis is less accurate than the matched GP. The constraint's contribution is a read-out that is always defined, a level crossing unique on every draw. The penalty is decisive, since it removes the reversal that the bare constraint suffers at high noise and keeps the penalized surrogate the most accurate arm at every noise level. Its order matters too, since the second-order form, which leaves a smoothly decaying rate unpenalized, centers the read-out of the saturation input with honest coverage, where a first-order penalty reads late. Fixing monotonicity after a GP fit recovers neither the accuracy nor the calibration, and a parametric fit that already knows the true form only ties. Figure~\ref{fig:scorecard} gathers the surrogates and the designs on one grid, where the penalized I-spline is the lightest cell in every column and the only surrogate whose own loop is its best design. On $250$ curves from an unseen distribution the frozen configuration keeps its accuracy, and its lead over the matched GP is modest in the mean and clearest in the tail.

Paired with ordinary uncertainty sampling, the platform reaches the output-noise level within twenty measurements in every seed, beats both random designs, and edges a log-spaced grid. The edge over the grid is small, though, and the decisive contribution is the surrogate. Recovering the input at a high saturation level is intrinsically ill conditioned, because near the plateau a small response error becomes a large input error whatever the model, so we report it as a cost-indexed quantile together with its under-saturation risk. Where that input is the quantity carried into a process, its accuracy and honest reporting matter more than any further gain in measurement efficiency.

The construction carries over to any monotone response that rises from a baseline to a plateau, adsorption isotherms and enzyme-kinetic assays among them, and a non-negative intercept would relax the pinned origin for a baseline offset without touching the monotonicity guarantee (Sec.~\ref{sec:limitations}). The sequential loop carries over more narrowly, to settings where measurements are costly and taken one at a time, which they are not in a multi-well plate assay or an automated volumetric run, so we claim the surrogate broadly and the loop narrowly. Applying the platform to a measured response curve, where the normalization is not an identity and the plateau anchor is a real instrument reading, is the next step, and the released implementation does this on a two-column data file without retuning. We limit the demonstration to synthetic ground truths, for which the true curve is known exactly and every claim can be checked against it. A measured curve whose reduction to a clean saturation profile is itself uncertain would test that reduction as much as the method.

\begin{acknowledgement}
The authors acknowledge the Department of Biomedical Engineering at the University of Illinois Chicago and the Kurt J. Lesker Company for their support. The authors used a generative artificial-intelligence assistant (Claude, Anthropic) to improve the readability of the text and to help write and check the code. The authors reviewed and edited all of the resulting material and take full responsibility for the content.
\end{acknowledgement}

\paragraph{Data and Software Availability.}
This is a simulation study, in which all data are generated by the ground-truth curves of Sec.~\ref{sec:kinetics} and the held-out generators of SI Sec.~\ref{sec:si-heldout} under the noise model of Sec.~\ref{sec:noise} and no experimental measurements are used. The Python implementation, the scripts that regenerate every figure, table, and quoted number in the main text and the SI, and the cached results needed to do so without re-running the campaign are openly available at \url{https://github.com/pouyna/SplineBAL} under the MIT license. The exact package versions are recorded with the code, and the hardware, sampler settings, and seed protocol are given in SI Sec.~\ref{sec:si-compute}.

\paragraph{Competing Interests.}
The authors declare no competing financial interest.

\begin{suppinfo}
The Supporting Information, appended after the references, gives the measurement setting (how the campaign differs from a hand-measured sweep, and the cost basis of the read-out quantile), a glossary of the statistical terms, the surrogate in full (its spline basis, the classical form of the roughness penalty, and the read-out formulas), the comparison models and how the hyperparameters were fixed, the study configuration (evaluation metrics, study design, and basis configuration), sampler and prior diagnostics, the five kinetic families in detail, the input read-out (its conditioning, window dependence, and band approximations), the comparison arms (GP failure modes, the matched-design control, and parametric baselines), the acquisition controls, the held-out generator and its change-of-measure control, the high-noise mechanism diagnostics, behavior outside the benchmark's assumptions, a notation table, and computational details.
\end{suppinfo}

\bibliography{refs}

\clearpage
\setcounter{page}{1}
\renewcommand{\thepage}{S-\arabic{page}}
\begin{center}
{\LARGE\bfseries Supporting Information}\\[8pt]
{\Large Shape-Constrained Bayesian Active Learning of\\ Self-Limiting Saturation Curves}\\[10pt]
{\large Pouyan Navabi and Christos G. Takoudis}
\end{center}
\vspace{12pt}

\setcounter{section}{0}
\setcounter{figure}{0}
\setcounter{equation}{0}
\setcounter{table}{0}
\renewcommand{\thesection}{S\arabic{section}}
\renewcommand{\thesubsection}{S\arabic{section}.\arabic{subsection}}
\renewcommand{\thefigure}{S\arabic{figure}}
\renewcommand{\theequation}{S\arabic{equation}}
\renewcommand{\thetable}{S\arabic{table}}

\section{The measurement setting and the cost of the recommended input}

\subsection{Atomic layer deposition as the setting, and what the loop changes}
\label{sec:si-cost}
The cost arguments in this paper only mean something in a setting where each measurement is expensive. Atomic layer deposition (ALD) is one such setting. For readers new to it, this subsection explains the measured curve, how a practitioner measures it today, and what the loop changes. Nothing in the method depends on this setting.

An ALD cycle exposes a substrate to a pulse of precursor vapor, purges, exposes it to a co-reactant, and purges again. Each surface reaction uses up a finite number of surface sites and stops when they are gone, so the process is self-limiting, and this gives ALD its atomic-scale thickness control.\cite{4,5} The response is the film thickness added per complete cycle, measured in \AA{} per cycle by ellipsometry or X-ray reflectivity, or as the mass gained per cycle on a quartz-crystal microbalance. Plotting this response against the length of one precursor pulse gives the saturation curve. It starts at zero (no input, no growth), climbs as more of the surface reacts, and flattens once the reaction has run to completion. The plateau is the signature of self-limiting behavior, and a process engineer needs to find where it begins.

Each point on that curve costs a separate experiment, and how much depends on how the film is measured. When the film is characterized outside the reactor, one point means running for hundreds of cycles at a fixed input, enough film to measure reliably, and then venting, unloading, and characterizing the sample, so a point takes hours of tool time plus precursor. When the film is measured in situ, by ellipsometry\cite{66} or a quartz-crystal microbalance\cite{67} inside the reactor, a point is read without unloading, and it costs the cycles needed for a reading the analysis can use. Precursors can be expensive and sometimes thermally fragile,\cite{4,5} especially for less common metals, and reactor time is often shared. A careful practitioner mapping a new precursor for the first time therefore runs a small sweep by hand, five to eight inputs at coarse geometric spacing, chosen and adjusted between runs by looking at the curve, and refines only if it looks unusual. Because the curve is read by eye, each point has to be low in noise, so in situ each is often run for tens of cycles\cite{66} to average the reading. The sweep aims only at locating the saturation input, and it ends with a single estimate of it and no statement of how uncertain it is.

The loop changes this in three ways. It chooses each input itself, so no operator has to stop between runs to look at the curve and decide where to measure next, and with in situ metrology the reactor, the measurement, and the loop close into one system, the setting the Introduction of the main text has in view. It stops when its band is tight enough, and on the benchmark its error falls below the output noise after a median of $5$ to $9$ points (Sec.~\ref{sec:acq_compare}), a count on a par with the hand sweep. The twenty points used throughout the benchmark are an evaluation budget, chosen so that every strategy, family, and surrogate is compared at the same point, and they overstate a campaign. The loop keeps returning to the largest input (Sec.~\ref{sec:acq}), so even a twenty-point design has about a dozen distinct inputs, repeats that a spacing rule could spread out or that can serve as replication for detecting drift (Sec.~\ref{sec:si-outside}). And it returns more than the sweep does. The result is the whole curve as a posterior with a calibrated band (Sec.~\ref{sec:calibration}), a read-out whose bias and coverage are measured (Table~\ref{tab:readout}), and the under-saturation risk at any input. On the benchmark the loop's placement of points is at least as accurate as a log-spaced grid of the same size and well ahead of random placement (Sec.~\ref{sec:acq_compare}), the grid being the fair stand-in for a hand-chosen geometric sweep. Each of its points can also be cheaper in situ. The surrogate pools all its points and infers the noise level from them (Sec.~\ref{sec:priors_sec}), and the benchmark runs at an output noise of a tenth of the plateau on every point, so an active point can be a noisier reading from a handful of cycles, only enough to reach steady state, and the campaign finishes in a fraction of the reactor time.

Whichever way the curve is measured, the number carried into practice is the input reaching a high fraction of the plateau, conventionally $95\%$, written $x_{\rm sat}(0.95)$ here. It goes into a recipe and is run every cycle for the life of the process. Setting it too short leaves the surface reaction slightly incomplete in every cycle, which degrades conformality\cite{68} and film quality, and because the same recipe runs for thousands of cycles a small shortfall per cycle adds up to a large defect in the finished film. Setting it too long wastes precursor and lengthens every cycle, so a $50\%$ overshoot on a two-second pulse is a real and recurring cost. Because the curve is nearly flat near the plateau, this input is also the hardest feature to pin down, which is why the main text reports it separately from overall curve accuracy, and why the next subsection weighs the two errors against each other.

\subsection{Choosing the read-out quantile}
\label{sec:si-quantile}
Once the curve is in hand, however it was measured, the process is committed to the recommended input, and that input can be wrong in the two directions set out above. Over-dosing costs precursor and cycle time on every cycle, a small amount that is known in advance and recurs for the life of the process. Under-saturating leaves a film that, grown for thousands of cycles, can fall outside its specification for conformality or quality and be scrapped, a cost that is large, since it is a whole production run, but uncertain, because a small shortfall may do no visible harm. The read-out quantile $q$ of Eq.~(\ref{eq:quantile}) sets how far the recommendation leans toward the long side to protect against the second error at the price of the first. This subsection puts numbers on the first, recurring cost in the ALD setting, so that the choice of $q$ can be made with its price in view. The model is deliberately rough, with three assumptions, and it aims at the right size to within a factor of a few without trying to be a full cost account.

\paragraph{Assumptions.} (i) Precursor is delivered at a roughly constant rate during a dose, as it is for a fixed-conductance or fixed-pressure dose, so the precursor consumed scales with the pulse length. (ii) Only a small share of the precursor is used in a self-limiting process, so the excess delivered during an overlong pulse is pumped away unused. The overshoot is pure waste, since it does not shorten the route to the same endpoint. (iii) A representative cycle is $2$~s dose, $10$~s purge, $1$~s co-reactant, $10$~s purge, totaling $23$~s.

\paragraph{The recurring cost of the median read-out.} The median-crossing read-out is $2.32$~s, the pooled fifty-seed value of $2.0+0.32$~s (the smaller perturbation and window sub-campaigns of Tables~\ref{tab:outside} and~\ref{tab:window} give $2.36$ and $2.25$~s). Against a required $2.0$~s that is $16\%$ longer, so by (i) and (ii) it uses about $16\%$ more precursor on every cycle. By (iii) it also lengthens the cycle by $0.32$~s, which is $1.4\%$ of the cycle time, or $320$~s (about $5$ minutes) over a thousand-cycle film. Both figures are quoted this way because they need no precursor price or delivery rate. The cost is small, and it stays small only because the read-out is accurate.

\paragraph{What the conservative read-out buys and costs.} At the median crossing ($q=0.5$) the posterior probability that the true curve is still short of the target is one half by construction. At the one-sided limit ($q=0.95$) it is one in twenty. On the benchmark the one-sided read-out is $4.05$~s against the median's $2.32$~s, so by (i) to (iii) it adds $2.05$~s per cycle, $103\%$ of the required pulse, in place of the median's $0.32$~s. Choosing $q=0.95$ therefore trades a known, recurring over-dose of about two seconds per cycle, roughly a doubling of the precursor used and a tenth added to every cycle, for a tenfold drop in the chance of running the process short of saturation for its whole life. That is a large recurring cost, so it is not one to take on by default. Whether it is worth taking depends on what a shortfall costs, and that cost is set by the process and lies outside the model. Where a short film is caught and scrapped, one lost production run costs more than many runs of over-dose, and the conservative read-out is cheap insurance. Where a small shortfall does no harm, or is caught early by in-line metrology, the median read-out is the better choice. The under-saturation risk $R(x)$ of Eq.~(\ref{eq:risk}) helps with that judgment, because it says how likely the process is to be short and by how much, and a shortfall to $0.93$ of the plateau is a different matter from one to $0.80$.

\paragraph{Eliciting the read-out.} Equation~(\ref{eq:quantile}) sets the quantile from one number, the cost ratio $c$, defined as the cost of under-saturating by one second divided by the cost of over-dosing by one second. It is a positive number with no upper limit. A value of $1$ means the two errors hurt equally, a value above $1$ means a shortfall hurts more, and a value below $1$ means an over-dose hurts more. The quantile follows as $q=c/(1+c)$, so $c=1$ gives the median ($q=0.5$), $c=10$ gives $q\approx0.9$, $c=19$ gives the one-sided limit ($q=0.95$), and $c$ far below $1$ would push the recommendation to the short side. The recurring cost supplies the denominator of that ratio. One extra second of pulse costs, per cycle, about $4.3\%$ of the cycle time and $50\%$ of the precursor of a $2$~s dose. A practitioner who can say how a one-second shortfall in saturation compares with that recurring second, as a multiple of it, has fixed $c$, and with it $q$. The model supplies everything except that one judgment, and the conclusion is that the margin should be priced in this way instead of being applied by default.

\section{Statistical terms in brief}
\label{sec:si-terms}

Because the method is Bayesian, a few statistical terms appear throughout the paper. For readers more familiar with deterministic curve-fitting, we define them in plain language here.

\begin{description}
\item[Probability distribution.] A description of how likely each possible value of a quantity is. Where a deterministic fit reports a single number, a Bayesian analysis reports a distribution whose spread represents the uncertainty.
\item[Gaussian (normal) distribution.] The familiar symmetric bell curve, specified by a mean (its center) and a standard deviation (s.d., its width). Measurement noise is commonly modeled as Gaussian.
\item[Half-normal distribution.] A Gaussian bell curve cut at zero and kept only on the non-negative side, so it assigns probability only to values $\ge 0$. A single positive scale parameter sets its width. We use it for the two scale parameters that cannot physically be negative, the roughness scale $\tau_\eta$ and the noise scale $\sigma$ of Eq.~(\ref{eq:priors}). The spline weights are kept positive by a different route, as exponentials of the random walk on their logarithms, and this construction makes monotonicity exact for every draw. Half-normal weights appear only in the two unpenalized comparison arms (Table~\ref{tab:si-surrogates}).
\item[Prior.] What is assumed about a quantity before looking at the data, for example that the weights are positive and change smoothly from one block to the next. It encodes physical knowledge into the fit.
\item[Likelihood.] How well a given candidate curve explains the measured data. A curve passing near all points has high likelihood.
\item[Posterior.] The updated distribution of a quantity after combining the prior with the data via the likelihood. It is the output of the analysis, a best estimate together with its uncertainty.
\item[Credible interval (CI) / band.] A range that contains the quantity with a stated probability (e.g.\ a $95\%$ credible band around the curve). It is the Bayesian analogue of an error bar. The bands plotted in this work are the symmetric summary $\mu_\ast\pm1.96\sigma_\ast$ of the posterior mean and s.d.\ (Sec.~\ref{sec:si-posterior}), which is an approximation to the exact pointwise CI because the non-negativity constraint boxes the plausible values of $f(x)$ in on one side near the onset and the plateau, so their spread is lopsided about the mean. The empirical coverage of the band is measured in Sec.~\ref{sec:calibration} of the main text.
\item[Markov-chain Monte Carlo (MCMC).] A computational method that draws representative samples from the posterior when it has no closed-form expression. The No-U-Turn Sampler (NUTS) is an efficient modern MCMC algorithm. In this work it produces the $S=500$ candidate curves distributed according to the posterior.
\end{description}

\section{The surrogate in full}

\subsection{The building blocks, M-splines and I-splines}
\label{sec:si-blocks}
The two kinds of block are central to the method, so we describe each one concretely. Fig.~\ref{fig:basis} of the main text shows both on a coarse version of the basis used in this work, and the panel references below are to that figure.

The M-splines are the rate blocks (Fig.~\ref{fig:basis}a). Each $M_m(x)$ is a single bump. It is zero everywhere except over one short window of input, where it rises and falls smoothly, and it is never negative. Think of one M-spline as a small, pre-shaped contribution to the rate of increase that is active only during its own window of input. The full set is a row of overlapping bumps that tile the input axis, so at any $x$ a few neighboring bumps are active. We make the bumps narrow and closely spaced near the origin, where the curve rises fast, and wide and sparse toward the plateau. This is why the bumps in Fig.~\ref{fig:basis}a are evenly spaced on its logarithmic axis, which crowds them toward the origin on a linear one. Each bump is non-negative and encloses unit area, $M_m(x)\ge 0$ and $\int_0^{x_{\max}}M_m(u)\,du=1$.

The bumps are defined by the knot sequence $t_1\le t_2\le\cdots\le t_{P+k}$ through the recursion of Ramsay,\cite{22} indexed by order $k$ (we use cubic splines, $k=4$). The order-1 bump is a rectangle on one knot interval,
\begin{equation}
M_m(x\mid 1)=
\begin{cases}
\dfrac{1}{t_{m+1}-t_m}, & t_m\le x< t_{m+1},\\[4pt]
0, & \text{otherwise},
\end{cases}
\label{eq:mspline_base}
\end{equation}
and higher orders follow the two-term recursion
\begin{equation}
\begin{aligned}
M_m(x\mid k)={}&\frac{k}{(k-1)(t_{m+k}-t_m)}\,\times{}\\
&\big[(x-t_m)\,M_m(x\mid k\!-\!1)\\
&\;+(t_{m+k}-x)\,M_{m+1}(x\mid k\!-\!1)\big]
\end{aligned}
\label{eq:mspline_rec}
\end{equation}
with the order-1 base case of Eq.~(\ref{eq:mspline_base}). Repeated knots, used at the two ends of the domain to clamp the basis, follow the usual convention that any term with a zero denominator is zero. Each step up in order gives a smoother, wider bump, differentiable $(k\!-\!2)$ times, and keeps it non-negative with unit area. The cubic ($k=4$) bumps in Fig.~\ref{fig:basis}a are the result.

As a concrete illustration, the order-2 (piecewise-linear) bump on three knots $t_m<t_{m+1}<t_{m+2}$ is a triangular tent,
\begin{equation}
M_m(x\mid 2)=\frac{2}{t_{m+2}-t_m}
\begin{cases}
\dfrac{x-t_m}{t_{m+1}-t_m}, & t_m\!\le\! x\!<\! t_{m+1},\\[10pt]
\dfrac{t_{m+2}-x}{t_{m+2}-t_{m+1}}, & t_{m+1}\!\le\! x\!<\! t_{m+2},\\[10pt]
0, & \text{else,}
\end{cases}
\label{eq:mspline_linear}
\end{equation}
which peaks at $2/(t_{m+2}-t_m)$ at the middle knot and has area exactly one. Its integral, the order-2 I-spline, is the running area under this tent. It is zero before $t_m$, rises as a smooth quadratic through the window, and holds at one for $x\ge t_{m+2}$.

The I-splines are the curve blocks (Fig.~\ref{fig:basis}b), obtained by integrating each bump from the origin,
\begin{equation}
I_m(x)=\int_0^{x} M_m(u\mid k)\,du .
\label{eq:ispline_def}
\end{equation}
Each order-$k$ bump is a piecewise polynomial of degree $k-1$, so its integral $I_m(x)$ is a piecewise polynomial of degree $k$ and is evaluated in closed form, with no numerical integration. Three properties matter, namely that $I_m(x)=0$ for $x<t_m$, that $I_m$ rises smoothly while $x$ crosses the window, and that $I_m(x)=1$ for $x\ge t_{m+k}$, once the unit area has all been collected. In words, $I_m(x)$ is the running total of bump $m$ up to input $x$. Each I-spline is therefore a smooth step that starts at zero, rises once, and levels off, and because it adds up a non-negative quantity it can only go up.

The staggered windows of the bumps become the staggered rise points of the steps in Fig.~\ref{fig:basis}b. Giving each step a weight $w_m\ge 0$ and adding them up, $f(x)=\sum_m w_m I_m(x)$, gives a curve that starts at zero, the \emph{pinned origin} of the main text, and only rises. Figure~\ref{fig:basis}c shows one such weighted sum. Since each step is the running total of its bump, the slope of the assembled curve is the weighted sum of the bumps,
\begin{equation}
f'(x)=\sum_{m=1}^{P} w_m\, M_m(x)\;\ge\;0,
\label{eq:deriv}
\end{equation}
which is never negative. The I-splines build the curve and the M-splines describe its slope, so keeping the weights non-negative keeps the slope non-negative.

\subsection{The P-spline penalty and its parameters}
\label{sec:si-pspline}
The basis of Sec.~\ref{sec:si-blocks} is deliberately over-complete, with $P=25$ blocks for $N=20$ points, so its coefficients must be held in check or they chase noise. We do this with the penalized-spline (P-spline) idea of Eilers and Marx,\cite{50} which keeps a fixed, rich basis and fits its coefficients by penalized likelihood, with the penalty placed on the differences between neighboring coefficients instead of on the coefficients themselves. In its classical least-squares form one minimizes
\begin{equation}
\sum_{i=1}^{N}\bigl(y_i - f(x_i)\bigr)^2 \;+\; \lambda \sum_{m} \bigl(\Delta^{d} c_m\bigr)^2
\label{eq:pspline_obj}
\end{equation}
over the basis coefficients $c_m$, where $\Delta^{d}$ is the $d$-th order difference operator ($\Delta^1 c_m = c_{m+1}-c_m$ and $\Delta^2 c_m=\Delta^1(\Delta^1 c_{m-1}) = c_{m+1}-2c_m+c_{m-1}$) and $\lambda\ge 0$ is a smoothing parameter. The construction has exactly two parameters, the order $d$ and the smoothing $\lambda$. The order is a modeling choice fixed in advance, and the smoothing is set from the data, classically by a separate tuning step and here inside the fit.

The order $d$ sets what the penalty leaves alone. Because $\Delta^{d}$ is zero on any polynomial of degree $d-1$ in the coefficient index, a $d$-th order penalty leaves a degree-$(d-1)$ trend in the coefficients free and pulls only departures from it back. We take $d=2$. The second difference penalizes curvature and leaves a straight-line trend in the log-weights free.

On a log-spaced knot grid this free trend has a specific and useful form. Take geometric knots $t_m = t_0\,\kappa^{m}$, so that $m=\log(t_m/t_0)/\log\kappa$. A straight-line log-weight trend $\log w_m = a + b\,m$ is then $w_m\propto t_m^{\,b/\log\kappa}$. Each unit-area rate bump $M_m$ has height $\propto 1/t_m$, so the rate near an input $x$ with $t_m\approx x$ is $f'(x)\approx w_m/t_m\propto x^{\,\xi-1}$, a power law with free exponent $\xi=b/\log\kappa$. The second-order penalty therefore leaves a power-law rate free, $f'(x)\propto x^{\,\xi-1}$. The fit can roll this smoothly into the plateau, since any $\xi<1$ gives a decaying rate, at whatever power the data imply, which is why the same fixed penalty follows the Elovich $1/x$ rate ($\xi=0$), the Kohlrausch--Williams--Watts (KWW) tail, and the Langmuir onset without retuning.

A rate that speeds up ($\xi>1$) is just as free under the penalty, but it has no plateau. The data, which hold the fit down at $x_{\max}$, rule it out even though the penalty does not, so the free trend cannot run away. A first-order penalty ($d=1$) instead leaves the level of the log-weights free, that is, $w_m$ constant, which by the same argument is a rate $f'(x)\propto 1/x$. That is a logarithmic rise, which reaches a plateau only by fighting the penalty, and this is why the first-order fit reaches saturation late (the $+1.8$~s read-out bias of Table~\ref{tab:si-model-ablation}). The penalty-order comparison of Sec.~\ref{sec:si-surrogates} measures this contrast.

The smoothing $\lambda$ sets how strongly the penalty acts. A large $\lambda$ forces the fit toward the free trend and a small $\lambda$ lets it follow the data. Classical P-splines choose $\lambda$ by cross-validation or an information criterion. We instead put the penalty into a prior with its own scale and infer that scale from the data. Writing the second difference of the log-weights as Gaussian steps of unknown scale $\tau_\eta$,
\begin{equation}
\Delta^{2}\!\left(\log w_m\right) \;=\; \tau_\eta\, z_m, \qquad z_m \sim \mathcal{N}(0,1),
\label{eq:pspline_bayes}
\end{equation}
is exactly the second-order random walk of Eq.~(\ref{eq:priors}) in the main text. Its penalty term is the P-spline difference penalty of Eq.~(\ref{eq:pspline_obj}) applied to the log-weights, with the smoothing $\lambda=\sigma^{2}/\tau_\eta^{2}$, so a smaller inferred $\tau_\eta$ means a stronger penalty. The data term is the nonlinear least-squares fit implied by Eq.~(\ref{eq:model}), because the response is a sum of exponentials of the log-weights, which is nonlinear in them, so the posterior mode is not the exact minimizer of the quadratic Eq.~(\ref{eq:pspline_obj}). This is the Bayesian P-spline of Lang and Brezger,\cite{51} the difference penalty recast as a random-walk prior with an inferred smoothing scale. Because $\tau_\eta$ has its own prior ($\mathrm{HalfNormal}(s_\tau)$ with $s_\tau=0.3$), the smoothing adapts to each data set inside the fit with no outside tuning. Heavy scatter pulls $\tau_\eta$ down and a sharp onset lets it rise.

Three features set our construction apart from a textbook P-spline. First, the penalty acts on the logarithms of the weights, so every weight stays positive and the monotonicity and pinned-origin guarantees hold for every draw (Sec.~\ref{sec:si-blocks}). A side effect is that it limits relative roughness instead of absolute roughness, which suits a log-spaced basis whose weights span orders of magnitude. Second, the blocks are integrated splines (I-splines), so the penalized object is the cumulative curve and the penalty smooths its rate. Third, Eilers and Marx use evenly spaced knots,\cite{50} while ours are log-spaced and the penalty still takes plain differences between neighbors, which is what makes the free trend a power law. Everything else follows the Bayesian P-spline recipe, namely a fixed over-complete basis, a difference penalty, and a single inferred smoothing scale.

\paragraph{The prior in full.} The complete prior, with the two starting values of the walk and the two half-normal scales $s_\tau=s_\sigma=0.3$, is Eq.~(\ref{eq:priors}) of the main text. Its walk is Eq.~(\ref{eq:pspline_bayes}) above, and Sec.~\ref{sec:si-compute} reports the $\sigma$ and $\tau_\eta$ the data return.

\paragraph{What the roughness-prior scale means on this grid.} The scale $s_\tau$ says how large a bend to expect before seeing the data. Consecutive interior knots are a factor $\kappa\approx1.33$ apart (Sec.~\ref{sec:si-metrics}), and by the argument above a bend of $\tau_\eta$ in the log-weights changes the local exponent $\xi$ by $\tau_\eta/\log\kappa\approx3.5\,\tau_\eta$ from one knot to the next. A half-normal with scale $0.3$ has its median at $0.20$ and its $95\%$ point at $0.59$, so the exponent is expected to drift by about $0.7$ per knot and by more than $2$ only one time in twenty. That is room enough to turn from the onset into the plateau over a few knots and too little for the weights to zigzag between neighbors. The value $0.3$ was used in the development comparison of Sec.~\ref{sec:si-surrogates} and kept. How far the data move $\tau_\eta$ from its prior at $N=20$ is reported in Sec.~\ref{sec:si-compute}, and the pull of $s_\tau$ on the fitted curve in Sec.~\ref{sec:si-priors}.

\subsection{Posterior summaries and the read-out formulas}
\label{sec:si-posterior}
This subsection collects the posterior summaries and read-out formulas that the main text states in words. The likelihood itself is Eq.~(\ref{eq:likelihood}) of the main text, and the Student-$t$ variant recommended for field data (Sec.~\ref{sec:si-outside}) replaces each of its Gaussian factors by a Student-$t$ density with four degrees of freedom and the same scale.

The sampler returns $S$ weight sets $\{\bm{w}^{(s)}\}_{s=1}^{S}$, each a monotone curve $f^{(s)}(x)$. The fitted curve is their point-by-point average and the uncertainty their s.d.,
\begin{equation}
\mu_\ast(x)=\frac{1}{S}\sum_{s=1}^{S} f^{(s)}(x),
\qquad
\sigma_\ast(x)=\sqrt{\frac{1}{S-1}\sum_{s=1}^{S}\bigl(f^{(s)}(x)-\mu_\ast(x)\bigr)^2},
\label{eq:postmean}
\end{equation}
and the plotted band is $\mu_\ast(x)\pm1.96\,\sigma_\ast(x)$, whose coverage is measured in Sec.~\ref{sec:calibration} of the main text and whose approximations are discussed in Sec.~\ref{sec:si-window}.

The recommended input of Eq.~(\ref{eq:quantile}) in the main text is the quantile that minimizes the pinball loss.\cite{57} For an error $e=x_{\rm sat}-x$ between the true saturation input and a recommendation $x$, the loss
\begin{equation}
\rho_q(e)=q\max(e,0)+(1-q)\max(-e,0)
\label{eq:pinball}
\end{equation}
charges a recommendation that is too short at rate $q$ and one that is too long at rate $1-q$, and its expected value over the crossing posterior is minimized by the $q$ quantile of that posterior, which is the recommendation $\hat{x}_q$.

The response-domain quantities are formed draw by draw. For an input $x$ and draw $s$, the attained fraction and the under-saturation risk are
\begin{equation}
\theta^{(s)}(x)=\frac{f^{(s)}(x)}{f^{(s)}(x_{\max})},
\qquad
R(x)=\Pr\!\bigl(\theta(x)<\alpha\bigr)=\frac{1}{S}\sum_{s=1}^{S}\mathbf{1}\!\left[\theta^{(s)}(x)<\alpha\right],
\label{eq:risk}
\end{equation}
where $\mathbf{1}[\cdot]$ is one when its argument holds and zero otherwise. Because each $\theta^{(s)}$ rises from $0$ to $1$ across the window, the event $\theta^{(s)}(x)<\alpha$ is the event that $x$ lies below draw $s$'s crossing $x_{\rm sat}^{(s)}(\alpha)$, so $R(x)=\Pr(x_{\rm sat}>x)$ is the crossing posterior read as a survival curve.

\section{Surrogates, hyperparameters, and study configuration}

\subsection{The comparison models, and how the hyperparameters were fixed}
\label{sec:si-surrogates}
The main text introduces the three unconstrained Gaussian-process (GP) comparators as a design space and the two unpenalized monotone bases beside them (Sec.~\ref{sec:comparators}). This section adds the models that appear only in the results, collects all of them in one place (Table~\ref{tab:si-surrogates}), and then records how the production surrogate's settings were fixed, which the main text states only in passing. The main point is that nothing is tuned per curve. The basis and the prior scales were chosen once, on development runs of the same five families kept separate from the reported campaigns, and then frozen for every result in the paper. The one quantity that does adapt to each data set, the effective smoothness, is inferred within the fit and never selected from it.

\paragraph{The models, in one place.} Table~\ref{tab:si-surrogates} lists all eleven models with their definitions and the feature each isolates. They are the penalized production surrogate (Sec.~\ref{sec:priors_sec}), its first-order-penalty and Student-$t$ variants, the two unpenalized monotone bases ($P=25$ and $P=11$), the heteroscedastic and plateau-parameter ablation arms, the unconstrained Eilers--Marx acquisition driver, and the three GP comparators of Sec.~\ref{sec:comparators}. The easiest to lose track of are the unpenalized bases. The $P=25$ basis is the key one, since its paired difference from the production surrogate isolates the penalty and its difference from GP-log-pin isolates monotonicity (Sec.~\ref{sec:whymono}).

\begin{table}[htbp]
\centering
\caption{The production surrogate and its comparison models. Each comparison model removes or changes one feature of the production surrogate, so a paired difference isolates that feature. Every I-spline arm keeps the pinned origin exact on every draw, and every one except the signed Eilers--Marx driver also keeps monotonicity exact. Full settings are in the sections referenced.}
\label{tab:si-surrogates}
\footnotesize
\begin{tabular}{@{}p{0.20\textwidth}p{0.44\textwidth}p{0.28\textwidth}@{}}
\toprule
model & definition & role (what it isolates) \\
\midrule
penalized monotone I-spline (production) & $P=25$ log-spaced I-splines on cubic M-splines; second-order log-weight random-walk roughness penalty, $\tau_\eta$ inferred & the surrogate used for all headline results (Sec.~\ref{sec:priors_sec}) \\[4pt]
first-order penalty & the production basis with a first-order (levels) log-weight penalty in place of the second-order (curvature) one & penalty-order ablation, isolating the order of the penalty (Table~\ref{tab:si-model-ablation}) \\[4pt]
Student-$t$ likelihood ($\nu=4$) & the production second-order prior with a Student-$t$ likelihood in place of the Gaussian & robustness to a corrupted anchor; recommended for field data (Table~\ref{tab:si-studentt}) \\[4pt]
unpenalized monotone I-spline, $P{=}25$ & the production over-complete basis with independent half-normal weights and no roughness penalty & isolates the penalty at fixed resolution (penalized vs.\ this) and monotonicity vs.\ GP-log-pin (Sec.~\ref{sec:whymono}) \\[4pt]
unpenalized monotone I-spline, $P{=}11$ & a coarse basis with seven interior knots, independent half-normal weights, no roughness penalty & the coarse unpenalized arm (Sec.~\ref{sec:whymono}) \\[4pt]
heteroscedastic likelihood & the production second-order prior with an errors-in-variables likelihood whose noise grows with the local slope & isolates the noise model (Table~\ref{tab:si-model-ablation}) \\[4pt]
plateau parameter & the production second-order prior with the plateau carried as an explicit parameter instead of read off $f(x_{\max})$ & isolates the plateau anchor (Table~\ref{tab:si-model-ablation}) \\[4pt]
penalized unconstrained spline (Eilers--Marx) & the production I-spline basis with signed weights and a first-order Gaussian random-walk penalty; pinned origin but not monotone & the acquisition driver of the unconstrained-driven acquisition control (Sec.~\ref{sec:si-controls}) \\[4pt]
GP-linear & stationary radial-basis-function GP on a linear input axis; unconstrained & unconstrained baseline (Sec.~\ref{sec:whymono}) \\[4pt]
GP-log & the same kernel on a logarithmic input axis & adds multi-scale resolution (Sec.~\ref{sec:whymono}) \\[4pt]
GP-log-pin & the log-axis kernel with the origin pinned at $(0,0)$ & adds the pinned origin, still no monotonicity (Sec.~\ref{sec:whymono}) \\[4pt]
\bottomrule
\end{tabular}
\end{table}

\paragraph{The penalty order and two additions, isolated.} The model choice that matters most is the order of the roughness penalty, and two natural additions to the model add nothing. To show this we refit three variants on the production surrogate's own $N=20$ designs, with the same collected data and the same acquisition, so that only the model changes, and compare them with the second-order production model ($50$ seeds $\times$ five families, paired). The first variant uses a first-order penalty on the log-weights. The other two keep the second-order penalty and change one further feature each, namely a heteroscedastic likelihood, which inflates the noise where the fitted rate is steep, and a plateau-parameter model, which gives the saturated value its own prior instead of reading it from the weight sum. The result (Table~\ref{tab:si-model-ablation}) is that the first-order penalty recovers the curve less well (median saturation error $0.044$ against $0.037$, paired $p=8\times10^{-17}$) and leaves the $95\%$ read-out badly biased ($+1.78$~s, $66\%$ interval coverage), because a penalty on the level of the log-weights favors a $1/x$ rate, which keeps the fit rising into the plateau and delays the crossing (Sec.~\ref{sec:si-pspline}). The second-order penalty, which leaves a smoothly decaying rate free, removes that bias ($+0.27$~s, $94\%$) with no other change to the model. The two additions do not improve on it. The heteroscedastic likelihood is numerically unstable on this problem. Its near-plateau variance term collapses on $42$ of the $250$ fits, and those failures alone account for its later read-out and lower coverage over all $250$ fits ($+0.53$~s, $76\%$), since on the $208$ fits that converge it matches the production model ($+0.28$~s, $92\%$). The plateau-parameter model ties the production surrogate on every column ($0.037$, $+0.25$~s, $94\%$, paired $p=0.9$ on saturation error). This is expected, because the weight sum already carries the plateau uncertainty draw by draw, so an explicit plateau parameter adds a degree of freedom that the $x_{\max}$ data pin anyway. The order of the penalty therefore settles the read-out bias, and the penalty itself closes the high-noise reversal of the unpenalized monotone basis (Sec.~\ref{sec:noisesweep}), whereas the likelihood and the plateau parameterization have little effect on either.

\begin{table}[htbp]
\centering
\caption{Model ablation by paired refit on one common set of $N=20$ designs (Sec.~\ref{sec:si-surrogates}), changing the penalty order, the likelihood, or the plateau parameterization one at a time. Saturation error and read-out bias are medians across runs, and coverage is the fraction of runs whose $95\%$ interval covers the true $x_{\rm sat}=2.0$~s. The production row is the same set of refits that Table~\ref{tab:readout} summarizes by the mean ($+0.32$~s at $94\%$ there).}
\label{tab:si-model-ablation}
\footnotesize
\begin{tabular}{lcccl}
\toprule
model & sat.\ error & $x_{\rm sat}(0.95)$ bias & coverage & note \\
\midrule
first-order penalty            & $0.044$ & $+1.78$~s & $66\%$ & reads late \\
heteroscedastic likelihood     & $0.042$ & $+0.53$~s & $76\%$ & unstable ($42/250$ fits) \\
plateau parameter              & $0.037$ & $+0.25$~s & $94\%$ & ties production \\
second-order penalty (production) & $0.037$ & $+0.27$~s & $94\%$ & adopted \\
\bottomrule
\end{tabular}
\end{table}

\paragraph{How the hyperparameters were fixed.} Beyond the basis, the production surrogate has only the two starting spreads of the walk, the two half-normal prior scales, and the sampler settings, and none was tuned on the campaign data reported here. The held-out benchmark of Sec.~\ref{sec:heldout} tests the frozen configuration on curves it never saw. The basis order and count ($k=4$ cubic, $P=25$) are set as described in Sec.~\ref{sec:si-metrics}. Many fixed log-spaced knots give the resolution to follow every shape, and the roughness penalty controls the flexibility that comes with them, the job a sparse knot count would otherwise do. That division of labor came out of a model comparison on development runs of the five families, separate from the reported campaigns, which tested two alternatives, adding fixed knots without a penalty and moving a few knots adaptively. Adding knots without a penalty lowers the representational error but overfits once noise is present, and adaptive knot placement, even when allowed to read the true curve, does no better than adding fixed knots. A roughness penalty over many fixed knots beat both and was adopted. The roughness-prior scale $s_\tau=0.3$ is the value used in that comparison and kept. What it means on the knot grid is spelled out in Sec.~\ref{sec:si-pspline}, the value the data then give $\tau_\eta$ is reported in Sec.~\ref{sec:si-compute}, and its remaining pull at the reported budget, together with the near-irrelevance of the noise scale, is measured in Sec.~\ref{sec:si-priors}. The noise scale $s_\sigma=0.3$ comfortably covers the generative output noise ($\sigma_y=0.1$) while discouraging the sampler from treating signal as noise, and the inferred $\sigma$ it yields is reported in Sec.~\ref{sec:si-compute}. The sampler settings (500 warmup and 500 retained draws, one chain in production, the non-centered roughness prior, NumPyro defaults otherwise) are stated in Sec.~\ref{sec:si-compute}, and their convergence is checked in Sec.~\ref{sec:si-diag}.

\paragraph{Why this does not incur a selection penalty.} A model that picked its basis, its knot positions, or its degree of smoothing from the data would owe a correction for that selection, because the same data would set both the model and the uncertainty quoted within it. The production surrogate avoids this by construction, since the knots never move and are not chosen from the data, and the number of basis functions is fixed in advance. The only quantity that adapts to each data set is the roughness scale $\tau_\eta$, an ordinary parameter inferred jointly with the weights instead of a model chosen by a separate criterion. The smoothing therefore adapts to the noise, small $\tau_\eta$ under heavy scatter and larger on a sharp onset, without any after-the-fact correction, and the credible bands keep their nominal meaning, subject only to the coverage measured in the main text (Sec.~\ref{sec:calibration}).

\subsection{Metrics, study design, and basis configuration}
\label{sec:si-metrics}
This section expands the evaluation setup of Sec.~\ref{sec:setup}, covering how the level-wise metrics locate true crossings near the left boundary, the design of the strategy comparisons, and the choice of basis order and count.

\paragraph{Crossing conventions at the boundary.} The two level-wise metrics of Sec.~\ref{sec:metrics} find the true crossing inputs $x_\alpha$ by root finding on the ground-truth function, so they resolve levels that fall near or below the first measurable input $x_{\min}=0.010$~s. For a curve whose $\alpha$-level crossing lies below $x_{\min}$ (for example the $5\%$ level of the headline KWW curve, whose true crossing is near $0.001$~s), the response error $\Delta y(\alpha)$ is read from the fit at that true input below the window, where the fit is a continuation from the pinned origin with no data behind it, whereas the input read-out $\hat{x}_\alpha$ searches only within the window and so sits near $x_{\min}$. The vanishing seed-to-seed spread of such entries in Table~\ref{tab:isotherms}(b) is the signature of this window limit and should not be read as precision.

\paragraph{Study design.} The acquisition comparison uses four strategies that share the same surrogate, priors, and budget. These are uncertainty sampling and three non-adaptive baselines, namely a uniform random design drawing each input uniformly over the domain, a log-uniform random design drawing each input uniformly in $\log x$ (the fair random comparison, since the domain and knots are log-scaled), and a log-spaced grid placing $N$ points geometrically across the window. All four observe the maximum input $x_{\max}$, the adaptive and random strategies through the shared $N_0=5$ survey and the grid as an endpoint of its geometric spacing. The noise-free sweeps cover each family's range of shapes. The one-parameter families sweep their single time scale from an onset-dominated curve to one that saturates at the window edge (Langmuir $\tau\in\{0.15,0.67,1.5\}$~s, and Michaelis--Menten $K\in\{0.03,0.11,0.25\}$~s, whose $95\%$ point sits at $x_{\rm sat}=19K$), while the two-parameter families re-solve the scale so that every variant keeps $f(2.0~\mathrm{s})=0.95A$ and only the shape changes (Sips $n\in\{1.3,2,4\}$, Elovich $\gamma\in\{1.2,1.69,2.4\}$, and KWW $\beta\in\{0.35,0.55,0.85\}$), with the headline value in the middle of each set.

\paragraph{Basis order and count.} With $k=4$ (cubic M-splines, so quartic I-splines) the rate $f'(x)$ is twice continuously differentiable across the interior knots and the curve itself three times. The lowest practical order, $k=2$, would leave corners in the rate, and higher orders add parameters with no benefit on this domain. The knots are repeated $k=4$ times at $x=0$, so that every I-spline vanishes at the origin, and again at $x_{\max}$. The $21$ interior knots are log-spaced between $x_{\min}$ and $x_{\max}$ with the two ends of the geometric sequence dropped. That sequence divides the ratio $x_{\max}/x_{\min}=500$ into $22$ equal factors, so consecutive knots are a factor $\kappa=500^{1/22}\approx1.33$ apart, and the innermost interior knot sits near $0.013$~s and the outermost near $3.8$~s. The geometric grid is anchored at $x_{\min}$ itself. Because the first knot interval, from $0$ to $0.013$~s, straddles the first measurable input $x_{\min}=0.010$~s, the rate below $x_{\min}$ has no data behind it and shows the narrow near-origin feature of Fig.~\ref{fig:final_fit}. This is purely a matter of basis layout with no effect on the results, since the response $f$, which starts from the pinned origin, is unaffected and no data lie below $x_{\min}$ under any knot placement. The number of basis functions, $P = n_{\rm interior} + k = 21 + 4 = 25$, is deliberately over-complete. Many log-spaced knots give the resolution to follow any of the shapes, including the sigmoidal Sips onset, the kinked Elovich, and the step-like Sips $n=4$, and the roughness penalty of Sec.~\ref{sec:priors_sec} controls the added flexibility so the extra knots do not fit noise. The penalty makes a basis this fine usable at small budgets. The coarser $P=11$ basis ($n_{\rm interior}=7$) is kept only as an unpenalized comparison arm.

\section{Sampler and prior diagnostics}

\subsection{MCMC convergence diagnostics}
\label{sec:si-diag}
The roughness penalty of Eq.~(\ref{eq:priors}) puts a second-order random walk with an inferred step scale $\tau_\eta$ on the log-weights. A model in which the steps share an inferred scale has the ``funnel'' shape\cite{54} that can trip up Hamiltonian Monte Carlo,\cite{69} so we sample it in the non-centered form\cite{70} and check convergence directly, with the potential scale reduction $\hat{R}$ and the effective sample size (ESS) as the diagnostics,\cite{69,71} together with the count of divergent transitions. In plain words, $\hat{R}$ compares chains started from different points and equals $1$ once they have all settled on the same distribution, so values within a few hundredths of $1$ mean the chains agree and larger values mean they have not met. The ESS is the number of independent draws a correlated chain is worth, so an ESS far below the number of draws means the chain moves slowly and its summaries carry more Monte Carlo error. A divergent transition is a step at which the sampler's numerical integration broke down, a sign of a region of the posterior it cannot explore properly. For each family's $N=20$ seed-0 data set we drew four independent chains of 500 samples after 500 warmup steps (Table~\ref{tab:si-diag}). The chains agree and the effective sample sizes are adequate, but the divergence rate is high, and we report that here. Across families $\hat{R}$ reaches $1.030$, the smallest ESS falls to $207$ of the $2000$ post-warmup draws, and $11\%$ to $22\%$ of those draws are divergent transitions. The problem is not confined to seed~$0$. Run as a single chain at each design's own sampler seed across all $250$ main-campaign designs, $100\%$ of production fits show at least one divergence, and the single-chain ESS has a median of $87$ of $500$ retained draws. These are refits of the campaign's stored designs, separate from the campaign's own fits. They sample a single chain on the same design, so they share the model, the data, and the settings, though not the production fits' exact arithmetic.

\paragraph{The cause is the starting point.} The second-order walk builds $\log w$ from a double running sum, so under the prior alone the last log-weight has an s.d.\ of about $23$ and exceeds $\pm75$ one time in a hundred, and $w=\exp(\log w)$ is astronomically large in the tails. Only the likelihood keeps the posterior sane, so NumPyro's default starting point, a uniform draw, can place a chain far outside the region where the posterior mass lies. At the default target acceptance of $0.8$ the steps are large enough to walk back, at the cost of the divergences reported above. Raising the target on its own makes things worse, because the smaller steps leave the chain stuck. At target $0.95$ from a uniform start the four chains fail to meet on four of the five families ($\hat{R}$ up to $20$, ESS about $2$, and a posterior-mean curve of order $10^{3}$ on a response normalized to one), quadrupling the warmup leaves the same four stuck, and at target $0.99$ all five are stuck (Table~\ref{tab:si-sampler}). Which families get stuck depends on where the uniform draw happens to land, which itself points to the starting point as the cause. The first-order penalty, whose single running sum gives far lighter prior tails, tolerates a uniform start, so this defect does not appear in its diagnostics.

\paragraph{The remedy, and why the reported results stand.} Starting the chains at the prior median stops them from getting stuck at no extra cost, and it cuts the divergences by roughly an order of magnitude. At target acceptance $0.95$ with a prior-median start and the same $500$ warmup steps, no family gets stuck and divergences fall to $0.5$--$4.2\%$ against the production configuration's $11\%$--$22\%$ (Table~\ref{tab:si-sampler}). The fix is not complete, since $\hat{R}$ improves on four families while Sips stays marginal at $1.027$ and its ESS falls. No setting we tried is clean everywhere, so this posterior is hard to sample and the production numbers should be read with that in mind. Pooled over $50$ paired fits on identical designs, the change alters nothing that is reported. The saturation error moves from $0.0387$ to $0.0385$ (paired Wilcoxon $p=0.45$, a shift of $0.01$ seed-to-seed s.d.) while divergences fall $6.4$-fold, and the largest pointwise change in the posterior-mean curve is $0.008$ on average and $0.027$ at worst, against $\sigma_y=0.1$. The divergences were therefore a price the sampler paid to escape the prior's tail, and they did not bias the answer. We report the campaigns as run, under the production configuration, since the alternative changes no reported quantity. The prior-median start is nonetheless the better setting, and it is worth adopting when the campaigns are next re-run. We recommend it for new work, and it is the default in the released \texttt{real\_data.py}. The main-text study uses a single chain of 500 draws per fit for speed, so $\hat{R}$ cannot be computed there.

\begin{table}[htbp]
\centering
\caption{MCMC convergence diagnostics for the $N=20$ seed-0 fit of each family under the production configuration (Sec.~\ref{sec:si-diag}). $\hat{R}$ and ESS are the worst values over all sampled parameters (largest $\hat{R}$, smallest ESS), and divergences are counted over the $2000$ post-warmup draws.}
\label{tab:si-diag}
\begin{tabular}{lccc}
\toprule
family & max $\hat{R}$ & min ESS & divergences \\
\midrule
Langmuir        & 1.022 & 207 & 322 \\
Michaelis--Menten & 1.012 & 266 & 316 \\
Sips ($n=2$)    & 1.024 & 360 & 325 \\
Elovich         & 1.010 & 349 & 224 \\
KWW             & 1.030 & 246 & 434 \\
\bottomrule
\end{tabular}
\end{table}

\begin{table}[htbp]
\centering
\caption{Sampler settings compared on the seed-0 fits of all five families (four chains, 500 draws after the stated warmup). ``Stuck'' counts families whose chains never reach the bulk of the posterior, seen as a posterior-mean curve orders of magnitude above the plateau of one.}
\label{tab:si-sampler}
\footnotesize
\begin{tabular}{lccccc}
\toprule
configuration & target & warmup & $\hat{R}$ range & ESS range & stuck \\
\midrule
production (uniform init) & $0.8$  & $500$  & $1.010$--$1.030$ & $207$--$360$ & $0/5$ \\
uniform init              & $0.95$ & $500$  & $1.011$--$20.0$  & $2$--$361$   & $4/5$ \\
uniform init              & $0.95$ & $2000$ & $1.027$--$22.2$  & $2$--$210$   & $4/5$ \\
uniform init              & $0.99$ & $500$  & $39.1$--$54.3$   & $2$--$2$     & $5/5$ \\
\textbf{prior-median init} & $0.95$ & $500$ & $1.002$--$1.027$ & $142$--$942$ & $0/5$ \\
prior-median init         & $0.99$ & $500$  & $1.003$--$1.047$ & $70$--$653$  & $0/5$ \\
\bottomrule
\end{tabular}
\end{table}

\subsection{Prior sensitivity}
\label{sec:si-priors}
The production surrogate fixes two prior scales in advance, $s_\tau=0.3$ on the roughness (random-walk step) scale $\tau_\eta$ and $s_\sigma=0.3$ on the observation noise $\sigma$, both from Eq.~(\ref{eq:priors}). This section asks how much the recovered curve would change if either had been set differently.

\paragraph{What is changed.} Each scale is multiplied by five and divided by five while the other is held at $0.3$, which gives four perturbations, and the two scales are changed together in the four possible combinations, which gives four more, eight perturbations in all. Everything else, the basis, the knots, the sampler settings, the designs, and the data, is held fixed, so each refit differs from the baseline in the prior scales alone.

\paragraph{What is measured.} For each perturbation we refit the stored designs and record the largest pointwise change in the posterior-mean curve, $\max_x|\mu_\ast(x)-\mu_\ast^{\rm base}(x)|$, in response units. Table~\ref{tab:si-priors} reports, per family and budget, the largest of these eight changes, as the mean over five seeds followed by the worst seed. Because the learning curves of Sec.~\ref{sec:si-perfamily} (panel (c) of Figs.~\ref{fig:langmuir}--\ref{fig:kww}) begin at $N=5$, where twenty-five basis functions face five observations, the check is run at three budgets, $N=5$, $10$, and $20$.

\paragraph{What we observe.} First, the dependence on the priors falls steadily as data arrive. A fivefold change moves the posterior mean by $0.14$ to $0.22$ at $N=5$, by $0.05$ to $0.15$ at $N=10$, and by $0.035$ to $0.066$ at $N=20$ (worst single seed $0.083$, on Sips), so by the headline budget the fit is set by the data and only weakly by the prior scales. Second, the shift is largest where the shape of the curve depends most on the smoothing, the cooperative Sips and capped Elovich onsets, and smallest on the smoothly saturating KWW. Third, the roughness-prior scale is responsible, and the noise scale has almost no effect. Changing $s_\sigma$ alone by a factor of five moves the curve by at most $0.011$ at $N=20$ (pooled means over families and seeds, here and for the two shifts that follow), whereas every perturbation that produces a visible shift changes $s_\tau$. Tightening $s_\tau$ to $0.06$, a much stiffer penalty, shifts the curve by $0.043$ at $N=20$, and tightening it while loosening $s_\sigma$ shifts it by $0.044$. The mechanism is direct. $s_\tau$ is the prior scale on the inferred roughness $\tau_\eta$, so tightening it over-smooths a genuinely structured onset, while $s_\sigma$ only rescales a noise term that the data and the $x_{\max}$ measurement already pin. Since $\tau_\eta$ itself is inferred, the smoothing still adapts to each data set, and the sensitivity here is to the prior scale on that adaptation. No smoothness is hard-coded.

\paragraph{What it means for the results.} The paired comparisons are protected. All four acquisition strategies, and all the ablation arms that share the surrogate, use the same priors at every budget, so a shared prior sensitivity moves their arms together and largely cancels from the differences between them. The absolute error at small $N$ should be read with care, while the rankings are largely unaffected. The one exposed number is the headline margin over the resolution- and origin-matched GP-log-pin, $0.039$ against $0.049$ at $N=20$ (Table~\ref{tab:gp}), a gap of $0.010$ that is smaller than the curve shift the priors alone can produce at that budget ($0.035$ to $0.066$ by family). Because the perturbation that matters, tightening $s_\tau$, over-smooths exactly the structured onsets where that margin lives, the margin holds only for $s_\tau$ within a factor of a few of its adopted value. The base $s_\tau=0.3$ is the value used in the model comparison of Sec.~\ref{sec:si-surrogates} and kept, and Sec.~\ref{sec:si-pspline} gives its reading on the knot grid. Anyone running the loop at fixed small budgets could reduce the remaining dependence further by setting the scale once to a family's typical roughness or by placing a further prior on $s_\tau$ itself.

\begin{table}[htbp]
\centering
\caption{Prior sensitivity across budgets (Sec.~\ref{sec:si-priors}). Each entry is the largest pointwise change in the posterior-mean curve, in response units, under the worst of the eight prior perturbations, as the mean over five seeds and as the worst seed.}
\label{tab:si-priors}
\footnotesize
\setlength{\tabcolsep}{5pt}
\begin{tabular}{lcccccc}
\toprule
 & \multicolumn{2}{c}{$N=5$} & \multicolumn{2}{c}{$N=10$} & \multicolumn{2}{c}{$N=20$} \\
\cmidrule(lr){2-3}\cmidrule(lr){4-5}\cmidrule(lr){6-7}
family & mean & worst & mean & worst & mean & worst \\
\midrule
Langmuir          & $0.218$ & $0.314$ & $0.093$ & $0.138$ & $0.046$ & $0.060$ \\
Michaelis--Menten & $0.144$ & $0.214$ & $0.046$ & $0.089$ & $0.037$ & $0.062$ \\
Sips ($n=2$)      & $0.215$ & $0.372$ & $0.151$ & $0.224$ & $0.066$ & $0.083$ \\
Elovich           & $0.204$ & $0.295$ & $0.089$ & $0.128$ & $0.057$ & $0.079$ \\
KWW               & $0.142$ & $0.217$ & $0.047$ & $0.070$ & $0.035$ & $0.049$ \\
\bottomrule
\end{tabular}
\end{table}

\section{The five kinetic families in detail}

\subsection{Disciplinary origins of the five families}
\label{sec:si-families}
The main text (Sec.~\ref{sec:kinetics}) gives each family's formula, shape parameter, and role. This section adds where each family comes from and what sets its shape apart from the others.

\paragraph{Langmuir.} The concave, ideal self-limiting form, an exponential approach to the plateau with a single time constant $\tau$. It is the first-order kinetics of Langmuir adsorption on a fixed number of sites\cite{1} and the reference shape against which the others are read.

\paragraph{Michaelis--Menten.} A rectangular hyperbola, the Michaelis--Menten law of enzyme kinetics ($x$ the substrate concentration, $A$ the maximum rate, $K$ the half-saturation constant),\cite{2,58} which is the Langmuir isotherm in pressure written in another field's notation. Its slow crawl to the asymptote for small $K$ sets it apart here. It leaves much of the curve just below saturation, which makes the family a demanding test of plateau recovery.

\paragraph{Sips (Hill).} The Hill equation of cooperative ligand binding,\cite{3} with $\tau_S$ the ligand concentration at half saturation (the half-maximal effective concentration, EC$_{50}$, when the same form is fitted to a concentration--response curve). The related Sips adsorption isotherm\cite{59} has the same algebra, written in pressure, but its derivation needs an exponent between zero and one, which gives a concave curve. We use the $n>1$ branch as a deliberate stress shape, because its zero initial slope, inflection, and cooperative rise form the one shape a concave-only model cannot represent.

\paragraph{Elovich.} The Elovich equation of chemisorption kinetics,\cite{60} often read as uptake onto a surface whose sites differ in energy, where the most reactive sites are used first and ever higher barriers remain, although other mechanisms give the same law. Its logarithmic rise (rate $\propto 1/x$) has no plateau of its own. The hard cap of Eq.~(\ref{eq:iso_elovich}) makes it self-limiting and puts a kink in the slope, the only kink among the five and a worst case for any smooth basis.

\paragraph{KWW (stretched exponential).} The stretched exponential found throughout glassy and disordered systems,\cite{61,62,63} commonly read as a spread of rate constants,\cite{62} which on a surface could come from sites with a range of activation energies. Smaller $\beta$ gives a fast initial rise followed by a long, slow tail, the opposite emphasis to the sigmoidal Sips case.

\subsection{Curve-by-curve recovery}
\label{sec:si-perfamily}
This section looks at each headline family on its own. The smoothing bias and the level-wise errors are in the main text (Sec.~\ref{sec:per_isotherm}, Table~\ref{tab:isotherms}). Each family has a four-panel figure with a shared layout (Figs.~\ref{fig:langmuir}--\ref{fig:kww}), and the same axis limits in every panel, with off-scale points labeled by their values. Panel (a) shows the posterior fit at $N=20$ for the median-error seed of the fifty, so the panel is typical instead of flattering, with the true curve in black, the posterior mean in blue inside its $95\%$ band, and the observations in red. Its inset gives the run's saturation error and the fifty-seed mean and s.d. Panel (b) shows the noise-free regime sweep, which tracks what the basis can represent, with the true curves thick and translucent, the posterior means dashed inside their $95\%$ bands, and each variant's saturation error in the legend. Panel (c) shows the learning curves at the fixed $x_{\rm sat}=2.0$~s as mean $\pm$ one s.d.\ over the fifty seeds on a logarithmic scale, with the output noise $\sigma_y=0.1$ as a dashed line, for uncertainty sampling (blue), uniform random (grey dotted), log-uniform random (orange), the log-spaced grid (green), and the resolution-matched unconstrained arm GP-log (purple dashed, Sec.~\ref{sec:whymono}) run through the same loop and shown at every budget. GP-log runs below the I-spline at the smallest budgets on the spread-rise shapes (Langmuir, Sips, and Elovich, whose rise is spread across the window), is overtaken between about $N=8$ and $N=10$, and ends $1.36\times$ above it pooled at $N=20$ (Table~\ref{tab:gp}). That late gap comes partly from the designs the GP's own loop chooses and partly from the missing pinned origin, since on identical data GP-log-pin trails the I-spline by only a few percent (Table~\ref{tab:si-gpmatched}). The constraint itself does not set curve accuracy (Sec.~\ref{sec:whymono}). Panel (d) shows the level-wise errors at $N=20$, the response error $\Delta y(\alpha)$ as circles on the left axis and the input error $\Delta x(\alpha)/x_\alpha$ as squares on the right axis, where positive means late. The input error is plotted as a fraction of the true crossing time so that the small errors at low levels stay visible next to the large $95\%$ value.

\subsubsection{Langmuir}

\begin{figure}[htbp]
\centering
\includegraphics[width=\textwidth]{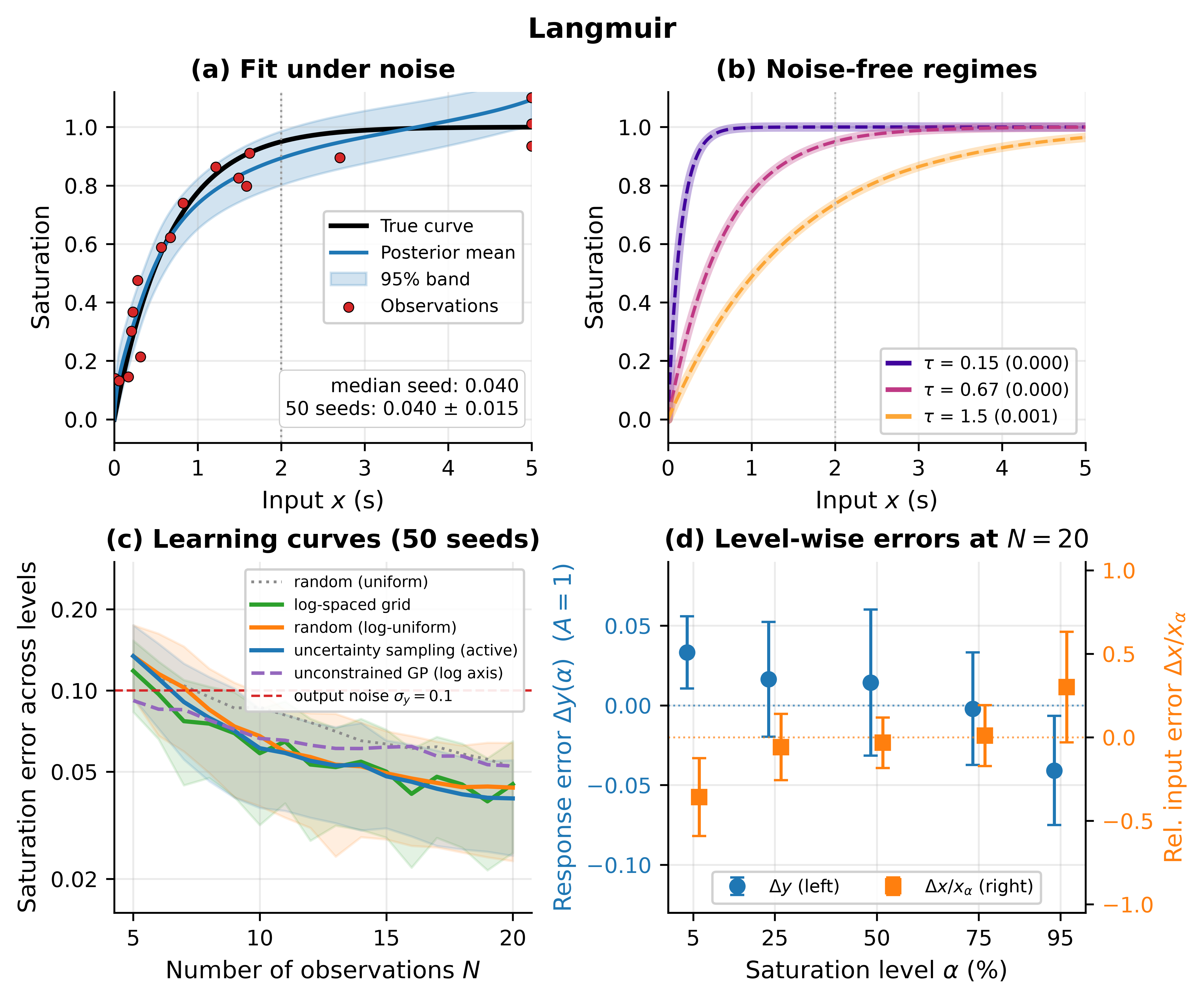}
\caption{Langmuir curve in detail, in the four-panel layout described in Sec.~\ref{sec:si-perfamily} and shared by Figs.~\ref{fig:langmuir}--\ref{fig:kww}. The noise-free sweep (b) spans the time constant $\tau$.}
\label{fig:langmuir}
\end{figure}

The exponential Langmuir rise is the textbook saturation shape, and the noise-free sweep reproduces it almost exactly at every time scale in reach (saturation error at most $0.001$). The log-spaced knots resolve both the onset-dominated $\tau=0.15$~s variant and the slow $\tau=1.5$~s variant without retuning (Fig.~\ref{fig:langmuir}b). Under noise it learns quickly, with the error crossing the output-noise level at a median of $N=7$ measurements and reaching $0.040$ at $N=20$, and in the randomized-location campaign of Table~\ref{tab:randxsat} uncertainty sampling leads the log baselines by about $1.2\times$ and uniform random by $1.8\times$. The level-wise errors show only the smoothing bias of the main text, positive at $5\%$, near zero mid-curve, and negative at $95\%$ (Table~\ref{tab:isotherms}(b)).

\subsubsection{Michaelis--Menten}

\begin{figure}[htbp]
\centering
\includegraphics[width=\textwidth]{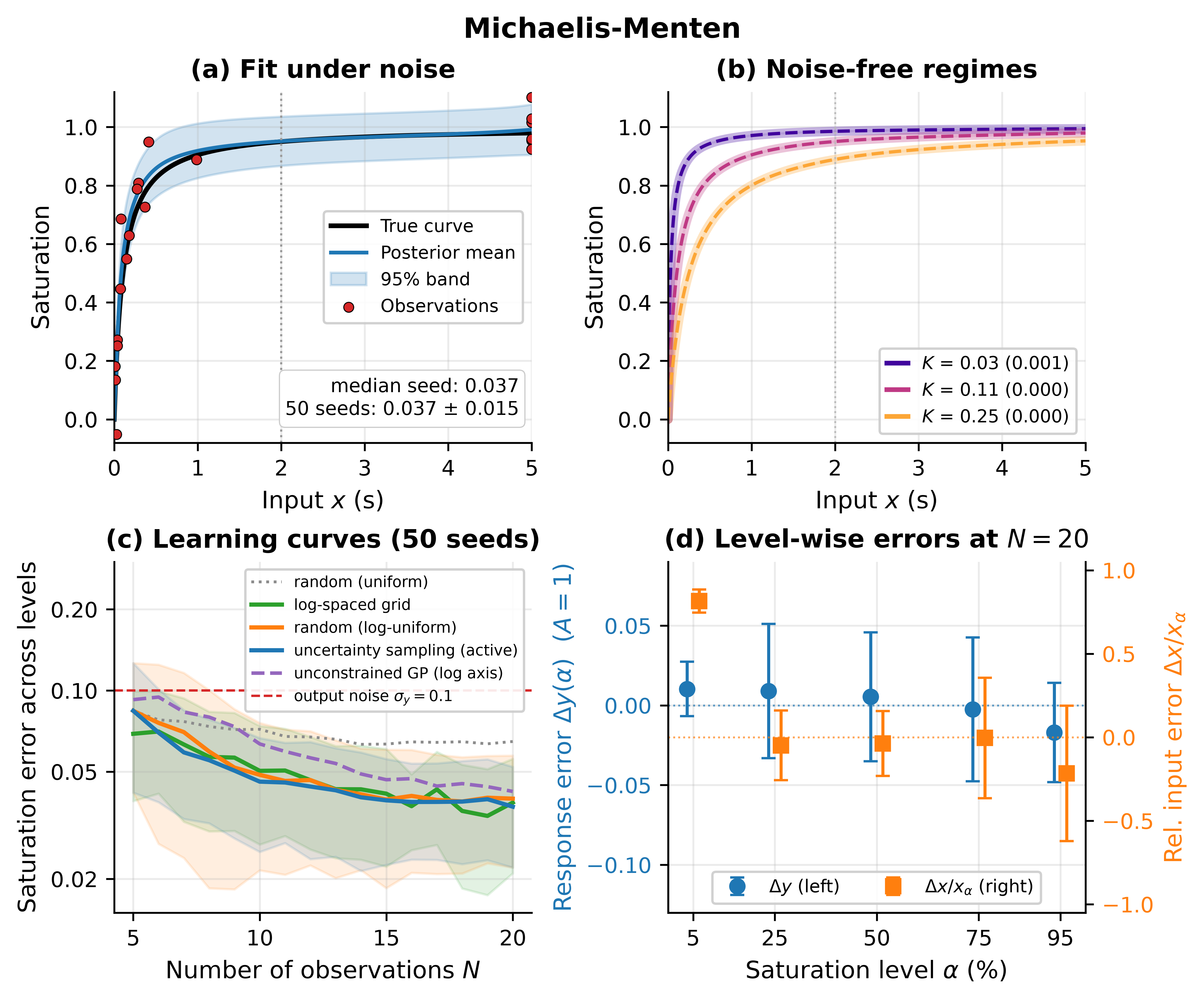}
\caption{Michaelis--Menten curve in detail, in the layout of Fig.~\ref{fig:langmuir}. The noise-free sweep (b) spans the half-saturation input $K$, whose $95\%$ point sits at $x_{\rm sat}=19K$.}
\label{fig:dissociative}
\end{figure}

The rectangular hyperbola is reproduced almost exactly in the noise-free sweep at all three half-saturation inputs (saturation error below $0.002$), including $K=0.25$~s, whose $95\%$ point sits near the window edge. There the inferred plateau settles at the true in-window value $0.952$, which shows that the data-anchored plateau adapts to whatever the window exposes and does not assume full saturation (Fig.~\ref{fig:dissociative}b). Despite the slow crawl toward the asymptote, the noisy plateau lands at $0.978$ on average, within $0.2\%$ of the value $0.979$ that the true curve reaches at $x_{\max}$ and about $2\%$ below its asymptote. The regime is among the easiest to learn (median crossing $N=5$, with the log-spaced grid fully as good) and its input read-outs are the tightest of the five from the $5\%$ through the $75\%$ level, although at $95\%$ the slow approach to the asymptote gives one of the widest spreads (Table~\ref{tab:isotherms}(b)).

\subsubsection{Sips (Hill)}

\begin{figure}[htbp]
\centering
\includegraphics[width=\textwidth]{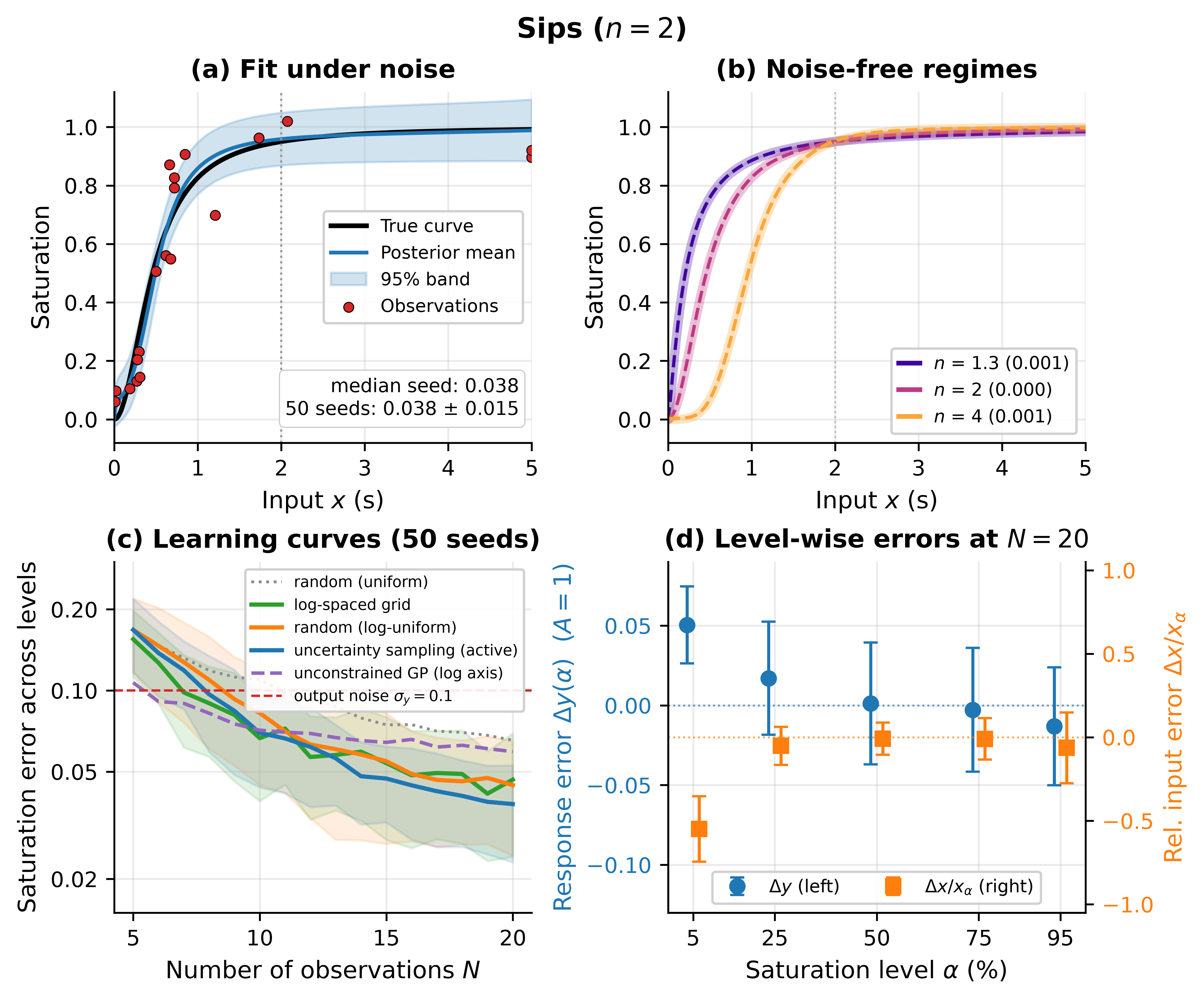}
\caption{Sips (Hill) curve in detail, in the layout of Fig.~\ref{fig:langmuir}. The noise-free sweep (b) spans the Hill coefficient $n$ at fixed $x_{\rm sat}=2.0$~s.}
\label{fig:sips}
\end{figure}

The sigmoidal family shows both the value and the limit of the platform. The noise-free sweep is near exact at all three onsets, including the step-like $n=4$ variant ($0.001$), whose rise falls in the outermost, coarsest knot intervals and is still resolved by the $P=25$ grid (Fig.~\ref{fig:sips}b). Under noise the headline $n=2$ curve is the slowest regime to learn (median crossing $N=9$), but it is also where uncertainty sampling pays most. Concentrating measurements in the narrow onset window keeps an edge of about $1.2\times$ over both log baselines at the fixed location (Fig.~\ref{fig:sips}c) and of $1.3$ to $1.4\times$ once the rise location is randomized (Table~\ref{tab:randxsat}). Its level-wise errors carry the family's signature, the largest onset overshoot and the largest early input bias, because the smooth fit begins rising before the true cooperative delay (Table~\ref{tab:isotherms}(b)).

\subsubsection{Elovich}

\begin{figure}[htbp]
\centering
\includegraphics[width=\textwidth]{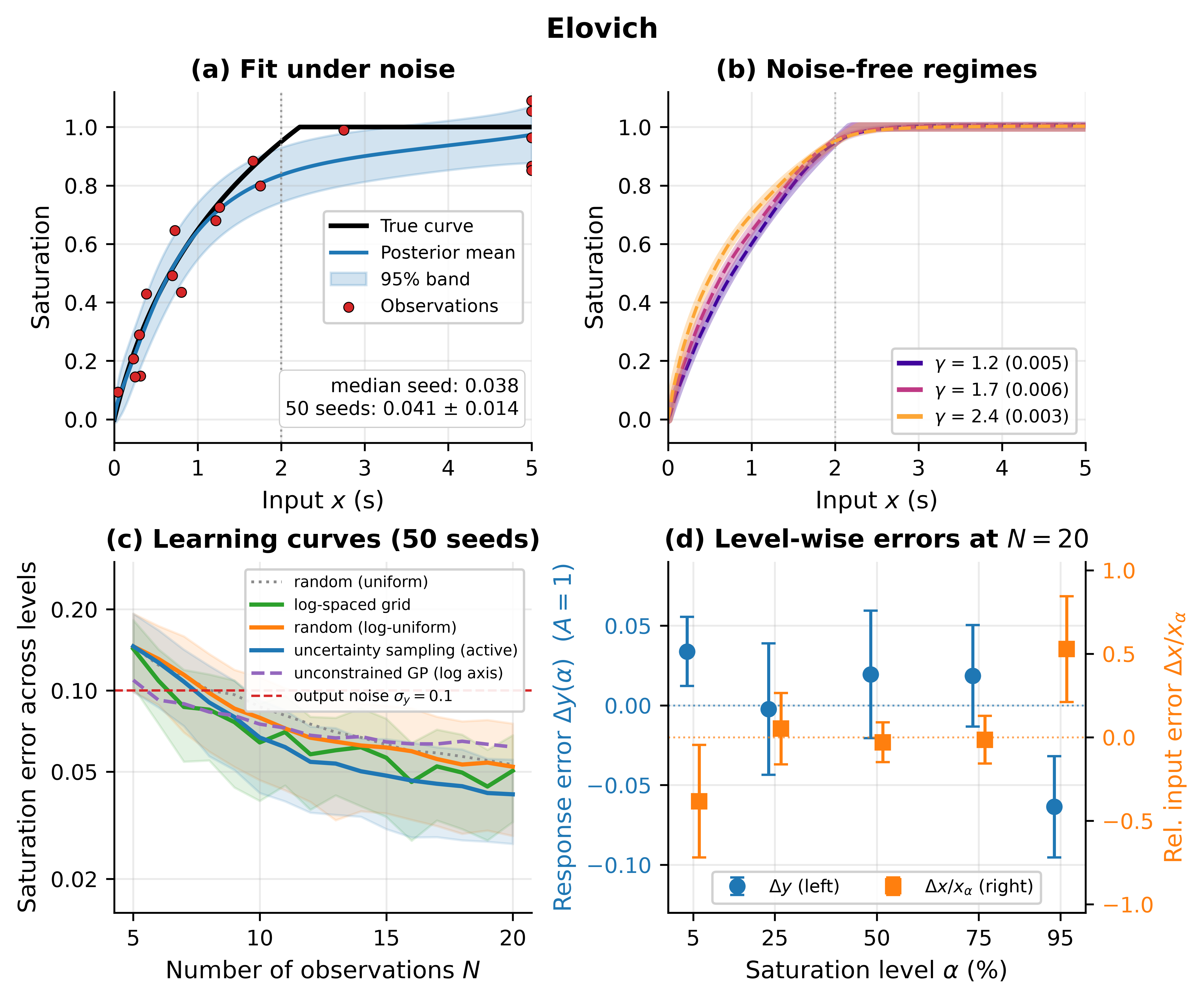}
\caption{Elovich curve in detail, in the layout of Fig.~\ref{fig:langmuir}. The noise-free sweep (b) spans the logarithmic slope $\gamma$ at fixed $x_{\rm sat}=2.0$~s, so the cap falls near $2.2$~s in every regime.}
\label{fig:elovich}
\end{figure}

Elovich is the only family with a noise-free residual in every regime ($0.003$ to $0.006$), because the smooth basis has to round off the kink where the logarithmic rise meets the cap. That leaves a small undershoot just below the plateau while the rest of the curve is captured cleanly (Fig.~\ref{fig:elovich}b). It makes Elovich nominally the least accurate regime under noise ($0.041$ at $N=20$, though within seed scatter of the others), with the largest plateau-end errors of the five (Table~\ref{tab:isotherms}(b)). Any smooth basis would have the same trouble at the kink. Even so, the inferred plateau stays within $1.3\%$ of the truth on average ($1.013$, Table~\ref{tab:isotherms}(a)).

\subsubsection{KWW}

\begin{figure}[htbp]
\centering
\includegraphics[width=\textwidth]{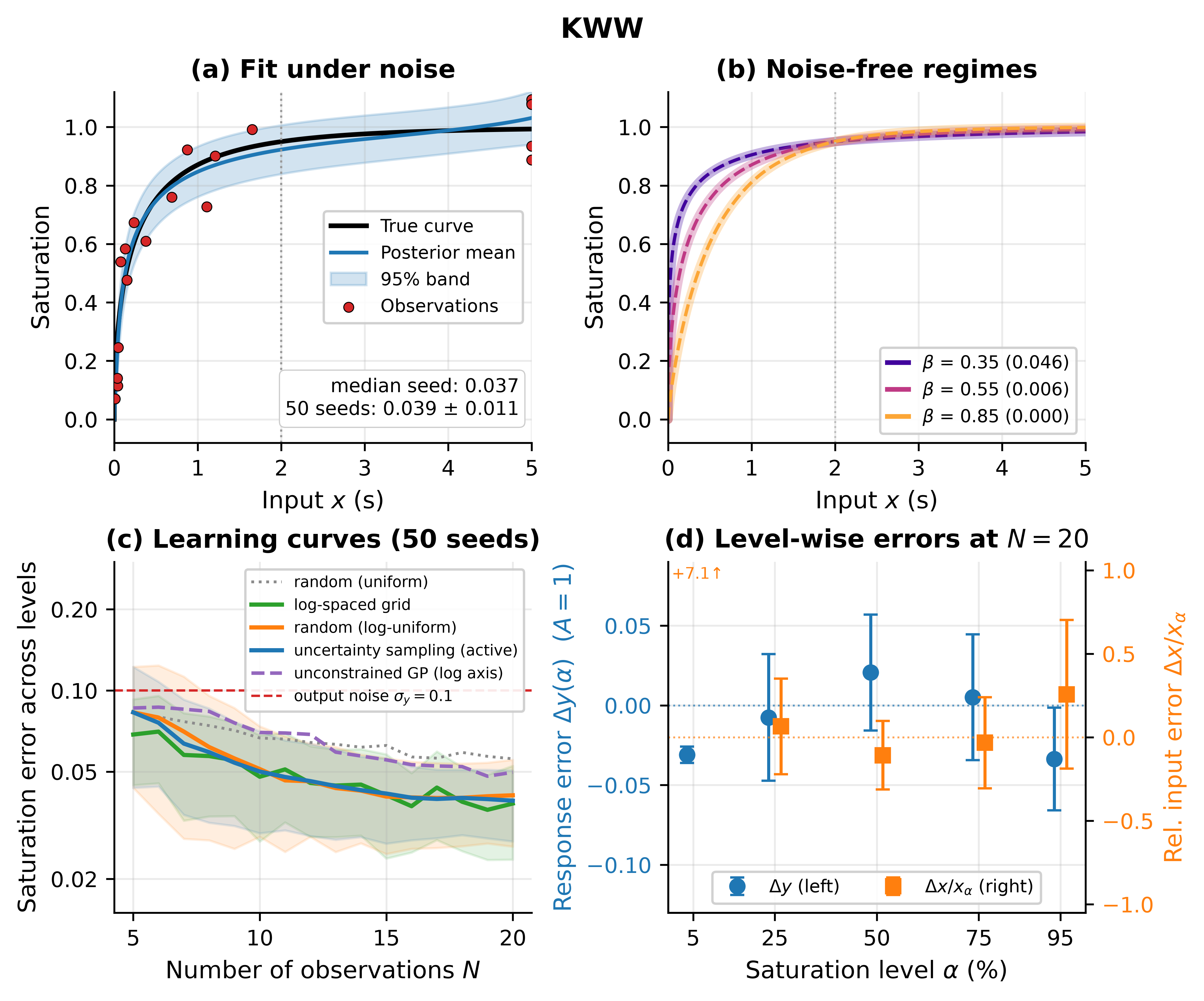}
\caption{KWW curve in detail, in the layout of Fig.~\ref{fig:langmuir}. The noise-free sweep (b) spans the stretch exponent $\beta$ at fixed $x_{\rm sat}=2.0$~s.}
\label{fig:kww}
\end{figure}

The dispersive family is among the easiest to learn, alongside Michaelis--Menten, because most of its variation sits near the origin, where the knots are densest. Every strategy crosses the output-noise level within a few measurements (median $N=5$), and the three log-aware designs finish close together ($0.038$ to $0.041$, with uniform random behind at $0.056$). Active selection helps least where the shape is most forgiving (Fig.~\ref{fig:kww}c). The noise-free sweep is near exact except at the strongest dispersion, $\beta=0.35$, whose $25\%$ level falls below the first measurable input $x_{\min}=0.010$~s and so cannot be recovered (sweep error $0.046$, with a $-0.09$ response offset at the unreachable $25\%$ level even without noise). Of that sweep error, $0.026$ is the best any non-negative combination of the basis can reach (Fig.~\ref{fig:knots}), so most of it is a limit of what can be represented below the window instead of a failure of the fit. The limit comes from the observation window, since the in-window levels of the same variant are exact (Fig.~\ref{fig:kww}b).

\section{Conditioning, window dependence, and band approximations of the input read-out}

\subsection{Predicted versus observed conditioning shift}
\label{sec:si-cond}
This section checks the claim of Secs.~\ref{sec:readout} and~\ref{sec:pulse_time} that the input error at the $95\%$ level is the amplification of small height errors by a nearly flat curve, by deriving the conditioning relation and testing it against the observed read-out biases. The relation follows from expanding the fitted mean about the crossing. Let $\Delta y=\mu_\ast(x_\alpha)-\alpha A$ be the vertical misfit at the true crossing $x_\alpha$ and $\Delta A=\mu_\ast(x_{\max})-A$ the plateau misestimate. The read-out crossing $\hat{x}_\alpha$ solves $\mu_\ast(\hat{x}_\alpha)=\alpha\,\mu_\ast(x_{\max})=\alpha(A+\Delta A)$, and expanding the left side about $x_\alpha$ gives $\mu_\ast(\hat{x}_\alpha)\approx\alpha A+\Delta y+\mu_\ast'\,\Delta x$, so to first order the read-out shift is
\begin{equation}
\Delta x(\alpha) \;\simeq\; \frac{\alpha\,\Delta A-\Delta y}{\mu_\ast'(\hat{x}_\alpha)},
\label{eq:cond}
\end{equation}
where $\mu_\ast'(\hat{x}_\alpha)$ is the slope of the fitted curve at the crossing, and the plateau term enters amplified by $\alpha$. Table~\ref{tab:si-cond} evaluates it at $\alpha=0.95$ from the level-wise errors of Table~\ref{tab:isotherms}(b) ($\Delta y$), the inferred plateaus of Table~\ref{tab:isotherms}(a) ($\Delta A=\mu_\ast(x_{\max})-1$), and the posterior amplification $1/\mu_\ast'$, the reciprocal of the fitted slope $\mu_\ast'(\hat{x}_\alpha)$ in Eq.~(\ref{eq:cond}) taken as the median over the same $50$ read-out refits per family, against the observed input bias of Table~\ref{tab:readout}. The plateau term turns the Michaelis--Menten crossing early. With $\Delta y=-0.017$ the vertical term alone predicts a late crossing, but $\alpha\,\Delta A=0.95\times(-0.022)$ outweighs it and gives the observed early ($\Delta x<0$) shift, which the plateau-free form of the relation cannot reproduce. The first-order expansion captures the sign in four of the five families (Sips, whose $-0.13$~s bias is within seed scatter, is the exception) and the order of magnitude throughout. On the three large shifts it overestimates the magnitude, as a linearization of an ill-conditioned inverse about the biased crossing will, and on Michaelis--Menten, where the two terms nearly cancel, it underestimates it. The read-out errors of Table~\ref{tab:readout} are therefore accounted for by noise-level height errors passed through a slope factor of $18$ to $36$, which is what makes the $95\%$ input intrinsically hard to pin down for any method, and the fitted slope gives that factor on real data, where the truth is unknown.

\begin{table}[htbp]
\centering
\caption{Predicted versus observed input shift $\Delta x(95\%)$ per family from the conditioning relation Eq.~(\ref{eq:cond}) at $\alpha=0.95$, with $\Delta y$, $\Delta A$, and the amplification $1/\mu_\ast'$ taken as described in Sec.~\ref{sec:si-cond}. ``Predicted'' is $(\alpha\Delta A-\Delta y)/\mu_\ast'$ and ``observed'' is each family's mean over the $50$ seeds of the median crossing minus $2.0$~s, the I-spline bias column of Table~\ref{tab:readout}.}
\label{tab:si-cond}
\footnotesize
\setlength{\tabcolsep}{5pt}
\begin{tabular}{lccccc}
\toprule
curve & $\Delta y$ & $\Delta A$ & $1/\mu_\ast'$ & predicted $\Delta x$ (s) & observed $\Delta x$ (s) \\
\midrule
Langmuir          & $-0.041$ & $-0.002$ & $25.9$ & $+1.01$ & $+0.60$ \\
Michaelis--Menten & $-0.017$ & $-0.022$ & $22.5$ & $-0.09$ & $-0.43$ \\
Sips ($n=2$)      & $-0.013$ & $-0.010$ & $17.7$ & $+0.06$ & $-0.13$ \\
Elovich           & $-0.064$ & $+0.013$ & $24.2$ & $+1.85$ & $+1.05$ \\
KWW               & $-0.034$ & $-0.012$ & $35.6$ & $+0.80$ & $+0.51$ \\
\bottomrule
\end{tabular}
\end{table}

\subsection{Window dependence and band approximations}
\label{sec:si-window}
This section gives the window-sweep detail behind the window dependence of the read-out (Sec.~\ref{sec:calibration}), the per-family coverage of the band and its relation to where the error falls, and the two further band approximations mentioned there.

\paragraph{The median read-out is robust to the window, and the one-sided limit grows with it.} The experimenter chooses the window edge $x_{\max}$, the read-out is referenced to the plateau inside it, and every headline result is taken at $x_{\max}=5$~s, so this section asks whether the recommended input and its coverage depend on that choice. Every draw reaches $\theta(x_{\max})=1$ exactly, so the crossing posterior lives on $[0,x_{\max}]$ and can move with the window even when the curve does not. Sweeping the window edge $x_{\max}$ from $1.5$ to $8.0$~s while holding the curves fixed separates the median from the conservative limit (Table~\ref{tab:window}). For any window that comfortably brackets saturation ($x_{\max}\ge2.5$~s, at least $1.25$ times the true $x_{\rm sat}$) the median crossing stays close to the truth ($1.82$ to $2.55$~s against $x_{\rm sat}=2.0$~s) and two-sided coverage holds near the nominal $95\%$ ($90$ to $93\%$), so the well-centered read-out of the I-spline is not an artifact of the default window. This sub-campaign runs twelve seeds per family at each window, so each of these two-sided coverage figures rests on $60$ runs, carries a binomial standard error\cite{64} of $3$ to $4$ percentage points, and cannot be told apart from nominal. More seeds would tighten the estimate. The one-sided $95$th-percentile limit does scale with the window, climbing from $1.4$ to $5.9$~s as $x_{\max}$ widens. A one-sided upper bound referenced to an in-window plateau has to grow with the window, and this growth is the price of its conservatism. The read-out degrades only when the window fails to bracket saturation ($x_{\max}\le2.0$~s), where the crossing is cut off from above and both coverages fall to zero. The practical risk is therefore a truncated window, since a generous one does not inflate the median, and the window-sufficiency statistic of Sec.~\ref{sec:readout} is designed to catch the truncated case (Sec.~\ref{sec:si-outside}). For a practitioner the guidance is therefore to choose $x_{\max}$ generously, since a wide window costs the median read-out nothing, to check $r$ against the window length, and, when using the one-sided limit, to remember that its value grows with the window.

\begin{table}[htbp]
\centering
\caption{Window sweep (Sec.~\ref{sec:si-window}), with the knot basis rebuilt on each window and $12$ seeds per family in each cell. ``True frac'' is the true saturation reached at $x_{\max}$, ``median $\hat{x}_{\rm sat}$'' the per-run median crossing averaged over the runs, ``$r$'' the window-sufficiency statistic, and ``flag@$0.15$'' the share of runs flagged by $r>0.15$ at that window, a false-alarm rate wherever the window brackets saturation ($x_{\max}\ge2.5$~s, Sec.~\ref{sec:si-outside}).}
\label{tab:window}
\footnotesize
\setlength{\tabcolsep}{4pt}
\begin{tabular}{cccccccc}
\toprule
$x_{\max}$ (s) & true frac & median $\hat{x}_{\rm sat}$ & one-sided limit & one-sided cov. & two-sided cov. & $r$ & flag@$0.15$ \\
\midrule
$1.5$ & $0.897$ & $1.24$ & $1.42$ & $0\%$   & $0\%$   & $0.204$ & $70\%$ \\
$2.0$ & $0.950$ & $1.55$ & $1.86$ & $0\%$   & $0\%$   & $0.170$ & $52\%$ \\
$2.5$ & $0.974$ & $1.82$ & $2.29$ & $98\%$  & $90\%$  & $0.141$ & $32\%$ \\
$3.5$ & $0.986$ & $2.16$ & $3.04$ & $97\%$  & $90\%$  & $0.095$ & $18\%$ \\
$5.0$ & $0.993$ & $2.25$ & $4.05$ & $100\%$ & $93\%$  & $0.054$ & $2\%$ \\
$8.0$ & $0.996$ & $2.55$ & $5.87$ & $100\%$ & $93\%$  & $0.034$ & $2\%$ \\
\bottomrule
\end{tabular}
\end{table}

\paragraph{Band coverage by family.} The coverage of the nominal $95\%$ band is the fraction of input points and seeds at which the true curve falls inside $\mu_\ast\pm1.96\sigma_\ast$. Pooled over families it is $94\%$ over the whole window, against $87\%$ for the unpenalized $P=11$ basis, and $96\%$ on the part of the window with data ($x\ge x_{\min}$). Over the full window four of the five families sit at $97$ to $98\%$ and the dispersive KWW at $80\%$. Restricted to $x\ge x_{\min}$, KWW is at $96\%$ and every family lies within $95$ to $98\%$, so there is no outlier. The whole shortfall sits below $x_{\min}$, where nothing is measured, every draw starts from the pinned origin with a band of almost zero width, and the steep early rise of the dispersive KWW response has already happened. The posterior s.d.\ $\sigma_\ast(x)$ runs about $1.5\times$ the seed-to-seed s.d.\ of the posterior mean, so the band is if anything wider than the actual scatter, and the coverage it loses comes from the smoothing bias, which no variance band can express.

\paragraph{Band localization.} The band says little about where the error is. We measure this with the Spearman rank correlation $\rho_s$ between band width and realized error, which is $1$ if the widest bands always sit where the errors are largest and $0$ if width and error are unrelated. Pooling every input point and seed gives $0.35$. Restricted to the region with data, $x\ge x_{\min}$, it falls to $0.06$, and correlating across the $250$ (family, seed) runs at each fixed $x$ in the window and then averaging over those $x$, which removes the shared near-origin profile that inflates the pooled figure, gives $0.08$. The convention matters more than the exact value, so we report all three, and none of them is large. The unpenalized $P=11$ basis does slightly better on this measure ($0.47$ pooled, $0.18$ in the window, $0.19$ across seeds), so the penalized band's advantage over it is an advantage in coverage, with no matching advantage in locating the error. This bears on the acquisition result of Sec.~\ref{sec:acq_compare}. The penalized surrogate edges a log-spaced grid, yet its band tracks its own error no better than the unpenalized band tracks its own, so whatever produces the edge, it is not the band pointing accurately at the error. The acquisition rule is not undercut by this, because it needs the band only to mark where the data leave the curve undetermined, the gaps it then fills, and a band can do that well while saying little about where the mean is wrong. A low correlation here is also expected on its own terms, since much of the error at $N=20$ is the smoothing bias, the same in every seed, and a single realized error scatters widely about any band width, so even a perfectly calibrated band would give only a modest rank correlation.

\paragraph{Two further band approximations.} Two further approximations also make the nominal level approximate. The likelihood assumes one noise level everywhere, yet the input jitter of Eq.~(\ref{eq:noise}) makes the true scatter depend on the input and largest where the curve is steep. For the Michaelis--Menten and KWW rises, whose slopes reach $7.6$ to $7.9$~s$^{-1}$ at the first measurable input, the jitter alone adds scatter of the same order as the output noise there ($\sigma_x\lvert f'\rvert\approx0.08$) and almost none on the plateau, so a single inferred $\sigma$ under-covers the onset and over-covers the plateau. The plotted band is also the symmetric summary $\mu_\ast\pm1.96\sigma_\ast$, whereas the non-negativity constraint boxes the plausible values of $f(x)$ in on one side near the onset and the plateau, so their spread is lopsided and the symmetric band is itself an approximation to the true pointwise CI, which posterior quantiles would report better. The nominal $95\%$ level should therefore be read as approximate, and calibrating it exactly, by inflating the noise term to absorb the smoothing bias or by reporting posterior quantiles, is a worthwhile refinement. The obvious route, a likelihood whose noise grows with the local slope, proved numerically unstable here (Table~\ref{tab:si-model-ablation}).

\section{The comparison arms in detail}

\subsection{GP failure modes}
\label{sec:si-gpfail}
Table~\ref{tab:gpfail} lists the structural failures that survive the resolution match of Sec.~\ref{sec:whymono}, and this section states them in full. The headline, that the constraint guarantees a well-defined read-out and does not by itself guarantee accuracy, is in the main text.

Matched for resolution and for the origin, the unconstrained posterior mean still falls somewhere in about $70\%$ of fits, and in $18\%$ of GP-log fits and $3\%$ of GP-log-pin fits the deepest of those dips exceeds the output noise, so it cannot be excused as scatter. Across $3750$ active selections per arm, the unconstrained loops placed a share of their measurements at an input where their own posterior mean was falling, $12\%$ for GP-log-pin, $20\%$ for GP-log, and $46\%$ for GP-linear on the raw axis. The sharper event, a selection within one candidate spacing of a local minimum of the mean, is rarer, $0.1$ to $0.6\%$. These misled selections do not cost curve accuracy on this problem class. Regressing each GP-log-pin run's saturation error on its count of falling-mean selections gives a slope of $-0.002$ per selection (standard error $0.001$), which is negligible and, if anything, of the opposite sign to the positive slope a real accuracy cost would produce. This agrees with Table~\ref{tab:gp}, where resolution sets that arm's accuracy. The misled selections do undermine the surrogate as a physical model, and the input read-out is separately more accurate for the monotone arm (Table~\ref{tab:readout}). The multi-valued read-out and the inferred $f(0)$ point the other way. The multi-valued read-out, which reaches $48\%$ on the raw axis with a mean upcrossing span of $2.5$~s on a $5$~s window, all but disappears once the axis is logarithmic (under $1\%$ for both resolution-matched arms, and $0$ to $1\%$ for GP-log at every level of the output-noise sweep). It is a resolution artifact, so a multi-valued read-out is a rare event for a well-configured unconstrained fit. The inferred $f(0)$, which the linear-axis arm misses by $0.134$ and GP-log by $0.076$, is zero by construction for GP-log-pin as it is for ours. Nothing downstream uses $f(0)$, since the read-out never evaluates below $x_{\min}$, so we note the miss without pricing it.

\begin{table}[htbp]
\centering
\caption{Structural failures that survive the resolution match, for the three unconstrained arms, pooled over five families and fifty seeds at $N=20$ (Sec.~\ref{sec:si-gpfail}). Total negative variation is the summed downward movement of the posterior mean across the window, in response units, and the upcrossing span is averaged over the multi-valued fits only ($121$, $2$, and $1$ fits for the three arms).}
\label{tab:gpfail}
\begin{tabular}{lcccc}
\toprule
 & GP-linear & GP-log & GP-log-pin & I-spline \\
\midrule
non-monotone posterior mean          & $100.0\%$ & $78.0\%$ & $70.0\%$ & $0\%$ \\
total negative variation             & $0.303$  & $0.064$  & $0.022$  & $0$ \\
deepest dip                          & $0.131$  & $0.050$  & $0.019$  & $0$ \\
deepest dip $>\sigma_y$              & $52.4\%$ & $17.6\%$ & $3.2\%$  & $0\%$ \\
multi-valued $95\%$ read-out         & $48.4\%$ & $0.8\%$  & $0.4\%$  & $0\%$ \\
upcrossing span (multi-valued fits)  & $2.46$~s & $2.39$~s & $2.86$~s & n/a \\
inferred $f(0)$ (true $0$)           & $0.134$  & $0.076$  & $0$      & $0$ \\
selections on a falling mean         & $46.1\%$ & $20.5\%$ & $11.9\%$ & $0\%$ \\
\bottomrule
\end{tabular}
\end{table}

\paragraph{Monotonizing the GP after the fit does not recover the constraint.} A natural objection is that monotonicity need not be built into the prior at all, since one could fit the unconstrained GP-log-pin and force it to be monotone afterward. We tested the two standard projections on the same loop ($50$ seeds $\times$ five families), a pool-adjacent-violators (isotonic) projection\cite{20} of the posterior mean and a Lin--Dunson\cite{21} projection of every posterior draw. Both fall short of what the prior constraint provides. The projected saturation error is $0.049$ for both, the same as the raw GP and $1.26\times$ the I-spline's $0.039$. The dips in the mean are shallow, so removing them barely changes the curve, and the projection adds none of the resolution or roughness-penalty accuracy the I-spline carries. The projection also does not repair the uncertainty, since the Lin--Dunson ``monotone posterior'' reads $x_{\rm sat}(0.95)$ with a bias of $+0.49$~s but covers the truth in only $61\%$ of intervals, against the I-spline's $94\%$. Projection keeps the overconfident spread of the raw GP-log-pin posterior, which covers only $64\%$, so the projected spread is not calibrated. Fixing monotonicity after the fit therefore gives a monotone point estimate no more accurate than the raw fit and a read-out whose interval is not calibrated. For this reason the constraint is imposed in the prior, where it holds for every draw at no cost in accuracy or calibration.

\subsection{The matched-design control for the GP arms}
\label{sec:si-gpmatched}
Every arm in Table~\ref{tab:gp} drives its own loop, so that table compares two things at once, how well each surrogate fits the data it has and how well its band chooses the next input. This section separates them by fitting GP-log-pin and GP-log to the designs the I-spline campaign stored, the same inputs and the same noisy readings, for every acquisition strategy and all $50$ seeds per family, and scoring the fits with the same saturation error (Table~\ref{tab:si-gpmatched}). On identical data the surrogate advantage at matched resolution and origin is small. On the uncertainty-sampling designs GP-log-pin scores $1.07\times$ the I-spline's error, a consistent difference over the $250$ paired runs (Wilcoxon $p=3\times10^{-4}$), and on the log-spaced grid designs the two are indistinguishable. GP-log, without the pinned origin, trails by $1.1$ to $1.2\times$ on every design. Most of the $1.26\times$ gap of Table~\ref{tab:gp} is therefore the difference between the designs the two loops chose.

The reason is the shape of the GP band, which is nearly flat because the marginal likelihood picks long lengthscales on this problem, a median of about half the warped domain for GP-log-pin, so the posterior s.d.\ varies little across the input axis. Re-running the GP loops and recording the band at each of the fifteen acquisition steps (Table~\ref{tab:si-gpdesign}), we find that on average $40\%$ of the candidate inputs lie within $10\%$ of the widest band value and the narrowest band is $0.58$ of the widest, so the maximum that uncertainty sampling chases is barely distinguished from the rest. The loop then cycles among a few inputs. A finished GP-log-pin design has $10.1$ distinct inputs out of $20$ against the I-spline's $12.9$, and puts $6.3$ readings at $x_{\max}$ against $4.7$. The GP's own uncertainty sampling ends at a pooled error of $0.049$ (Table~\ref{tab:gp}), above the $0.045$ the same surrogate reaches on the I-spline's log-spaced grid designs, so for the GP the band does not help the acquisition. The flat band is a property of an unconstrained stationary kernel fitted by marginal likelihood to twenty points on this problem, and the I-spline band, whose width is the spread of the monotone draws, does not have it. The read-out comparison of Table~\ref{tab:readout} also lets GP-log-pin drive its own loop, so the same caveat applies to it.

For the scorecard of the main text (Fig.~\ref{fig:scorecard}) GP-linear and the two unpenalized bases were fitted to the same stored designs as well. Like the log-axis GPs, none of them does better under its own uncertainty sampling than on a log-spaced grid, GP-linear scoring $0.095$ against $0.073$ and the unpenalized bases $0.078$ against $0.072$ for $P=25$ and $0.059$ against $0.058$ for $P=11$, so the penalized I-spline is the only surrogate in that figure whose own loop is its best design.

The same split was run for the noise sweep and the held-out benchmark, whose own-loop comparisons in Tables~\ref{tab:si-noise} and~\ref{tab:heldout} have the structure of Table~\ref{tab:gp} (Table~\ref{tab:si-gpmatched-sweeps}). At every noise level the surrogate advantage on identical data is a few percent, and at the highest level it is not significant, whereas GP-log-pin driving its own loop trails by $1.2$ to $1.3\times$ throughout, so the acquisition also explains the noise-sweep margin. GP-log, without the pinned origin, trails on identical data by a margin that grows with the noise, from $1.18\times$ at $\sigma_y=0.05$ to $1.31\times$ at $0.15$, so at high noise the origin matters more than the constraint. On the held-out curves the surrogate advantage on identical data is larger than on the development set, $1.18\times$ under the log-uniform draw and $1.14\times$ under the uniform one, against $1.27$ and $1.33\times$ for the GP's own loop, and part of the reliability edge survives the matching, since GP-log-pin on the I-spline's designs exceeds the output-noise level on about twice as many curves as the I-spline ($6\%$ against $3\%$, and $2\%$ against none under the uniform draw) and its largest error over the $250$ curves is twice the I-spline's ($0.34$ against $0.17$).

\begin{table}[htbp]
\centering
\caption{The matched-design control (Sec.~\ref{sec:si-gpmatched}), as the saturation error at $N=20$ pooled over five families and $50$ seeds. The ratio is GP-log-pin to I-spline, and $p$ is a paired Wilcoxon test over the $250$ runs.}
\label{tab:si-gpmatched}
\footnotesize
\begin{tabular}{lccccc}
\toprule
design & I-spline & GP-log-pin & GP-log & ratio & paired $p$ \\
\midrule
uncertainty sampling & $0.039$ & $0.042$ & $0.048$ & $1.07$ & $3\times10^{-4}$ \\
log-spaced grid      & $0.044$ & $0.045$ & $0.050$ & $1.02$ & $0.4$ \\
log-uniform random   & $0.044$ & $0.048$ & $0.051$ & $1.08$ & $4\times10^{-3}$ \\
uniform random       & $0.058$ & $0.060$ & $0.066$ & $1.03$ & $0.05$ \\
\bottomrule
\end{tabular}
\end{table}

\begin{table}[htbp]
\centering
\caption{Design clustering and band flatness when each surrogate drives its own loop, as means over the $250$ runs at $N=20$ (Sec.~\ref{sec:si-gpmatched}). Readings within $0.02$~s of each other count as one input, and within $0.02$~s of $x_{\max}$ as at $x_{\max}$. Flat share is the share of candidate inputs whose band is within $10\%$ of the widest, and both band columns are averaged over the fifteen acquisition steps of the re-run GP loops. The I-spline loops were not re-run.}
\label{tab:si-gpdesign}
\footnotesize
\begin{tabular}{lcccc}
\toprule
surrogate (own loop) & distinct inputs & readings at $x_{\max}$ & flat share & narrowest/widest \\
\midrule
I-spline   & $12.9$ & $4.7$ & n/a & n/a \\
GP-log-pin & $10.1$ & $6.3$ & $0.40$ & $0.58$ \\
GP-log     & $10.2$ & $4.8$ & $0.34$ & $0.73$ \\
\bottomrule
\end{tabular}
\end{table}

\begin{table}[htbp]
\centering
\caption{The matched-design control on the noise sweep and the held-out benchmark (Sec.~\ref{sec:si-gpmatched}), as the saturation error at $N=20$, pooled over five families and $50$ seeds for the noise sweep and over the $250$ held-out curves for each draw of $x_{\rm sat}$. Ratios are to the I-spline, and $p$ is a paired Wilcoxon test on the same-design fits.}
\label{tab:si-gpmatched-sweeps}
\footnotesize
\setlength{\tabcolsep}{4pt}
\begin{tabular}{lcccccc}
\toprule
 & & \multicolumn{3}{c}{GP-log-pin, same designs} & \multicolumn{2}{c}{GP-log-pin, own loop} \\
\cmidrule(lr){3-5}\cmidrule(lr){6-7}
case & I-spline & error & ratio & paired $p$ & error & ratio \\
\midrule
noise sweep, $\sigma_y=0.05$ & $0.023$ & $0.024$ & $1.05$ & $5\times10^{-5}$ & $0.030$ & $1.28$ \\
noise sweep, $\sigma_y=0.10$ & $0.039$ & $0.042$ & $1.07$ & $3\times10^{-4}$ & $0.049$ & $1.26$ \\
noise sweep, $\sigma_y=0.15$ & $0.054$ & $0.055$ & $1.02$ & $0.9$ & $0.065$ & $1.20$ \\
held-out, $x_{\rm sat}$ log-uniform & $0.045$ & $0.053$ & $1.18$ & $3\times10^{-8}$ & $0.058$ & $1.27$ \\
held-out, $x_{\rm sat}$ uniform & $0.040$ & $0.045$ & $1.14$ & $2\times10^{-7}$ & $0.053$ & $1.33$ \\
\bottomrule
\end{tabular}
\end{table}

\subsection{Parametric baselines and the cost of assuming a form}
\label{sec:si-parametric}
The surrogate assumes no kinetic model, and the natural objection is that a practitioner who knows the mechanism could fit its two- or three-parameter form directly and do better. We measure both sides by fitting, on each family's own $N=20$ design, all five parametric forms of Sec.~\ref{sec:kinetics} by least squares and reading the saturation error and $x_{\rm sat}(0.95)$ off each fit. The true-form fit is the oracle, available only to someone who already knows the mechanism, and a wrong-form fit is the realistic case when they do not (Table~\ref{tab:si-parametric}).

Two things follow. First, knowing the form buys no curve accuracy at this budget. The oracle fit's pooled saturation error is $0.038$, no different from the nonparametric surrogate's $0.039$, because at twenty noisy points the fit is limited by parameter uncertainty instead of by the model class. Its read-out is also no more reliable, since the oracle $x_{\rm sat}$ bias swings from $-0.6$ to $+2.2$~s across families (recomputed per family from the cached fits, worst on the capped Elovich form, which is poorly conditioned for least squares), against the surrogate's $+0.3$~s pooled on the same designs (Table~\ref{tab:readout}). Second, guessing the form wrong is expensive, since a wrong form averages $0.053$ and reaches $0.078$ pooled, and up to $0.118$ for a concave form forced onto the sigmoidal Sips onset, which is $1.4$ to $3$ times the nonparametric error. The parametric route can therefore at best match the nonparametric surrogate, and only when the mechanism is both known and well conditioned. When it is not, the parametric fit is worse, so assuming no form gives up nothing and removes the risk of guessing wrong.

\begin{table}[htbp]
\centering
\caption{Parametric least-squares fits on the same $N=20$ designs, pooled over five families and $50$ seeds. ``Oracle'' fits each family's true form. The two misspecified rows take, for each run, the mean and the worst of the saturation error over the four wrong forms, pooled like the rest.}
\label{tab:si-parametric}
\footnotesize
\begin{tabular}{lc}
\toprule
parametric least-squares fit & saturation error \\
\midrule
oracle (true form)        & $0.038$ \\
mean misspecified form    & $0.053$ \\
worst misspecified form   & $0.078$ \\
\midrule
I-spline (reference) & $0.039$ \\
\bottomrule
\end{tabular}
\end{table}

\section{Acquisition robustness controls}
\label{sec:si-controls}
This section gives the full acquisition comparison behind Sec.~\ref{sec:acq_compare} and three controls, two that ask whether the edge of uncertainty sampling over a log-spaced grid depends on how the loop is set up and one that asks whether the monotone constraint costs anything while the data are collected.

\paragraph{Full acquisition comparison.} In the fixed-location comparison ($x_{\rm sat}=2.0$~s for all five curves), uncertainty sampling improves the $N=20$ error by about $1.5\times$ over uniform random and by $1.12$ to $1.13\times$ over both log-uniform random and a log-spaced grid. It wins on the spread-rise families (Langmuir $1.13\times$, Sips $1.23\times$, Elovich $1.22\times$) and trails slightly on the fastest (KWW $0.98\times$). Randomizing the saturation location (Table~\ref{tab:randxsat}) leaves the picture much the same. Adaptive selection beats log-uniform random ($1.19\times$ pooled) and uniform random ($1.79\times$) and edges the log-spaced grid ($1.13\times$, $p=0.002$). Drawing $x_{\rm sat}$ uniformly in $[0.2,4.0]$~s instead (median $2.1$~s, so more curves rise late, where the log-spaced grid is sparse) shrinks that edge only slightly while keeping it significant (active over grid $1.08\times$, $p=0.02$), so at this noise the advantage does not come from the grid's fixed geometry mismatching the distribution of rise locations. It also does not come from the band locating the error better than the unpenalized basis's band, because it does not (Sec.~\ref{sec:si-window}). We report the edge as an observation whose mechanism we have not isolated. The one caveat is at the left end of the learning curves, where at $N=5$ and $N=10$ the absolute error still depends on the roughness-prior scale (Sec.~\ref{sec:si-priors}), but all four strategies share it, so the dependence shifts them together and largely cancels.

\begin{table}[htbp]
\centering
\caption{Acquisition comparison with the saturation location randomized (Sec.~\ref{sec:si-controls}), as the saturation error across levels at $N=20$, mean $\pm$ one s.d.\ over the $R=50$ seeds, under uncertainty sampling and the three non-adaptive baselines. Ratios quoted in the text use the unrounded means.}
\label{tab:randxsat}
\footnotesize
\setlength{\tabcolsep}{4pt}
\begin{tabular}{lcccc}
\toprule
curve & uncertainty sampling & log-spaced grid & log-uniform random & uniform random \\
\midrule
Langmuir          & $0.039\pm0.016$ & $0.045\pm0.019$ & $0.047\pm0.022$ & $0.069\pm0.058$ \\
Michaelis--Menten & $0.040\pm0.018$ & $0.041\pm0.023$ & $0.039\pm0.018$ & $0.063\pm0.043$ \\
Sips ($n=2$)      & $0.038\pm0.014$ & $0.049\pm0.024$ & $0.054\pm0.030$ & $0.089\pm0.061$ \\
Elovich           & $0.040\pm0.014$ & $0.050\pm0.018$ & $0.052\pm0.024$ & $0.072\pm0.062$ \\
KWW               & $0.044\pm0.014$ & $0.044\pm0.017$ & $0.047\pm0.015$ & $0.066\pm0.039$ \\
\bottomrule
\end{tabular}
\end{table}

\paragraph{Candidate-set geometry.} The candidate set over which $\sigma_\ast$ is maximized is linear at the $0.001$~s input resolution of Sec.~\ref{sec:acq}, which already resolves the onset below $0.02$~s finely, so the loop is free to propose inputs anywhere the steepest curves are rising. Replacing it with a geometric candidate set of the same span leaves the outcome unchanged. The $N=20$ error moves by at most $0.005$ on any family, with the largest shift on Michaelis--Menten and the others smaller, all within the seed-to-seed scatter. The designs themselves show why, because even though it is free to, the loop places almost none of its fifteen active measurements below $0.0175$~s ($0.03$ per run pooled, at most $0.10$ on the fast KWW rise, and never more than $2$ in a single run). The survey point at $x_{\min}$ already pins the posterior there, and $\sigma_\ast$ is largest in the wider gaps further out. The conclusions of this section therefore do not depend on the geometry of the candidate set.

\paragraph{Survey size.} A second control turns from the candidate set to the starting survey. The survey is fixed at $N_0=5$ of $N=20$ throughout, so a quarter of every design is non-adaptive by construction, and the explanation just given for the candidate-set result leans on the survey point at $x_{\min}$. If the survey were doing work that the active-versus-grid comparison credits to the acquisition rule, then enlarging it should push uncertainty sampling toward a space-filling design and shrinking it should pull the other way. It does neither, since sweeping $N_0\in\{3,5,8\}$ at fixed total budget on the randomized-location benchmark leaves the pooled $N=20$ error at $0.040$, $0.040$, and $0.041$, against $0.046$ for the log-spaced grid, with every family moving by at most $0.006$. Over a survey share from $15\%$ to $40\%$ of the budget the result is flat, so the edge over the grid belongs to the problem and does not depend on the survey size.

\paragraph{Unconstrained-driven acquisition.} A third control asks whether the monotonicity constraint, useful for the read-out and the guarantees, might still cost something during learning by narrowing where the loop looks, so that one would do better to collect the data with an unconstrained spline and impose monotonicity only at the end. We tested that split directly by letting a penalized but unconstrained signed-weight spline (the Eilers--Marx driver of Table~\ref{tab:si-surrogates}, with monotonicity removed) drive the same uncertainty-sampling loop and refitting the monotone surrogate to the points it collects at each budget. Two results follow ($50$ seeds $\times$ five families). First, the unconstrained driver's own error and the monotone refit of the same data track each other closely from $N=8$ onward (pooled ratio $0.94$ to $1.02$ across $N=8$ to $20$). The monotone fit is visibly more constrained only at $N=5$ ($1.21\times$), where five points barely determine a shape. Second, the split does not pay, since at $N=20$ the unconstrained-driven design, refit by the monotone surrogate, reaches a pooled saturation error of $0.043$, slightly worse than the $0.039$ of the monotone surrogate driving its own loop (paired $p=0.006$). Collecting the data with the unconstrained spline therefore buys nothing, and it gives up the dip-free acquisition and the guaranteed read-out at every budget along the way. The monotone surrogate is at least as good a guide to the next measurement as the unconstrained one, so there is no reason to separate the two roles.

\section{Held-out benchmark generator construction and controls}
\label{sec:si-heldout}
This section specifies the distribution that the held-out benchmark of Sec.~\ref{sec:heldout} draws from and the change-of-measure control mentioned there.

\paragraph{Generators.} Two generators cover the two kinds of shape a self-limiting response can take. Every saturating response whose rate is completely monotone is a mixture of exponentials, by Bernstein's theorem,\cite{18} so the first generator draws
\begin{equation}
\begin{gathered}
f(x)=\sum_{k=1}^{K} a_k\left(1-e^{-x/\tau_k}\right),\\[3pt]
K\sim\mathrm{Uniform}\{1,\dots,8\},\quad
\bm{a}\sim\mathrm{Dirichlet}(\bm{1}),\quad
\log\tau_k\sim\mathrm{Uniform}[\log 0.01,\,\log 3].
\end{gathered}
\label{eq:gen_mix}
\end{equation}
Here $K$ is the number of exponential components, drawn uniformly from one to eight. The weights $\bm{a}=(a_1,\dots,a_K)$ come from a flat Dirichlet distribution, so they are non-negative and sum to one, which puts the plateau at one. The time constants $\tau_k$ are drawn log-uniformly between $0.01$ and $3$~s, so fast and slow components are equally likely. A single component ($K=1$) has the Langmuir shape up to its time scale, and several components with spread time constants give the dispersive, Elovich-like shapes, so the set is not entirely new, though no draw repeats a headline curve with its headline rise location. Cooperative onsets are not completely monotone, so the sigmoidal branch needs its own generator,
\begin{equation}
f(x)=\frac{u^{n}}{1+u^{n}},\qquad u=\frac{x}{\tau_S},\qquad
\log n\sim\mathrm{Uniform}[\log 1,\,\log 4],
\label{eq:gen_hill}
\end{equation}
which is the Hill form of Eq.~(\ref{eq:iso_sips}) with $A=1$. Here $\tau_S$ is the half-maximal input, set by the rescaling below, and $n$ is the Hill coefficient, drawn log-uniformly between $1$, a concave curve, and $4$, a step-like onset. Each draw of either generator is then rescaled in $x$ so that $f(x_{\rm sat})=0.95$, with $x_{\rm sat}$ drawn log-uniformly on $[0.2,4.0]$~s as in the randomized-location campaign of Sec.~\ref{sec:acq_compare}, so neither the shape nor the position of the rise is fixed. We drew $200$ mixtures and $50$ Hill curves. All $250$ are monotone, are exactly zero at the origin, and saturate inside the window. Every held-out curve is observed through the noise model of the main text (Sec.~\ref{sec:noise}), with the same input jitter and output noise as the development set, so the benchmark changes the ground truth and nothing else. Most of what separates a held-out curve from the nearest of the five development curves is its randomized rise location, and its shape accounts for less, since rescaling the input axis, which a time-scale-agnostic surrogate does not notice, removes most of that separation. The generator therefore tests generalization in shape only partially and mostly tests it in rise location. We read the held-out result with that in mind.

\paragraph{Change-of-measure control.} We control one confound, the alignment between the draw of $x_{\rm sat}$ and the log-spaced basis. The generator draws $x_{\rm sat}$ log-uniformly, which is the measure the log-spaced basis is matched to, so a share of the I-spline's edge could come from that alignment. The main text reports the benchmark against GP-log-pin alone, the strongest unconstrained arm. The resolution-only GP-log was run as well and trails further. Under the log-uniform draw its pooled error is $0.092$ against the I-spline's $0.045$ (median error ratio $1.69$, the I-spline better on $77\%$ of curves), $30\%$ of its curves lie above the output-noise level, and its worst curve scores $0.400$. Repeating the whole benchmark with $x_{\rm sat}$ drawn uniformly on the same interval (Table~\ref{tab:heldout}, lower block) shrinks that advantage over GP-log to $0.067$ against $0.040$ (median ratio $1.48$, the I-spline better on $77\%$), as the alignment argument predicts, whereas the advantage over the origin-matched GP-log-pin is unchanged at $0.053$ against $0.040$ (median ratio $1.27$). The reliability edge survives the change of measure. No I-spline curve exceeds the output-noise level and its largest error is $0.092$, whereas GP-log-pin exceeds the level on $7\%$ of the curves with a largest error of $0.184$, and GP-log on $15\%$ with $0.390$. The I-spline's lead is more robust in reliability than in mean accuracy, which is the conclusion the main text draws (Sec.~\ref{sec:heldout}). The tail statistics rest on a single fit per curve. Three seeds per curve would tighten them, at the cost of $500$ more fits that a reliability claim this consistent across two measures does not need.

\section{Mechanism diagnostics for the high-noise reversal}
\label{sec:si-noise}
Section~\ref{sec:noisesweep} reports that the coarse unpenalized monotone basis, the one unpenalized arm that beats GP-log at low noise, loses to it once the output noise approaches a tenth of the plateau (the lines cross near $\sigma_y=0.08$, Fig.~\ref{fig:noise}), and puts that reversal down to a gap in smoothness adaptation after clearing the plateau anchor and the noise prior. This section records the diagnostics behind that attribution, run on the coarse unpenalized $P=11$ basis, and confirms that the roughness penalty of Eq.~(\ref{eq:priors}), adopted in this work, is the remedy. Table~\ref{tab:si-noise} lists the pooled errors behind Fig.~\ref{fig:noise}. The penalized surrogate and GP-log were also run at $\sigma_y=0.01$, where their errors are $0.009$ and $0.015$. The other arms were not run at that level, so it is left out of the comparison.

\begin{table}[htbp]
\centering
\caption{Saturation error at $N=20$ against output noise, pooled over five families and fifty matched seeds, for the arms of Table~\ref{tab:gp} other than GP-linear (Sec.~\ref{sec:si-noise}). Ratios are each arm's error over the penalized arm's, from the unrounded means, and the $p$ values are paired Wilcoxon tests over the $250$ matched runs. The lower block counts non-monotone draws and means.}
\label{tab:si-noise}
\footnotesize
\begin{tabular}{lccc}
\toprule
 & $\sigma_y=0.05$ & $\sigma_y=0.1$ & $\sigma_y=0.15$ \\
\midrule
penalized (monotone, $P{=}25$)      & $0.023$ & $0.039$ & $0.054$ \\
unpen.\ $P{=}25$ (no penalty)       & $0.041$ & $0.078$ & $0.101$ \\
unpen.\ $P{=}11$ (coarse)           & $0.029$ & $0.059$ & $0.087$ \\
GP-log (resolution-matched)         & $0.032$ & $0.053$ & $0.074$ \\
GP-log-pin (\,$+$\,origin)          & $0.030$ & $0.049$ & $0.065$ \\
\midrule
ratio unpen.\ $P{=}25$\,/\,penalized & $1.78$  & $2.00$  & $1.88$ \\
ratio unpen.\ $P{=}11$\,/\,penalized & $1.25$  & $1.52$  & $1.61$ \\
ratio GP-log\,/\,penalized          & $1.40$  & $1.36$  & $1.37$ \\
ratio GP-log-pin\,/\,penalized      & $1.28$  & $1.26$  & $1.20$ \\
paired $p$ (penalized vs GP-log)    & $2\times10^{-22}$ & $5\times10^{-16}$ & $4\times10^{-18}$ \\
\midrule
monotone arms, non-monotone draws   & $0\%$   & $0\%$   & $0\%$ \\
GP-log non-monotone mean            & $82\%$  & $78\%$  & $71\%$ \\
GP-log deepest dip                  & $0.032$ & $0.050$ & $0.073$ \\
\bottomrule
\end{tabular}
\end{table}

\paragraph{The plateau anchor is not the cause.} The read-out is referenced to $f(x_{\max})$, a boundary value that a stationary kernel smooths against its neighbors and a monotone fit pinned at the boundary cannot, so the anchor is the obvious suspect. Testing it clears it, since rescaling each fitted curve to its true plateau, which removes the plateau error exactly, does not close the gap and slightly widens it, from $0.087$ to $0.096$, pooled over the twenty seeds of this diagnostic run on the unpenalized basis itself (the same arm scores $0.087$ over the fifty-seed sweep in Table~\ref{tab:si-noise}, so the two agree). On the fast-rising families the plateau overshoot was partly making up for a shape error instead of causing one.

\paragraph{The noise prior is not misspecified.} A second suspect, a noise-prior scale that fits poorly at $\sigma_y=0.15$, is also cleared. The inferred $\sigma$ is $0.169$, if anything above the truth, so the prior has not pulled it below the truth. The diagnostic also brackets the scale, and run at $s_\sigma=0.2$ and at $0.6$ the pooled error moves by $0.002$ ($0.087$ against $0.085$), which is neither a fix nor a diagnosis.

\paragraph{Smoothness adaptation.} The remaining suspect is a difference in how the two surrogates adapt their smoothness to the noise. The GP re-estimates its lengthscale and noise by maximum marginal likelihood at every step, so at $\sigma_y=0.15$ it explains more of the data as noise and returns a smoother mean. The unpenalized monotone basis has no such control. Its noise scale $\sigma$ and weight scale $\tau_w$ are both inferred, but neither buys smoothness, because the prior shrinks all weights toward zero together, which sets the height of $f$, already pinned by the measurement at $x_{\max}$, and leaves its roughness untouched. There is no penalty on differences between neighboring weights, so the posterior mean is free to wiggle within the monotone constraint however large the inferred noise becomes. The number of basis functions is fixed, and the constraint restricts shape without restricting smoothness. The fix, adopted in this work, is the roughness penalty of Eq.~(\ref{eq:priors}). Its inferred scale $\tau_\eta$ adapts the smoothness to the noise as the GP's marginal likelihood does, and it closes the reversal (Sec.~\ref{sec:noisesweep}). Tightening the weight prior instead would not help, because it would shrink the plateau the data already pin without smoothing the rise.

\paragraph{The reversal by family.} At $\sigma_y=0.1$ the coarse unpenalized $P=11$ arm loses most on the spread-rise shapes (Sips $0.74\times$, Langmuir $0.85\times$, Elovich $0.87\times$) and least or not at all on the fast-rising ones (Michaelis--Menten $0.91\times$, KWW $1.33\times$). Each number is the GP-log error divided by this arm's error, so a value below one means the arm loses. The spread-rise shapes are the ones where the GP's smoothness adaptation pays most, and the penalty lifts every family well above the GP ($1.14$ to $1.56\times$). The ratios use the unrounded family means behind Table~\ref{tab:gp}.

\paragraph{An incidental observation.} The same runs show that the loop takes several of its $20$ measurements at $x_{\max}$, and more of them as the data get noisier (in the fifty-seed main campaign, $4.7$ on average at $\sigma_y=0.1$, Sec.~\ref{sec:acq}), so the loop repeats the plateau measurement on its own, without that being built into the design.

\section{Details of behavior outside the benchmark's assumptions}
\label{sec:si-outside}
This section gives the four perturbation arms and their diagnostics behind the summary in Sec.~\ref{sec:limitations}. Every ground truth in the benchmark is exactly monotone, exactly zero at the origin, and fully saturated inside the window, and every measurement is clean, with Gaussian scatter and no outlier or drift. The benchmark therefore measures error when every assumption holds and never tests what happens when one fails. Four perturbations probe that, each paired with a check that can be run on real data, where the truth is unknown (Table~\ref{tab:outside}). All use the headline curves and the standard noise model and change one thing at a time.

\paragraph{A window that does not bracket saturation.} This is the failure that matters most in practice and the only one where the method is wrong without any sign of trouble in the fit itself. The plateau is defined as $f(x_{\max})$, so if the response has not saturated by $x_{\max}$ the method reports a plateau, a crossing, and a narrow interval regardless. Cutting the window to $x_{\max}=0.5$~s (where the true Langmuir curve has itself reached only $0.53$) drives the inferred plateau to $0.608$ and the recommended input to $0.46$~s, with a CI about one-twentieth as wide as normal, pooled across the five families, and that input delivers only $0.592$ of the true saturated value. The window-sufficiency statistic $r$ of Sec.~\ref{sec:readout} catches it. The share of the plateau gained over the last doubling of input is $0.057$ on average when the window is generous and $0.335$ at $x_{\max}=0.5$~s, so a cutoff of $0.15$ flags $84\%$ of the badly truncated runs ($99\%$ at $0.10$). The cutoff is not a fixed number, though, because the fit's gradual approach to the plateau inflates $r$ on its own, so the healthy value of $r$ rises as the window tightens. When $x_{\max}$ is swept from $5.0$ down to $2.5$~s, all still bracketing $x_{\rm sat}=2.0$~s, the false-alarm rate at a cutoff of $0.15$ climbs from $2\%$ to $32\%$ (Table~\ref{tab:window}). The cutoff therefore has to change with the window. The released tool prints $r$ together with this calibration, and the practitioner who chooses an efficient, tight window is also the one for whom a fixed cutoff false-alarms most.

\paragraph{A contaminated plateau anchor.} The read-out rests on the measurements at $x_{\max}$, so we corrupted one of them by $+3\sigma_y$. The inferred plateau rises from $0.997$ to $1.049$ and the recommended input from $2.36$ to $3.08$~s. The leverage $\mathrm{d}x_{\rm sat}/\mathrm{d}f(x_{\max})$, which can be computed at run time, is $0.25$~s per percent, and it sets the direction and the order of magnitude of the shift. The realized $+0.72$~s is smaller than the $0.25\times5.2\approx1.3$~s that the leverage applied to the full plateau move would give. The difference is not further dilution, since the $5.2$-percentage-point plateau shift is already the diluted effect of one $+3\sigma_y$ outlier spread across the several $x_{\max}$ readings. It comes from the leverage relation being a local linearization of an ill-conditioned inverse, which overestimates the size here as it does on the large shifts of Table~\ref{tab:si-cond} (Fig.~\ref{fig:readout}b). We also tried the obvious remedy, repeating the anchor measurement outside the active budget (the next row), and could not test it cleanly. Forcing three measurements at $x_{\max}$ changes nothing ($1.048$, $3.10$~s), but only because the loop already places $4.7$ readings there on average of its own accord (Sec.~\ref{sec:acq}). This arm therefore adds two anchors to a design that already has several, and it does not test repetition against a true single-anchor baseline. The null result does show that a monotone fit at its right edge is not rescued by a few more clean readings when one is contaminated, because the fit is pulled toward the largest of them. A likelihood that tolerates outliers is therefore the remedy that helps, whereas more anchor points leave the pull in place. A Student-$t$ likelihood in place of the Gaussian of Eq.~(\ref{eq:likelihood}) is our recommendation for field data (Sec.~\ref{sec:limitations}), and the refit is reported below. A clean test of repetition would keep the acquisition away from $x_{\max}$ to force a true one-anchor arm, which we did not run.

\paragraph{Instrument drift, turned into a detector.} The clustering that Sec.~\ref{sec:acq} counts as redundancy is also unplanned repetition at nearly the same input at different points in the campaign, which is exactly the design one would build to detect drift. Adding a straight-line drift of $2\sigma_y$ over the twenty measurements raises the saturation error to $0.115$, since the drift enters as extra apparent scatter, but leaves the read-out almost unmoved ($x_{\rm sat}=2.27$~s against a baseline $2.36$~s, delivering $0.952$ of true saturation). The drift is partly absorbed into the inferred noise, a benign failure for the recommended input. More usefully, regressing the response difference of every near-duplicate pair on how far apart they were taken in the campaign recovers the drift rate with no extra measurements, $+0.0091$ per measurement against a true $0.0105$, whereas the same estimator returns $-0.0014$ (consistent with zero) on the undrifted baseline. A design feature otherwise counted only as a cost thus provides a drift check at no cost in extra measurements.

\paragraph{An exact duplicate.} The Gaussian likelihood handles a repeated measurement at one input as it is, so a forced duplicate is a non-event, and every reported quantity matches the baseline to within seed scatter. Contrary to a common GP intuition, the acquisition here does not avoid near-duplicates. At $x_{\max}$ the uncertainty $\sigma_\ast$ does not collapse, because $x_{\max}$ is a boundary point whose plateau value depends on every weight, so repeating a measurement there does not tighten it the way it would at an interior point, and this is why the loop keeps returning to $x_{\max}$ (Sec.~\ref{sec:acq}). The case that deserves attention is duplicates that disagree, which is the drift and outlier detector described above.

\begin{table}[htbp]
\centering
\caption{Behavior under four perturbations of the headline curves, pooled over five families and $25$ seeds at $N=20$, against the unperturbed baseline (Sec.~\ref{sec:si-outside}). ``$\theta_{\rm true}$'' is the delivered fraction of the true saturated value at the recommended input, ``$\hat{x}_{\rm sat}$'' the per-run median crossing averaged over the runs, ``CI width'' the width of the nominal $95\%$ CI on $x_{\rm sat}$, ``$r$'' the window-sufficiency statistic, and ``leverage'' $\mathrm{d}x_{\rm sat}/\mathrm{d}f(x_{\max})$ in seconds per percent.}
\label{tab:outside}
\footnotesize
\setlength{\tabcolsep}{4pt}
\begin{tabular}{lccccccc}
\toprule
perturbation & sat.\ err & $f(x_{\max})$ & $\hat{x}_{\rm sat}$ (s) & CI width & $\theta_{\rm true}$ & $r$ & leverage \\
\midrule
none (baseline)          & $0.039$ & $0.997$ & $2.36$ & $3.18$ & $0.955$ & $0.057$ & $0.23$ \\
window truncated to $0.5$~s & $0.146$ & $0.608$ & $0.46$ & $0.16$ & $0.592$ & $0.335$ & $0.01$ \\
plateau $+3\sigma_y$     & $0.042$ & $1.049$ & $3.08$ & $3.22$ & $0.976$ & $0.087$ & $0.25$ \\
\quad anchor replicated  & $0.045$ & $1.048$ & $3.10$ & $3.17$ & $0.974$ & $0.088$ & $0.24$ \\
linear drift $2\sigma_y$ & $0.115$ & $1.095$ & $2.27$ & $3.30$ & $0.952$ & $0.054$ & $0.24$ \\
exact duplicate          & $0.039$ & $0.996$ & $2.39$ & $3.19$ & $0.956$ & $0.058$ & $0.23$ \\
\bottomrule
\end{tabular}
\end{table}

\paragraph{A Student-$t$ likelihood downweights a corrupted anchor.} The corrupted anchor above, a single $x_{\max}$ reading pushed up by $+3\sigma_y$, is the case a heavy-tailed likelihood is meant to absorb. We refit the stored $N=20$ designs of the main campaign with the same corruption, replacing the Gaussian of Eq.~(\ref{eq:likelihood}) by a Student-$t$ with four degrees of freedom,\cite{65} changing nothing else, and measure how far the outlier moves the fit under each (Table~\ref{tab:si-studentt}). The Student-$t$ reduces the outlier's effect on the plateau by about a third and on the read-out by nearly half, holding the inferred plateau to $1.03$ against the Gaussian's $1.05$ (a pull of $+0.038$ from the clean fit instead of $+0.058$) and cutting the read-out shift from $+0.73$ to $+0.42$~s. These are medians of the per-run shifts, which for the read-out differ from the differences between the medians over the runs in Table~\ref{tab:si-studentt} because the crossing posterior is lopsided, with a longer tail toward late inputs. The fix is partial, since the corrupted anchor still pulls the read-out late, but it costs almost nothing when there is no outlier (clean-data saturation error $0.041$ against $0.039$, with plateau and read-out almost unchanged), and no non-monotone draw occurred in any fit.

\begin{table}[htbp]
\centering
\caption{Student-$t$ against Gaussian likelihood under a corrupted plateau anchor (Sec.~\ref{sec:si-outside}), pooled over five families and $50$ seeds. Saturation error and $f(x_{\max})$ (true $1.0$) are pooled means over the $250$ runs, and $x_{\rm sat}$ (true $2.0$~s) is the median over the $250$ runs of each run's median crossing, since the crossing posterior has a longer tail toward late inputs.}
\label{tab:si-studentt}
\footnotesize
\begin{tabular}{llccc}
\toprule
likelihood & data & sat.\ err & $f(x_{\max})$ & $\hat{x}_{\rm sat}$ (s) \\
\midrule
Gaussian    & clean        & $0.039$ & $0.993$ & $2.27$ \\
Gaussian    & $+3\sigma_y$ anchor & $0.039$ & $1.052$ & $3.20$ \\
Student-$t$ ($\nu=4$) & clean        & $0.041$ & $0.992$ & $2.23$ \\
Student-$t$ ($\nu=4$) & $+3\sigma_y$ anchor & $0.039$ & $1.028$ & $2.82$ \\
\bottomrule
\end{tabular}
\end{table}

\section{Notation}
\label{sec:si-notation}
The symbols used in the main text and the SI, several of them in the $\sigma$ and $\tau$ families, are collected here to be read at a glance.

\begin{table}[htbp]
\centering
\caption{Symbols used in this work.}
\label{tab:si-notation}
\footnotesize
\begin{tabular}{lp{0.68\textwidth}}
\toprule
symbol & meaning \\
\midrule
$f(x)$                 & the saturating response, normalized to plateau $A=1$ \\
$A$                    & saturated (plateau) value; $=1$ throughout \\
$x_{\min},x_{\max}$    & measurement window, $0.010$ and $5.0$~s \\
$P$                    & number of basis functions ($=25$) \\
$k,\;d$                & spline order ($=4$) and penalty difference order ($=2$) \\
$M_m,\;I_m$            & M-spline (rate block) and I-spline (curve block) $m$ \\
$t_m$                  & knot sequence defining the block windows (Sec.~\ref{sec:si-blocks}) \\
$w_m$                  & non-negative weight on $I_m$ \\
$\tau_\eta$            & roughness (random-walk step) scale on the log-weights, inferred \\
$\tau_w$               & weight scale of the unpenalized monotone basis (Sec.~\ref{sec:si-noise}) \\
$s_\tau,\;s_\sigma$    & fixed prior scales on $\tau_\eta$ and $\sigma$, both $0.3$ \\
$\sigma$               & inferred observation-noise scale (one per fit) \\
$\sigma_x,\;\sigma_y$  & generative input jitter and output noise, $0.01$~s and $0.1$ \\
$\mu_\ast(x),\;\sigma_\ast(x)$ & posterior mean and standard deviation of $f$ at $x$ \\
$S$                    & posterior draws per fit ($=500$) \\
$N,\;N_0,\;N_a$        & total, survey, and active budget ($20$, $5$, $15$) \\
$R$                    & Monte Carlo seeds per condition ($=50$ unless stated; distinct from the risk $R(x)$ below) \\
$\alpha$               & saturation fraction; $x_\alpha$ its true crossing \\
$x_{\rm sat}(\alpha),\;\hat{x}_{\rm sat}(\alpha)$ & saturation input, the smallest $x$ at which the response reaches the fraction $\alpha$ of its plateau ($\alpha A$ on the true curve, $\alpha f^{(s)}(x_{\max})$ on a posterior draw), and its median read-out, also written $\hat{x}_\alpha$; $x_{\rm sat}$ alone means $x_{\rm sat}(0.95)$, $2.0$~s on the headline curves \\
$\Delta y(\alpha),\;\Delta x(\alpha),\;\Delta A$ & level-wise response and input error at level $\alpha$, and the plateau misestimate $\mu_\ast(x_{\max})-A$ \\
$\hat{x}_q,\;q$        & cost-indexed read-out (the $q$-th crossing quantile) and its cost-ratio index \\
$\theta(x)$            & attained fraction $f(x)/f(x_{\max})$ \\
$\theta_{\rm true}(x)$   & delivered fraction of the true plateau, $f(x)/A$ (validation only) \\
$R(x)$                 & under-saturation risk $\Pr(\theta(x)<\alpha)$ \\
$r$                    & window-sufficiency statistic, the plateau share gained over the last doubling of input (Sec.~\ref{sec:readout}) \\
$\lambda$              & P-spline smoothing, $\lambda=\sigma^2/\tau_\eta^2$ (Sec.~\ref{sec:si-pspline}) \\
$\xi$                  & free power-law rate exponent in the penalty null space (Sec.~\ref{sec:si-pspline}) \\
$\kappa$               & ratio of consecutive interior knots ($\approx1.33$) (Sec.~\ref{sec:si-pspline}) \\
$c$                    & cost ratio setting the read-out quantile, $q=c/(1+c)$ (Sec.~\ref{sec:si-quantile}) \\
$\rho_s$               & Spearman rank correlation of band width with realized error (Sec.~\ref{sec:si-window}) \\
$\hat{R},\;\mathrm{ESS}$ & MCMC convergence diagnostics (Sec.~\ref{sec:si-diag}) \\
$\tau,K,\tau_S,n,\gamma,\tau_E,\tau_p,\beta$ & kinetic parameters of Eqs.~(\ref{eq:iso_langmuir})--(\ref{eq:iso_kww}); $K$ is reused in Eq.~(\ref{eq:gen_mix}) for the number of mixture components \\
\bottomrule
\end{tabular}
\end{table}

\section{Computational details and reproducibility}
\label{sec:si-compute}
This section records what is needed to reproduce the results, namely the software and hardware, the configuration of the GP arms, the inventory of campaigns, the sampler settings, the noise model, the seed protocol, the scales the data return, and a map from each reported result to the command and cache file behind it.

\paragraph{Software and hardware.} The implementation is pure Python, with NumPy\cite{72} and SciPy\cite{73} building the spline basis and evaluating the read-out, and JAX with NumPyro\cite{53} running the sampler. The exact package versions used for the published results are recorded with the released code. All results were produced on CPU (no GPU), on a six-core desktop processor at $2.90$~GHz. Monte Carlo runs are distributed over worker processes, each pinned to a single compute thread. Every run is independently seeded, so results do not depend on the worker count or on completion order.

\paragraph{Configuration of the GP arms.} Because Sec.~\ref{sec:whymono} rests on those arms being fairly configured, their settings are given in the same detail as the sampler's. All three use a constant mean and a radial-basis-function kernel, with the empirical mean of the observations as the constant (for the pinned arm that mean includes the pseudo-observation at the origin). The lengthscale $\ell$, amplitude $s_f$, and noise $s_n$ are re-estimated at every step by maximizing the exact log marginal likelihood,\cite{19} evaluated through a Cholesky factorization with a $10^{-10}$ jitter, over a discrete grid instead of by gradient ascent, namely $26$ logarithmically spaced lengthscales spanning $1/400$ to $2$ times the width of the (warped) domain, eight amplitudes from $0.2$ to $3.5$, and eight noise scales from $0.01$ to $0.35$, giving $1664$ evaluations per fit. We use a grid because optimizing three marginal-likelihood parameters on as few as five points is a known failure mode,\cite{19} prone to degenerate optima at zero noise or at a domain-scale lengthscale depending on the starting point, and an exhaustive grid removes starting-point and local-optimum artifacts from the comparison entirely. The grid is verified not to bind, since with this range the optimum is interior in $100.0\%$ of GP-log and GP-log-pin fits and $81.6\%$ of GP-linear fits, and the remaining GP-linear cases are genuine short-lengthscale (noise-chasing) optima instead of artifacts of truncation. A narrower range, $\ell\in[0.1,3.0]$, would pin the Elovich optimum at the ceiling in $56\%$ of fits, which is why the range is as wide as it is. For the pinned arm the pseudo-observation at $(0,0)$ carries a fixed noise variance of $10^{-8}$ while the real observations carry $s_n^2$. Where a read-out is taken off a GP arm (in Table~\ref{tab:readout} and in the projection control of Sec.~\ref{sec:si-gpfail}), it uses $500$ exact draws from this closed-form posterior, matching the $500$ MCMC draws behind the monotone arm's read-outs, so every quoted quantile summarizes the same number of posterior draws. Widening the grid leaves the linear-axis arm's pooled error and non-monotone-mean rate essentially unchanged, so its conclusions do not depend on the search range and are due to the choice of axis.

\paragraph{Campaign inventory.} The results in this paper come from the following runs, all at $N_0=5$ and $N=20$ unless stated. The fixed-location comparison of Figs.~\ref{fig:langmuir}--\ref{fig:kww} ran 5 families $\times$ 4 strategies $\times$ 50 seeds. Two randomized-location campaigns (log-uniform and uniform $x_{\rm sat}$) ran the same dimensions. The candidate-set control, the noise-free regime sweeps (15 variants), and the prior-sensitivity refits (5 families $\times$ 3 budgets $\times$ 5 seeds $\times$ 9 prior configurations) ran at the same budgets. The read-out analysis refits the stored $N=20$ designs to recover posterior draws (250 fits, verified to reproduce the cached posterior means bitwise). The survey-fraction sweep ran $N_0\in\{3,8\}$ for 5 families $\times$ 50 seeds. The output-noise sweep ran $\sigma_y\in\{0.01,0.05,0.1,0.15\}$ for 5 families $\times$ 50 seeds, with the $\sigma_y=0.1$ arm being the main campaign, together with the penalty-decomposition arms over $\sigma_y\in\{0.05,0.1,0.15\}$ (the unpenalized monotone basis at $P=11$ and $P=25$, and the origin-matched GP-log-pin). The held-out benchmark fit 250 generated curves. The matched-design control fitted the three GP arms to the $1000$ stored designs of the main campaign and the two log-axis arms to the $750$ stored I-spline designs of the noise sweep and the $500$ of the held-out benchmark, and re-ran the three arms' own loops ($750$ runs), all without MCMC, and the two unpenalized bases were refit on the $1000$ stored designs of the main campaign ($2000$ MCMC fits) for the scorecard figure. The I-spline arms of the noise sweep at $\sigma_y=0.05$ and $0.15$ and both held-out benchmarks were re-run to store their designs and reproduced the cached errors exactly. The behavior-outside-assumptions arms of Sec.~\ref{sec:si-outside} ran a baseline plus the four perturbations and the replicated-anchor variant of Table~\ref{tab:outside}, each over 5 families $\times$ $25$ seeds, and the noise-mechanism diagnostic ran 3 conditions $\times$ 5 families $\times$ $20$ seeds. The recommended-input and window sweeps of Secs.~\ref{sec:calibration} and \ref{sec:limitations}, and the three GP arms (saturation error and the $x_{\rm sat}$ read-out), which use no MCMC, complete the list. The total is roughly five thousand active-learning runs and eighty thousand posterior fits.

\paragraph{Sampler settings.} NUTS\cite{52} runs with 500 warmup and 500 retained draws per fit, one chain in production and four for the diagnostics of Sec.~\ref{sec:si-diag}, at the NumPyro defaults, namely a target acceptance probability of $0.8$, a maximum tree depth of $10$, a diagonal mass matrix adapted during warmup, and a uniform starting point. The roughness-penalty prior is written in non-centered form (a standard-normal step scaled by the inferred $\tau_\eta$), which tames the funnel shape of the log-weight random walk, though a high divergence rate remains (Sec.~\ref{sec:si-diag}). A single fit takes about $2.5$~s for the penalized $P=25$ basis. The full campaign, listed above, runs in about a day on this machine.

\paragraph{Noise model.} The noise model is Eq.~(\ref{eq:noise}) of the main text. A jittered input that falls past either edge of the window is clamped to that edge instead of redrawn, which matters mainly for the survey point at $x_{\min}$ and the readings at $x_{\max}$, where about half of the jitters are clamped, and the recorded input is the commanded one, so the surrogate sees the jitter only as extra scatter, largest where the curve is steep (Sec.~\ref{sec:si-window}).

\paragraph{Seeds.} A single integer seed per run fixes both the data and the sampler. It seeds the NumPy generator that draws the survey design and every noise realization, and it is used as the JAX random-number key for NUTS. The four acquisition strategies are run at the same set of seeds, and in the randomized-location studies the saturation location $x_{\rm sat}$ is drawn from an independent generator keyed on the seed alone, so at a given seed all four strategies and all five families see the same $x_{\rm sat}$. The three survey-based strategies (uncertainty sampling, uniform random, log-uniform random) additionally consume the same noise stream. Each draws the same five-point survey and observes it before acquiring, so their survey data are identical and their later draws come from the same stream. The log-spaced grid arm instead rebuilds its design at every budget, as a practitioner would, so it shares the seed but draws its own realizations. Per-seed differences between strategies are therefore paired on the seed, and for the three survey-based arms on the data as well, and they are analyzed as paired differences in Sec.~\ref{sec:acq_compare}.

\paragraph{Inferred scales.} The two scales of Eq.~(\ref{eq:priors}) are sampled as free parameters, and Table~\ref{tab:si-sigma} reports their posteriors at $N=20$. In every family the inferred noise scale $\sigma$ is a little above the generative output noise $\sigma_y=0.1$, by $4$ to $10\%$, because the single noise term absorbs everything the mean curve cannot represent, namely the input jitter and the basis's own misfit. The calibration analysis reaches the same conclusion by another route (Sec.~\ref{sec:calibration}), and the reported credible bands are already somewhat widened for it. Unlike $\sigma$, the inferred roughness scale $\tau_\eta$ has a posterior mean close to the prior median in every family, highest on the structured Sips onset and lowest on the KWW and Michaelis--Menten curves, and its posterior s.d.\ within a single fit is between $0.4$ and $0.7$ of that mean. Twenty points therefore leave the roughness only partly determined and the prior scale $s_\tau$ fills in the rest, which is why $s_\tau$ is the one prior scale with measurable pull at this budget (Sec.~\ref{sec:si-priors}). Section~\ref{sec:si-pspline} gives the reading of $s_\tau$ on the knot grid.

\begin{table}[htbp]
\centering
\caption{Posterior means of the two inferred scales of Eq.~(\ref{eq:priors}) at $N=20$ under uncertainty sampling, as mean $\pm$ one s.d.\ over ten seeds. The excess is over the generative output noise $\sigma_y=0.1$, the roughness scale is to be read against its prior median of $0.202$, and the last column is the posterior s.d.\ of $\tau_\eta$ within one fit, averaged over the same seeds.}
\label{tab:si-sigma}
\begin{tabular}{lcccc}
\toprule
family & posterior $\sigma$ & excess over $\sigma_y$ & posterior $\tau_\eta$ & within-fit s.d.\ of $\tau_\eta$ \\
\midrule
Langmuir          & $0.104\pm0.015$ & $+4\%$ & $0.228\pm0.054$ & $0.110$ \\
Michaelis--Menten & $0.108\pm0.015$ & $+8\%$ & $0.177\pm0.038$ & $0.105$ \\
Sips ($n=2$) & $0.107\pm0.016$ & $+7\%$ & $0.337\pm0.040$ & $0.137$ \\
Elovich           & $0.107\pm0.017$ & $+7\%$ & $0.251\pm0.066$ & $0.116$ \\
KWW               & $0.110\pm0.021$ & $+10\%$ & $0.161\pm0.055$ & $0.109$ \\
\bottomrule
\end{tabular}
\end{table}

\paragraph{Where each result comes from.} Every number in the main text and the SI is computed from a cached result file in the released repository, and every cache is written by one subcommand of one script. The reader-facing entry point is \texttt{recompute\_stats.py}, which recomputes every quoted number, table entry, ratio, and percentage from the caches, section by section in the order the paper uses them, in about ten seconds and without re-running any campaign. Running \texttt{reproduce.py} regenerates the caches themselves, running the campaigns, the figures, and the tests in dependency order, and \texttt{make\_figures.py} draws every figure from the caches. Table~\ref{tab:si-provenance} gives the map. The two change-of-measure controls reuse a command with an environment variable that switches the draw of $x_{\rm sat}$ from log-uniform to uniform, as noted in the table, and Figs.~\ref{fig:gp_compare} and~\ref{fig:basis} need no cache, since one runs a short single-seed illustration loop at render time and the other draws the basis functions directly.

\begin{table}[htbp]
\centering
\caption{Where each result comes from. Each row gives the command that produced the cache and the cache file, all in the \texttt{code/} folder of the released repository, with caches in \texttt{code/data/}. The \texttt{.json} suffix is dropped from the cache names.}
\label{tab:si-provenance}
\footnotesize
\setlength{\tabcolsep}{4pt}
\begin{tabular}{@{}p{0.28\textwidth}p{0.36\textwidth}p{0.32\textwidth}@{}}
\toprule
result & command & cache \\
\midrule
Table~\ref{tab:isotherms}, Fig.~\ref{fig:final_fit}, Figs.~\ref{fig:langmuir}--\ref{fig:kww}, the noise-free sweeps, the learning-curve medians & \texttt{recovery.py main} & \texttt{results} \\[3pt]
Fig.~\ref{fig:knots} & \texttt{design\_choices.py knots} & \texttt{results\_knots} \\[3pt]
Table~\ref{tab:gp}, Table~\ref{tab:gpfail} (GP columns) & \texttt{gp\_arms.py}, \texttt{recovery.py unpen25}, \texttt{recovery.py ablation} & \texttt{results\_gparms}, \texttt{results\_unpen25}, \texttt{results\_unpen11} \\[3pt]
Fig.~\ref{fig:noise}, Table~\ref{tab:si-noise} & \texttt{recovery.py noise}, \texttt{recovery.py ablation}, \texttt{recovery.py unpen25}, \texttt{recovery.py gppin} & \texttt{results\_noise}, \texttt{results\_unpen11},\newline \texttt{results\_unpen25}, \texttt{results\_gppin\_noise} \\[3pt]
Table~\ref{tab:readout}, Fig.~\ref{fig:readout}, Table~\ref{tab:si-cond}, the read-out costs of Sec.~\ref{sec:si-quantile} & \texttt{readout\_study.py readout}, \texttt{readout\_study.py gpxsat} & \texttt{results\_readout}, \texttt{results\_gpxsat} \\[3pt]
Fig.~\ref{fig:scorecard} & \texttt{gp\_arms.py matched}, \texttt{refits.py unpen}, with the caches of Table~\ref{tab:gp} & \texttt{results\_gp\_matched}, \texttt{results\_unpen\_matched} \\[3pt]
Tables~\ref{tab:si-gpmatched}--\ref{tab:si-gpmatched-sweeps} & \texttt{gp\_arms.py matched} & \texttt{results\_gp\_matched} \\[3pt]
Table~\ref{tab:heldout}, Sec.~\ref{sec:si-heldout} & \texttt{generalization.py heldout} (again with \texttt{HELDOUT\_XSAT=uniform}),\newline \texttt{generalization.py heldoutgppin} & \texttt{results\_heldout},\newline \texttt{results\_heldout\_uniform},\newline \texttt{results\_heldout\_gppin},\newline \texttt{results\_heldout\_gppin\_uniform} \\[3pt]
Table~\ref{tab:randxsat}, the uniform control of Sec.~\ref{sec:si-controls} & \texttt{generalization.py randxsat} (again with \texttt{RANDXSAT\_XDIST=uniform}) & \texttt{results\_randxsat}, \texttt{results\_randxsat\_uniform} \\[3pt]
Candidate-set, survey-size, and unconstrained-driven controls (Sec.~\ref{sec:si-controls}) & \texttt{design\_choices.py acqgrid}, \texttt{generalization.py survey}, \texttt{controls.py decoupled} & \texttt{results\_acqgrid}, \texttt{results\_survey}, \texttt{results\_decoupled} \\[3pt]
Table~\ref{tab:si-model-ablation} & \texttt{refits.py penalty} & \texttt{results\_model\_ablation} \\[3pt]
Table~\ref{tab:si-diag} & \texttt{design\_choices.py diag} & \texttt{results\_diag} \\[3pt]
Table~\ref{tab:si-sampler}, the paired sampler comparison of Sec.~\ref{sec:si-diag} & \texttt{design\_choices.py targets}, \texttt{design\_choices.py sampler} & \texttt{results\_target\_accept}, \texttt{results\_sampler\_settings} \\[3pt]
Table~\ref{tab:si-priors} & \texttt{design\_choices.py priors} & \texttt{results\_priors} \\[3pt]
Table~\ref{tab:window} & \texttt{robustness.py window} & \texttt{results\_window} \\[3pt]
Band coverage and localization (Sec.~\ref{sec:si-window}), Table~\ref{tab:si-sigma} & \texttt{controls.py band} & \texttt{results\_band} \\[3pt]
Post-hoc monotonization (Sec.~\ref{sec:si-gpfail}) & \texttt{controls.py monoproj} & \texttt{results\_mono\_projection} \\[3pt]
Table~\ref{tab:si-parametric} & \texttt{refits.py parametric} & \texttt{results\_parametric} \\[3pt]
Mechanism diagnostics (Sec.~\ref{sec:si-noise}) & \texttt{robustness.py anchor},\newline \texttt{robustness.py anchor --model=unpen11} & \texttt{results\_anchor}, \texttt{results\_anchor\_unpen11} \\[3pt]
Table~\ref{tab:outside} & \texttt{robustness.py failure} & \texttt{results\_failure} \\[3pt]
Table~\ref{tab:si-studentt} & \texttt{refits.py studentt} & \texttt{results\_studentt} \\[3pt]
Figs.~\ref{fig:gp_compare} and~\ref{fig:basis} & \texttt{make\_figures.py gp\_compare}, \texttt{make\_figures.py basis} & none \\
\bottomrule
\end{tabular}
\end{table}

\end{document}